\documentclass[12pt]{article}

\pdfoutput=1

\usepackage{cite}
\usepackage[dvipsnames]{xcolor}
\usepackage{amssymb,amsmath,amscd}
\usepackage{epsfig}
\usepackage{epstopdf}
\usepackage{latexsym}
\usepackage{graphicx}
\usepackage{booktabs}
\usepackage{bbm}
\usepackage[margin=15pt,small]{caption}
\usepackage{subcaption}
\usepackage{enumitem}
\usepackage{float}
\usepackage[toc]{appendix}
\usepackage[numbers,compress]{natbib}
\usepackage{tikz}
\usepackage[labelfont=bf]{caption}
\usepackage[export]{adjustbox}
\usepackage{braket,cellspace}
\usepackage{mathrsfs}
\usepackage{braket}
\usepackage{chngpage}
\usepackage{tabu}
\usepackage{mathtools}

\usetikzlibrary{positioning}
\definecolor{darkspringgreen}{rgb}{0.09, 0.45, 0.27}

\usepackage{datetime}
\usepackage{hyperref}
\hypersetup{
      colorlinks=true,
      linkcolor=darkspringgreen,  
      urlcolor=darkspringgreen,    
      filecolor=darkspringgreen,     
      citecolor=darkspringgreen,
      linktocpage=true,
      pdfstartview=FitV,
      bookmarksopen=true    
      }

\setlist{nolistsep}
\ifpdf
\fi

\let\oldbibliography\thebibliography
\renewcommand{\thebibliography}[1]{\oldbibliography{#1}
\setlength{\itemsep}{0pt}} 

\numberwithin{equation}{section} 

\begin{document}  

\begin{titlepage}

\begin{center} 

\vspace*{23mm}

{\LARGE \bf 
Achronal averaged null energy condition 
\\[7pt]
for extremal horizons and (A)dS}

\bigskip
\bigskip
\bigskip
\bigskip

{\bf Felipe Rosso\\ }
\bigskip
Department of Physics and Astronomy \\
University of Southern California \\
Los Angeles, CA 90089, USA  \\
\bigskip
\tt{felipero@usc.edu}  \\
\end{center}

\bigskip

\begin{abstract}
\noindent We prove the achronal averaged null energy condition for general quantum field theories in the near horizon geometry of spherical extremal black holes (\textit{i.e.}~${{\rm AdS}_2\times S^{d-2}}$), de Sitter and anti-de Sitter. The derivation follows from monotonicity of relative entropy after computing the modular hamiltonian of a null deformed region. For incomplete (but maximally extended) achronal null geodesics in ${\rm AdS}_2\times S^{d-2}$, we prove the positivity of a different light-ray operator for arbitrary CFTs. This agrees with a constraint recently derived for the Lorentzian cylinder. 
\end{abstract}

\vfill

\end{titlepage}


\newpage

\setcounter{tocdepth}{3}
\tableofcontents

\section{Introduction}

Given a particular matter distribution characterized by the stress tensor $T_{\mu \nu}$, Einstein's gravitational equations determine the space-time metric $g_{\mu \nu}$. Early on it was realized that this logic can be inverted, meaning that for \textit{any} smooth metric, the equations of motion determine the appropriate matter required to produce it. This is problematic, given that there are plenty of unphysical space-times (\textit{e.g.} metrics with closed time-like curves) that we do not want to have as solutions of our theory. As a result, gravitational theories must be supplemented with ``reasonable" constraints (referred as Classical Energy Conditions \cite{Fewster:2012yh}) on the allowed matter distributions. Arguably the most interesting of these constraints is the Null Energy Condition, given by $T_{\mu \nu}k^\mu k^\nu\ge 0$ with $k^\mu$ null. Restricting to matter satisfying the Null Energy Condition has led to important results in classical gravity, most notably Hawking area \cite{Hawking:1971vc} and Penrose singularity theorems \cite{Penrose:1964wq} (see \cite{Witten:2019qhl} for a review).

The situation gets more interesting in the context of semiclassical gravity, where the classical stress tensor $T_{\mu \nu}$ is replaced by the expectation value of the QFT operator $\langle T_{\mu \nu} \rangle$. An old result due to Epstein, Glaser and Jaffe \cite{Epstein:1965zza} shows that local operators in a QFT satisfying Wightman axioms cannot be positive definite, implying that \textit{all} Classical Energy Conditions are violated as soon as we incorporate quantum effects.\footnote{See appendix A of \cite{Cordova:2017zej} for a modern presentation of the argument.} Instead of being discouraged by this general result we should become excited, as it opens up a small window into the realm of quantum gravity. The violation of the Null Energy Condition disqualifies Hawking's area theorem and allows black holes to evaporate \cite{Hawking:1974sw}, giving rise to the black hole information paradox \cite{Hawking:1976ra,Mathur:2009hf}, without question the most intriguing result in quantum gravity. 

The spectacular consequences of the violation of the Null Energy Condition, makes essential the search of a replacement which applies in the presence of quantum correlations. In doing so we must keep in mind the result of \cite{Epstein:1965zza}, which implies
$${\rm locality}\,\,+\,\,
  {\rm positivity}\,\,\neq
  \,\,{\rm QFT}\ .$$
A quantum generalization of the Null Energy Condition can retain either locality or positivity, but not both. The Quantum Null Energy Condition is a proposal which sacrifies positivity in favor of a bound on the local stress tensor, see \cite{Bousso:2015mna,Koeller:2015qmn,Balakrishnan:2017bjg,Ceyhan:2018zfg,Leichenauer:2018obf,Balakrishnan:2019gxl}. The Averaged Null Energy Condition takes the alternative route, its most refined version conjectured by Graham and Olum \cite{Graham:2007va} and stated as follows.\footnote{See \cite{Freivogel:2018gxj,Leichenauer:2018tnq} for a proposal of a non-local bound along null geodesics that is different from the achronal ANEC.}

\paragraph{Achronal Averaged Null Energy Condition:} Every complete achronal null geodesic on a self-consistent\footnote{See \cite{Graham:2007va} for a discussion regarding the self-consistency condition.} solution in semiclassical gravity satisfies
\begin{equation}\label{eq:147}
\int_{-\infty}^{+\infty}d\lambda\,
  T_{\lambda \lambda}
  \ge 0\ ,
\end{equation}
where $\lambda$ is an affine parameter and $T_{\lambda \lambda}\equiv k^\mu k^\nu T_{\mu \nu}$, with $k^\mu$ the geodesic tangent vector. 

\vspace{13pt}

Integrating the Null Energy Condition along a complete null geodesic is the most natural way of obtaining a non-local operator. A complete achronal null geodesic is one for which the affine parameter $\lambda$ can be extended to all real values and cannot be intersected twice by any time-like path. The achronality condition is imposed to avoid a trivial violation due to a negative Casmir energy contribution in space-times with a compact spatial direction, \textit{e.g.} $\mathbb{R}\times S^{d-1}$. While this might appear as an innocent addition to (\ref{eq:147}), it turns out to be highly restrictive, given that complete achronal null geodesics are extremely special and non-generic (see section 8 of \cite{Witten:2019qhl}). The main known examples of space-times with complete achronal null geodesics are: Minkowski and (A)dS, where all complete null geodesics are achronal, and geodesics along the horizon of typical black hole solutions. Let us stress that we are \textit{not} implying these to be the only cases with complete achronal null geodesics, they simply correspond to the main known examples. It is for QFTs defined on these space-times that there is hope we might be able to explicitly prove the achronal ANEC for arbitrary QFTs, as recently done for Minkowski \cite{Faulkner:2016mzt,Hartman:2016lgu,Longo:2018obd}.\footnote{See \cite{Wald:1991xn,Fewster:2006uf,Wall:2009wi,Wall:2011hj,Kontou:2012ve,Kontou:2015yha,Rosso:2019txh} for previous work on the ANEC in curved space-times.} 

This is the observation motivating this paper. We show it is possible to adapt the flat space calculations of \cite{Faulkner:2016mzt} and prove the achronal ANEC for arbitrary QFTs in de Sitter, anti-de Sitter and ${\rm AdS}_2\times S^{d-2}$. For \textit{incomplete} (but maximally extended\footnote{By this we mean that if the geodesic were to be extended any further, the achronality condition would be violated.}) achronal null geodesics in ${\rm AdS}_2\times S^{d-2}$ and arbitrary CFTs, we prove a bound similar to (\ref{eq:147}), which coincides with a recent result in $\mathbb{R}\times S^{d-1}$ derived in \cite{Rosso:2019txh} (see (\ref{eq:148}) below). We hope this provides solid evidence in favor of the achronal ANEC as a true statement of semiclassical gravity. The achronal ANEC has been used as a hypothesis to derive a variety of important results: topological censorship \cite{Friedman:1993ty}, conformal collider bounds \cite{Hofman:2008ar} and related results in gravity \cite{Tipler:1978zz,Borde:1987qr} and quantum theories \cite{Hofman:2009ug,Chowdhury:2012km,Cordova:2017zej,Cordova:2017dhq,Delacretaz:2018cfk,Belin:2019mnx}.

\subsection{Summary of results}

We start in \textbf{section \ref{sec:2}} by considering the achronal ANEC for an arbitrary QFT in the fixed space-time ${\rm AdS}_2\times S^{d-2}$. There are several reasons this background geometry is an interesting setup to study the ANEC: it arises from the near horizon limit of extremal black holes, plays a central role in the transversable wormhole constructed in \cite{Maldacena:2018gjk} and contains both achronal and chronal null geodesics. This provides a simple yet rich enough setup to fully test the achronality condition proposed in \cite{Graham:2007va}. We give three independent proofs of the achronal ANEC in ${\rm AdS}_2\times S^{d-2}$ with distinct regimes of validity: for general QFTs, general CFTs and free scalars in subsections \ref{sub:2.2}, \ref{sub:2.3} and \ref{sub:2.4} respectively. Let us briefly review the most salient features of each derivation.

The proof for general QFTs in subsection \ref{sub:2.2} follows the same approach used in~\cite{Faulkner:2016mzt} to prove the ANEC in Minkowski. More precisely, we consider a suitable half-space $A_0$ in ${\rm AdS}_2\times S^{d-2}$ (see (\ref{eq:87}) and (\ref{eq:88})) and compute the vacuum modular hamiltonian associated to a null deformation of $A_0$, parametrized by a vector $\zeta^\mu=\zeta^-\delta^\mu_-$. We show that the full vacuum modular hamiltonian $\hat{K}_A\equiv K_A-K_{A^c}$ to first order in $\zeta^\mu$ is given by
\begin{equation}\label{eq:146}
\hat{K}_A=\hat{K}_{A_0}+
  \pi r_h^{d-2}
  \int_{S^{d-2}}d\Omega(\vec{x}_\perp)
  \zeta^-(\vec{x}_\perp)
  \mathcal{E}(\vec{x}_\perp)+
  \mathcal{O}(\zeta^\mu)^2\ ,
\end{equation}
where $\vec{x}_\perp \in \mathbb{R}^{d-2}$ parametrizes the unit sphere $S^{d-2}$ of radius $r_h$. We have defined $\mathcal{E}(\vec{x}_\perp)$ as the ANEC operator
\begin{equation}\label{eq:149}
\mathcal{E}(\vec{x}_\perp)\equiv
  \int_{-\infty}^{+\infty}
  d\lambda\,
  T_{\lambda \lambda}
  (\lambda,\vec{x}_\perp)\ .
\end{equation}
The integral is over the only complete achronal null geodesics in ${\rm AdS}_2\times S^{d-2}$, which travel between the two boundaries of ${\rm AdS}_2$ at a fixed coordinate in $S^{d-2}$. From this result, a simple calculation using monotonicity of relative entropy \cite{Blanco:2013lea,Faulkner:2016mzt} gives the ANEC $\mathcal{E}(\vec{x}_\perp)\ge 0$.

In subsection \ref{sub:2.3} we apply the methods used in \cite{Rosso:2019txh} and independently derive the positivity of $\mathcal{E}(\vec{x}_\perp)$ for CFTs. The basic idea is to start from the null deformed modular hamiltonian for the Minkowski half-space (computed in \cite{Casini:2017roe,Balakrishnan:2020lbp} to all orders in the deformation) and apply a conformal transformation to ${\rm AdS}_2\times S^{d-2}$ (using a recent observation made in \cite{Galloway:2020xfz}). In this way we derive (\ref{eq:146}) for any CFT, to every order in $\zeta^\mu$, \textit{i.e.} we can drop $\mathcal{O}(\zeta^\mu)^2$ from (\ref{eq:146}). Monotonicity of relative entropy again implies the ANEC, in this case for arbitrary CFTs.

We finish in subsection \ref{sub:2.4} with our final proof of the achronal ANEC in ${\rm AdS}_2\times S^{d-2}$ for a free scalar field non-minimally coupled to the metric. We show $\mathcal{E}(\vec{x}_\perp)$ is positive by writing it as $\mathcal{E}(\vec{x}_\perp)=WW^\dagger\ge 0$ for some operator $W$. The explicit proof in this simple setup provides a sanity check for the previous more general and abstract derivations.

\textbf{Section \ref{sec:3}} gives a proof of the achronal ANEC for arbitrary QFTs in de Sitter and anti-de Sitter, establishing the ANEC for general QFTs in Lorentzian maximally symmetric space-times. This extends the recent derivation for conformal theories in (A)dS given in \cite{Rosso:2019txh}. The proofs follow after computing the vacuum modular hamiltonian associated to a null deformed region to first order in the deformation parameter and obtaining a result analogous to (\ref{eq:146}) (see~(\ref{eq:158}) and (\ref{eq:159})). In both cases the details of the computations are very similar to those leading to~(\ref{eq:146}) in ${\rm AdS}_2\times S^{d-2}$. For de Sitter, the modular hamiltonian coincides with the one obtained in (1.3) of \cite{Rosso:2019txh} for CFTs, to all orders in the deformation parameter.

In \textbf{section \ref{sec:4}} we derive a constraint for more general null geodesics in ${\rm AdS}_2\times S^{d-2}$, that are achronal but not complete. Starting from the ANEC in Minkowski and applying a conformal transformation we show that a large family of geodesics satisfy the following condition
\begin{equation}\label{eq:148}
\int_{-\pi/2}^{\pi/2}
d\lambda\,
\cos^d(\lambda)
\big(
T_{\lambda \lambda}-
\langle T_{\lambda \lambda} \rangle_0
\big)\ge 0\ ,
\end{equation}
where $\langle T_{\lambda\lambda} \rangle_0$ is the vacuum expectation value of the stress tensor. This ensures the inequality is not violated by a negative Casimir contribution. The null geodesics with affine parameter~$\lambda$ are incomplete, travel between the antipodal points of the $S^{d-2}$ and can have non-trivial motion along all the directions in ${\rm AdS}_2\times S^{d-2}$. These geodesics are achronal and maximally extended, meaning that going beyond $|\lambda|=\pi/2$ would result in a chronal curve. See figure \ref{fig:3} for a plot of several trajectories in the coordinates describing the ${\rm AdS}_2$ factor. The condition in~(\ref{eq:148}), which holds for arbitrary CFTs, is essentially equivalent to the one derived in~\cite{Rosso:2019txh} for the Lorentzian cylinder $\mathbb{R}\times S^{d-1}$.\footnote{See also \cite{Iizuka:2019ezn} for an independent derivation of the constraint in $\mathbb{R}\times S^{d-1}$ for holographic CFTs in $d=3,4,5$.}

The appendices contain some details and extensions not included in the main text. In particular, \textbf{appendix \ref{zapp:1}} provides a generalization of the conformal transformation recently noted in~\cite{Galloway:2020xfz}, which relates $\mathbb{R}\times S^{d-1}\rightarrow {\rm AdS}_2\times S^{d-2}$. We show it can be extended to the following conformal transformation
$$\mathbb{R}\times \Sigma^{(k)}
  \qquad \longrightarrow \qquad
  {\rm AdS}_n^{(k)}\times S^{d-n}\ ,$$
for $n\ge 2$, where $\Sigma^{(k)}$ is $\mathbb{R}^{d-1}$, $S^{d-1}$ or $\mathbb{H}^{d-1}$ for $k=0,\pm 1$ respectively. Accordingly, the slicing of the ${\rm AdS}_n^{(k)}$ factor is given in terms of these spaces. Since ${\rm AdS}_n\times S^{d-n}$ space-times appear a lot in the AdS/CFT correspondence, we suspect this transformation might be useful to simplify bulk calculations which incorporate quantum effects.

\subsection{Future directions}

There are several interesting future directions that might be worth pursuing in future work. As briefly mentioned in the introduction, there is an alternative proposal to the achronal ANEC, referred as the Quantum Null Energy Condition (QNEC), that gives a bound on the local stress tensor. This constraint has been proven for general QFTs in Minkowski in~\cite{Balakrishnan:2017bjg,Ceyhan:2018zfg}. The derivation of \cite{Ceyhan:2018zfg} is particularly interesting since it shows the suprising fact that the QNEC and the ANEC are implied by each other in Minkowski. Having established the ANEC for general QFTs in Lorentzian maximally symmetric space-times, this raises the question whether a general proof of the QNEC in (A)dS might be possible using the approach of~\cite{Ceyhan:2018zfg}.

In this work we have computed the vacuum modular hamiltonian associated to the null deformed half-space in de Sitter (\textit{i.e.} the dS static patch) to first order in the deformation. The final result agrees with the computation of \cite{Rosso:2019txh}, performed for CFTs to every order in the deformation parameter. It would be interesting to generalize the computation presented in subsection \ref{sub:2.2} to arbitrary orders and analyze whether the series can be resumed to give the result in \cite{Rosso:2019txh} for CFTs. For null deformation of the Minkowski half-space, this calculation was recently performed in \cite{Balakrishnan:2020lbp}.

Finally, it would be interesting to pursue the study of more general bounds on null geodesics that are not complete and achronal. To start, we could determine whether the bound~(\ref{eq:148}) for $\mathbb{R}\times S^{d-1}$ and ${\rm AdS}_2\times S^{d-2}$ applies to QFTs that are not conformal (see footnote \ref{foot:1}). This can be considered by studying the simple case of a free non-conformal scalar field.

\bigskip
\smallskip
\leftline{\bf Acknowledgements}
\smallskip
\noindent 
This work is partially supported by the DOE grant DE-SC0011687.

\section{Achronal ANEC and extremal horizons}
\label{sec:2}

In this section we study the ANEC for a QFT in the near horizon geometry of an extremal black hole, \textit{i.e.} ${\rm AdS}_2\times S^{d-2}$. We give three independent derivations with different regimes of validity: for generals QFTs, general CFTs and free scalar fields in subsections~\ref{sub:2.2},~\ref{sub:2.3} and~\ref{sub:2.4} respectively. 

Let us start by briefly reviewing how the ${\rm AdS}_2\times S^{2}$ geometry arises from near horizon geometry of an extremal black hole in four dimensional Einstein-Maxwell theory
$$I[g_{\mu \nu},A_\mu]=\frac{1}{16\pi G}
  \int d^4x\sqrt{-g}\,
  \mathcal{R}-
  \frac{1}{4}
  \int d^4x\sqrt{-g}\,
  F_{\mu \nu}F^{\mu \nu}
  \ .$$
The extremal and spherically symmetric black hole solution is given by \cite{Myers:1986un}
\begin{equation}\label{eq:28}
ds^2=-f(r)dt^2+\frac{dr^2}{f(r)}+
  r^2d\Omega^2_{2}\ ,
  \qquad {\rm where} \qquad
  f(r)=\left(
  \frac{r-r_h}{r}
  \right)^2\ ,
\end{equation}
and $r_h>0$ is an integration constant which determines the horizon radius (related to the mass and charge of the black hole). The line element on the unit sphere $S^2$ is given by $d\Omega_2$. To analyze the causal structure we first define the tortoise coordinate $r^\ast(r)$
\begin{equation}\label{eq:29}
\frac{dr^\ast}{dr}=\frac{1}{f(r)}
  \qquad \Longrightarrow \qquad
  \frac{r^\ast}{r_h}=
  \left(
  \frac{r_h}{r-r_h}
  \right)+2
  \ln\left(
  \frac{r_h}{r-r_h}
  \right)-\left(
  \frac{r}{r_h}
  \right)
  \  ,
\end{equation}
where $r^*(r)\in \mathbb{R}$. The advantage of this coordinate is that the $(t,r^\ast)$ sector is conformally flat. Further defining the coordinates $(\sigma,\theta)$ as $r^\ast\pm t=r_h\tan(\theta_\pm/2)$ with $|\theta_\pm|=|\theta\pm\sigma/r_h|\le \pi$, we find
\begin{equation}\label{eq:30}
ds^2=
  \frac{f(r(r^*(\theta_\pm)))}
  {4\cos^2(\theta_+/2)
  \cos^2(\theta_-/2)}
  \left(-d\sigma^2+
  r_h^2d\theta^2
  \right)+
  r(r^*)^2d\Omega^2_2
  \ .
\end{equation}
\begin{figure}
\centering
\includegraphics[scale=0.25]{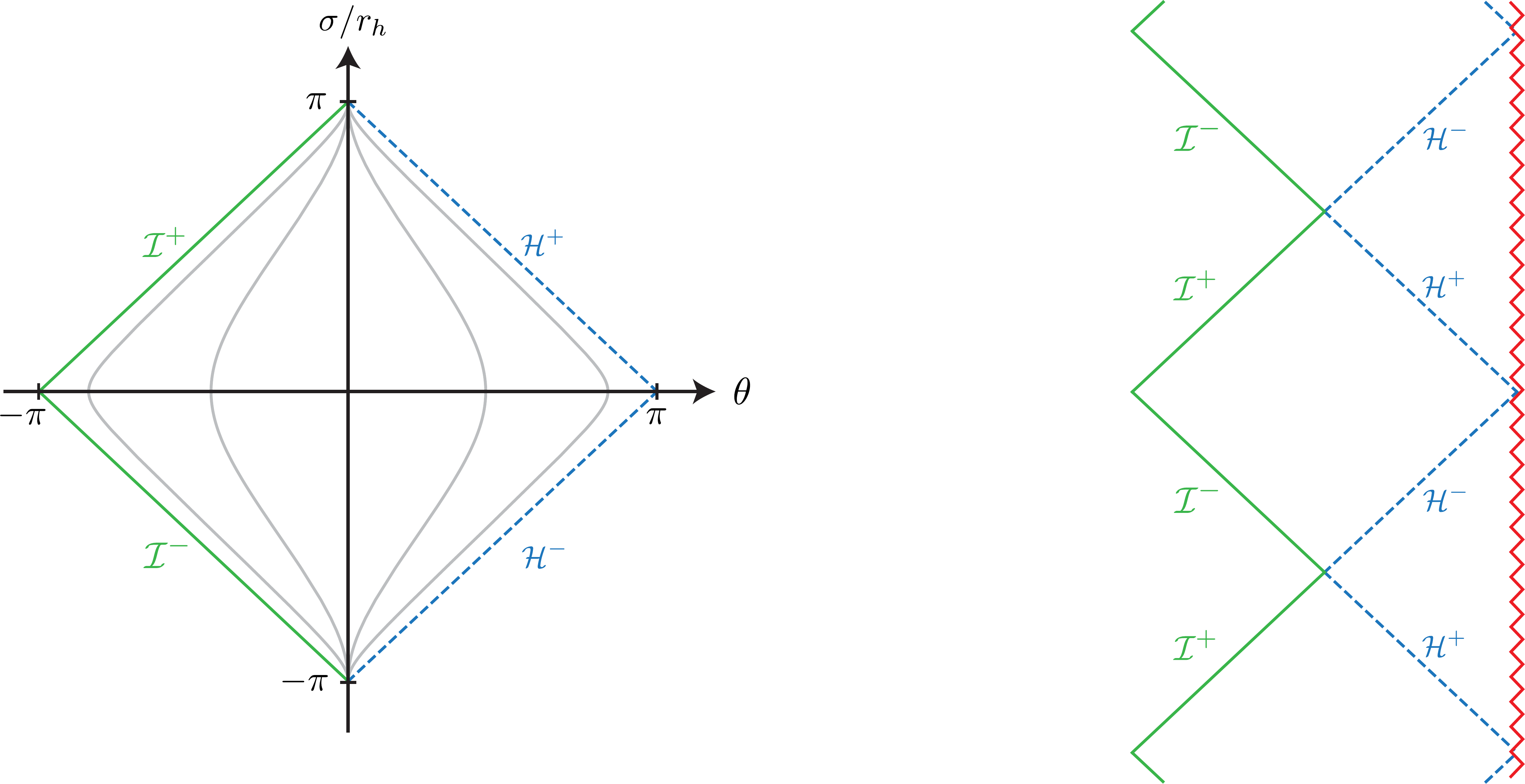}
\caption{The left diagram corresponds to the Penrose diagram of the exterior region of the extremal black hole in the coordinates $(\sigma,\theta)$ (\ref{eq:30}). The Minkowski null infinity $\mathcal{I}^\pm$ is indicated in green, while the future and past horizon of the black hole $\mathcal{H}^\pm$ in blue. Several fixed $r$ trajectories are plotted in gray. The right diagram gives the maximally extended space-time, with the time-like singularity in red.}\label{fig:1}
\end{figure}

The Penrose diagram of the outer region of the extremal black hole is obtained by taking fixed values in the $S^2$ and disregarding the conformal factor in $(\sigma,\theta)$. This give the diamond in the $(\sigma,\theta)$ plane seen in the left diagram of figure \ref{fig:1}. The future and past black hole horizons~$\mathcal{H}^\pm$ are given by $\theta_\pm=\pi$, while the asymptotic null infinity of Minkowski by $\theta_\pm=-\pi$. In gray we plot some constant $r$ curves. The maximal extension of the space-time is shown in the right diagram of this figure.

The near horizon limit is obtained by taking $(r-r_h)\ll r_h$, where the tortoise coordinate~$r^\ast(r)$ in (\ref{eq:29}) is particularly simple~${r^\ast(r)=r_h^2/(r-r_h)}$ and the black hole metric in (\ref{eq:30}) becomes
\begin{equation}\label{eq:34}
ds^2\simeq
  \frac{-d\sigma^2+
  r_h^2d\theta^2
  }
  {\sin^2(\theta)}
  +
  r_h^2d\Omega^2_2
  =
  {\rm AdS}_2\times S^{d-2}
  \ ,
\end{equation}
where we recognize an ${\rm AdS}_2$ factor in global coordinates. Since in this limit $r^\ast(r)$ only takes positive values, the coordinate $\theta$ is now restricted to $\theta\in(0,\pi)$, with $\theta=0,\pi$ corresponding to the two ${\rm AdS}_2$ boundaries.

\begin{figure}
\centering
\includegraphics[scale=0.25]{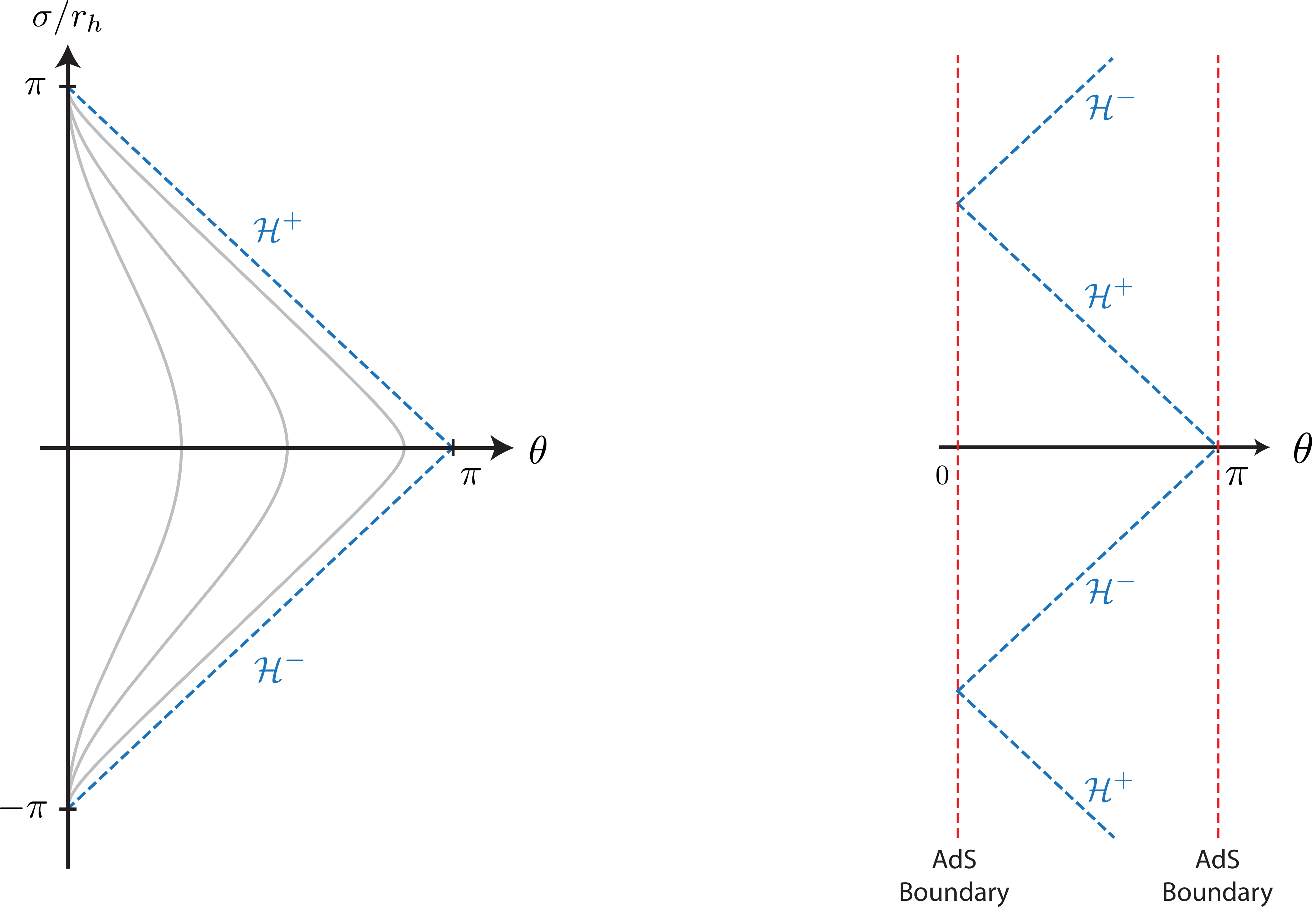}
\caption{Penrose diagram of the near horizon limit of the exterior region of an extremal black hole, which corresponds to the Poincare patch of AdS$_2$. The boundaries are located at $\theta=0,\pi$ while the black horizon $\mathcal{H}^\pm$ at $\theta_\pm=\pi$. We have plotted several constant $r$ curves in gray. To the right we have the maximal extension, corresponding to global AdS$_2$.}\label{fig:2}
\end{figure}

In the left diagram of figure \ref{fig:2} we plot the AdS$_2$ factor of the metric, with some constant~$r$ trajectories in gray. The near horizon geometry of the extremal black hole corresponds to the Poincare patch of AdS$_2$, with the black hole horizon $\mathcal{H}^\pm$ given by the Poincare horizon. The near horizon geometry can be maximally extended to the right diagram of figure \ref{fig:2}, which is global AdS$_2$. Comparing with the full black hole Penrose diagram in figure \ref{fig:1}, we see the near horizon limit corresponds to cutting the diagram in half, placing one AdS$_2$ boundary at the singularity and the other in the middle.\footnote{In a slightly different way, the space-time ${\rm AdS}_2\times S^{d-2}$ also arises from the near horizon limit of \textit{near} extremal black holes, see \cite{Spradlin:1999bn}.} Since the study of QFTs in the full black hole background in (\ref{eq:28}) is complicated, it is useful to consider the simpler near horizon instead, hoping the most relevant quantum aspects of the horizon are captured in this approximation. 

An analogous calculation for the extremal black hole in arbitrary dimensions shows the near horizon metric is given by
\begin{equation}\label{eq:87}
\frac{ds^2}{r_h^2}=
  \frac{-d\sigma^2+d\theta^2}
  {\sin^2(\theta)}+
  d\Omega_{d-2}(\vec{v}\,)\ ,
  \qquad \qquad
  d\Omega_{d-2}^2(\vec{v}\,)=
  \frac{4d\vec{v}.d\vec{v}}
  {(1+|\vec{v}\,|)^2}\ ,
\end{equation}
where the unit sphere $S^{d-2}$ is parametrized in stereographic coordinates\footnote{We can write this in the usual angles of $S^{d-2}$ by describing the vector $\vec{v}\in \mathbb{R}^{d-2}$ in spherical coordinates and parametrizing its radius according to $|\vec{v}\,|=\tan(\phi/2)$ with $\phi\in [0,\pi]$.} $\vec{v}\in \mathbb{R}^{d-2}$. We have also conveniently rescaled the time coordinate $\sigma\rightarrow \sigma/r_h$. Complete achronal null geodesics in this space-time correspond to paths going between the two ${\rm AdS}_2$ boundaries with fixed coordinates on $S^{d-2}$. These geodesics can be described in terms of an affine parameter $\lambda$ as
\begin{equation}\label{eq:103}
x^\mu(\lambda,\vec{x}_\perp)=
  (\theta_+,
  \theta_-,
  \vec{v}\,)=
  \left(
  \pi/2,
  2\,{\rm arccot}(\lambda)-\pi/2
  ,\vec{x}_\perp
  \right)
  \ ,\quad
  \qquad
  (\lambda,\vec{x}_\perp)
  \in \mathbb{R}\times
  \mathbb{R}^{d-2}
  \ .
\end{equation}
It is for these geodesics that we prove the achronal ANEC. The null surface obtained from $\vec{x}_\perp \in \mathbb{R}^{d-2}$ goes between the ${\rm AdS}_2$ boundaries at $\sigma=\pm\pi/2$ and coincides with $\mathcal{H}^+$ in figure~\ref{fig:2} after a rigid time translation.

\subsection{Relative entropy and null deformed modular hamiltonian}
\label{sub:2.2}

\begin{figure}
\centering
\includegraphics[scale=0.52]{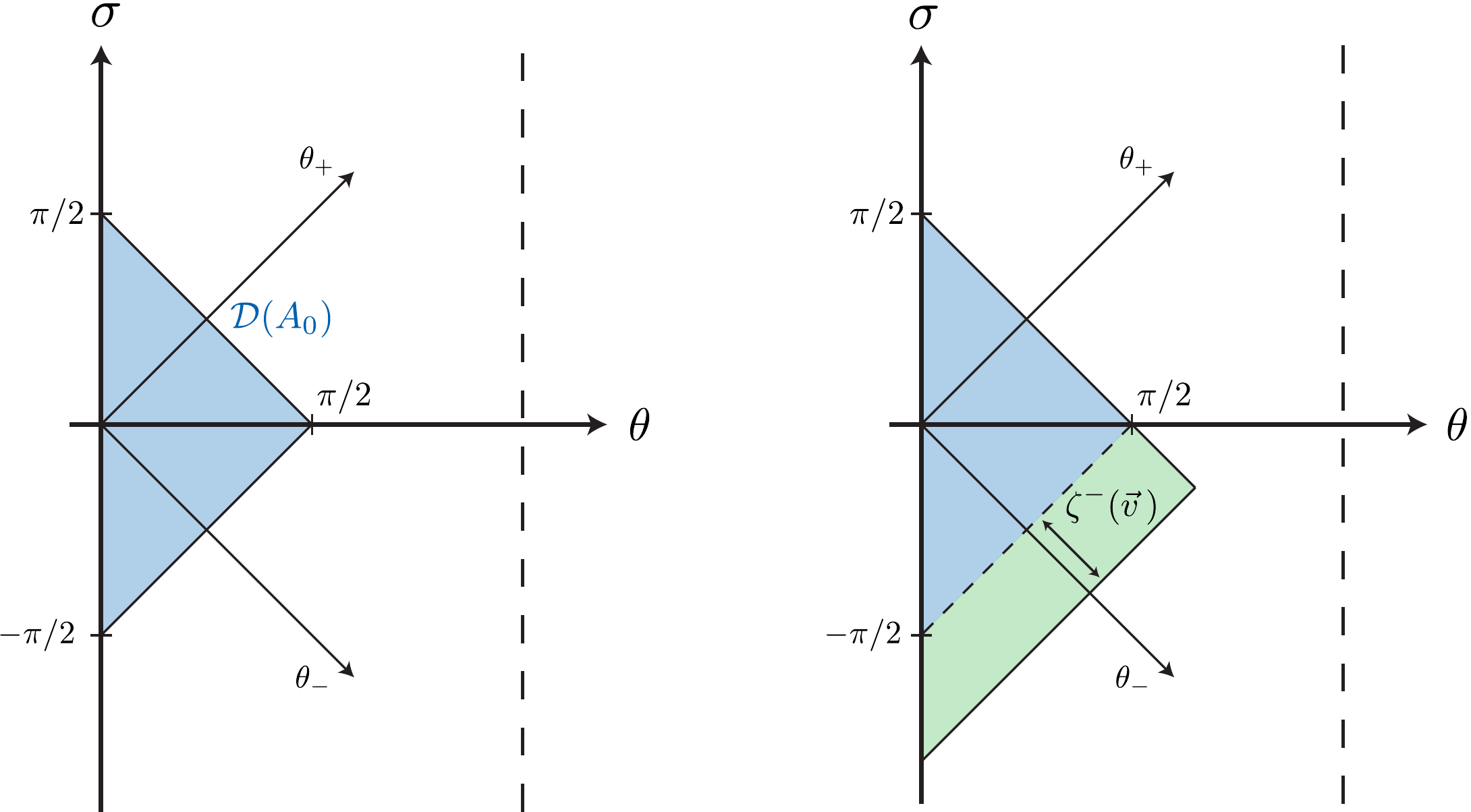}
\caption{On the left diagram we sketch the region $\mathcal{D}(A_0)$ described in (\ref{eq:88}). The deformed region $\mathcal{D}(A)$ shown in the right diagram contains the additional green section of size $\zeta^-(\vec{v}\,)$.}\label{fig:8}
\end{figure}

In this subsection we prove the achronal ANEC for general QFTs, where the null geodesics move between the ${\rm AdS}_2$ boundaries according to (\ref{eq:103}). Our approach follows the derivation of the ANEC in Minkowski presented in \cite{Faulkner:2016mzt}. The analogous region to the Rindler wedge in Minkowski for ${\rm AdS}_2\times S^{d-2}$ is given by
\begin{equation}\label{eq:88}
\mathcal{D}(A_0)=
  \left\lbrace
  (\sigma,\theta,\vec{v}\,)
  \in \mathbb{R}\times (0,\pi)\times
  \mathbb{R}^{d-2}:\quad
  \theta_+< \pi/2\ ,
  \quad
  \theta_-< \pi/2\,
  \right\rbrace\ ,
\end{equation}
plotted in the left diagram of figure \ref{fig:8}. More generally, we consider a deformation of this region in the null direction~$\theta_-$, parametrized by~${\zeta^\mu(\vec{v}\,)=\zeta^-(\vec{v}\,)\delta^\mu_-}$ as
\begin{equation}\label{eq:91}
\mathcal{D}(A)=
  \left\lbrace
  (\sigma,\theta,\vec{v}\,)
  \in \mathbb{R}\times (0,\pi)\times
  \mathbb{R}^{d-2}:\quad
  \theta_+< \pi/2\ ,
  \quad
  \theta_-< \pi/2+\zeta^-(\vec{v}\,)\,
  \right\rbrace\ ,
\end{equation}
plotted on the right diagram of figure \ref{fig:8}. For this setup, we compute the vacuum modular hamiltonian $K_{A}\equiv -\ln(\rho_A)$ in a perturbative expansion in $\zeta^\mu(\vec{v}\,)$. The technical result proven in the main part of this subsection is
\begin{equation}\label{eq:106}
K_{A}=
  K_{A_0}+
  \pi r_h^{d-2}
  \int_{S^{d-2}}d\Omega(\vec{x}_\perp)\,
  \zeta^-(\vec{x}_\perp)
  \int_{0}^{+\infty}
  d\lambda\,
  T_{\lambda \lambda}(\lambda,\vec{x}_\perp)+
  \mathcal{O}(\zeta^\mu)^2\ ,
\end{equation}
where $T_{\lambda \lambda}(\lambda,\vec{x}_\perp)$ is the stress tensor projected along the tangent vector $dx^\mu/d\lambda$ to the geodesic in (\ref{eq:103}). 

From this expression the ANEC follows very easily. The first order correction in the deformation vector $\zeta^\mu(\vec{v}\,)$ already contains ``half'' the ANEC operator. We can obtain the other half by constructing the full modular hamiltonian $\hat{K}_A\equiv K_A-K_{A^c}$, which gives 
\begin{equation}\label{eq:112}
\hat{K}_A=
  \hat{K}_{A_0}+
  \pi r_h^{d-2}
  \int_{S^{d-2}}d\Omega(\vec{x}_\perp)\,
  \zeta^-(\vec{x}_\perp)
  \mathcal{E}(\vec{x}_\perp)
  +
  \mathcal{O}(\zeta^\mu)^2\ ,
\end{equation}
where we have defined the ANEC operator $\mathcal{E}(\vec{x}_\perp)$ in (\ref{eq:149}). It is now that monotonicity of relative entropy (see \cite{Witten:2018lha} for a review) becomes useful. Some simple manipulations show that it implies the following constraint on the full modular hamiltonian~\cite{Blanco:2013lea}
$$\hat{K}_B-\hat{K}_C\ge 0
  \qquad \Longleftrightarrow \qquad
  C\subseteq B\ ,$$
for any two regions $B$ and $C$. Fixing $B=\mathcal{D}(A)$ and $C=\mathcal{D}(A_0)$, together with (\ref{eq:112}) gives the ANEC
\begin{equation}\label{eq:109}
\hat{K}_{A}-\hat{K}_{A_0}\ge 0
\quad \Longleftrightarrow \quad
\mathcal{D}(A_0)\subseteq \mathcal{D}(A)
\quad \Longleftrightarrow \quad
\zeta^-(\vec{v}\,)\ge 0
\quad \Longleftrightarrow \quad
\mathcal{E}(\vec{x}_\perp)\ge 0\ .
\end{equation} 
In the remaining parts of this subsection we show that the modular hamiltonian is given by~$K_A$ in~(\ref{eq:106}) for arbitrary QFTs, completing the proof of the ANEC in ${\rm AdS}_2\times S^{d-2}$.

\subsubsection{Undeformed region}

Let us start by computing the vacuum modular hamiltonian associated to the undeformed region $\mathcal{D}(A_0)$ in (\ref{eq:88}). We do this by following the same approach as the calculation for the Rindler region in Minkowski, pedagogically explained in \cite{Witten:2018lha}. Since the computation requires a path integral formulation we first analytically continue the metric (\ref{eq:87}) to Euclidean time~$\sigma_E=i\sigma$. This gives a useful representation of the vacuum state in terms of a path integral over the lower half $\sigma_E<0$ of the Euclidean space
\begin{equation}\label{eq:72}
\ket{0}=
  \int_{\sigma_E<0} D\Phi\,
  e^{-I_E[\Phi,\sigma_E<0]}=
\includegraphics[scale=0.4, valign=c]{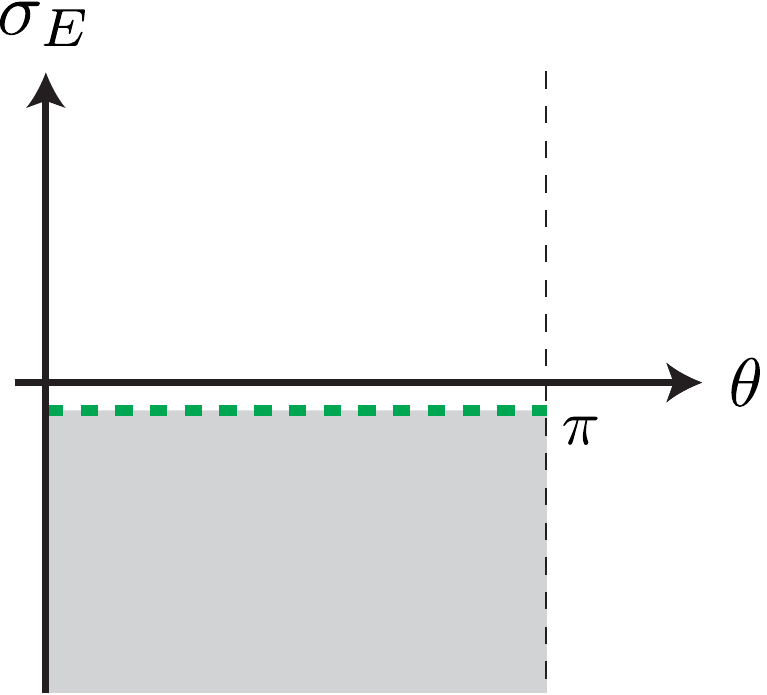}
\end{equation}
where $\Phi(\sigma_E,x)$ represents all the fields in the theory.\footnote{The extra spatial directions parametrizing the $S^{d-2}$ are implicit in this pictorial description of the path integral. In describing the ground state as in (\ref{eq:72}) we are setting the vacuum energy to zero (by shifting the hamiltonian) and assuming no degeneracy.} The state $\ket{0}$ is a functional, which provides a number once we specify the boundary conditions of the path integral at $\sigma_E=0^-$ (green dashed line) according to $\braket{\Phi(\sigma_E=0^-,x)|0}$.\footnote{There is no need to specify boundary conditions at $\theta=0,\pi$ since this is not a real boundary of the manifold, but a \textit{conformal} boundary of Euclidean AdS$_2$. Redefining the radial coordinate according to $\varrho=-\cot(\theta)$ the regions $\theta=0,\pi$ are pushed to infinity $\varrho \rightarrow \pm \infty$.} Analogously, the hermitian conjugate $\bra{0}$ is given by the integral over the future region $\sigma_E>0$. 

Following standard arguments \cite{Nishioka:2018khk,Rangamani:2016dms}, the reduced density operator corresponding to the vacuum reduced to $A_0$ is obtained from the path integral by sewing the regions $\sigma_E<0$ and $\sigma_E>0$ but only for $\theta>\pi/2$. This gives the following path integral representation of  $\rho_{A_0}$ 
\begin{equation}\label{eq:69}
\rho_{A_0}=
\includegraphics[scale=0.4, valign=c]{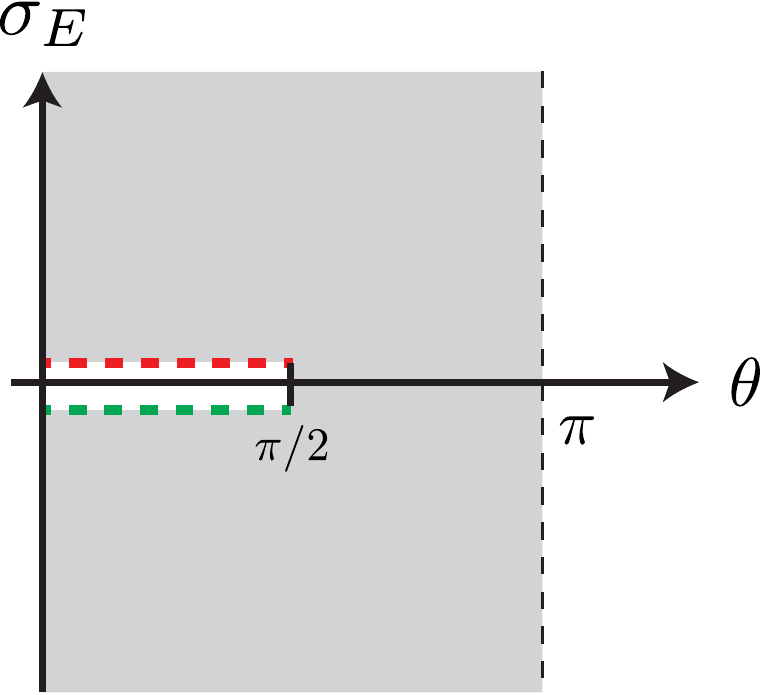}
\end{equation}
Specifying the boundary condition at $\sigma_E=0^+$ (red dotted line) gives the ket ${\rho_{A_0}\ket{\Phi(\sigma_E=0^+,x)}}$, which after fixing the field configuration at $\sigma_E=0^-$ (green dotted line) gives a number ${\bra{\Phi(\sigma_E=0^-,x')}\rho_{A_0}\ket{\Phi(\sigma_E=0^+,x)}}$. The path integral in (\ref{eq:69}) ``propagates" the boundary conditions from either side of the cut along the Euclidean manifold, from the top red boundary to the bottom green one. 

\begin{figure}
\centering
\includegraphics[scale=0.35]{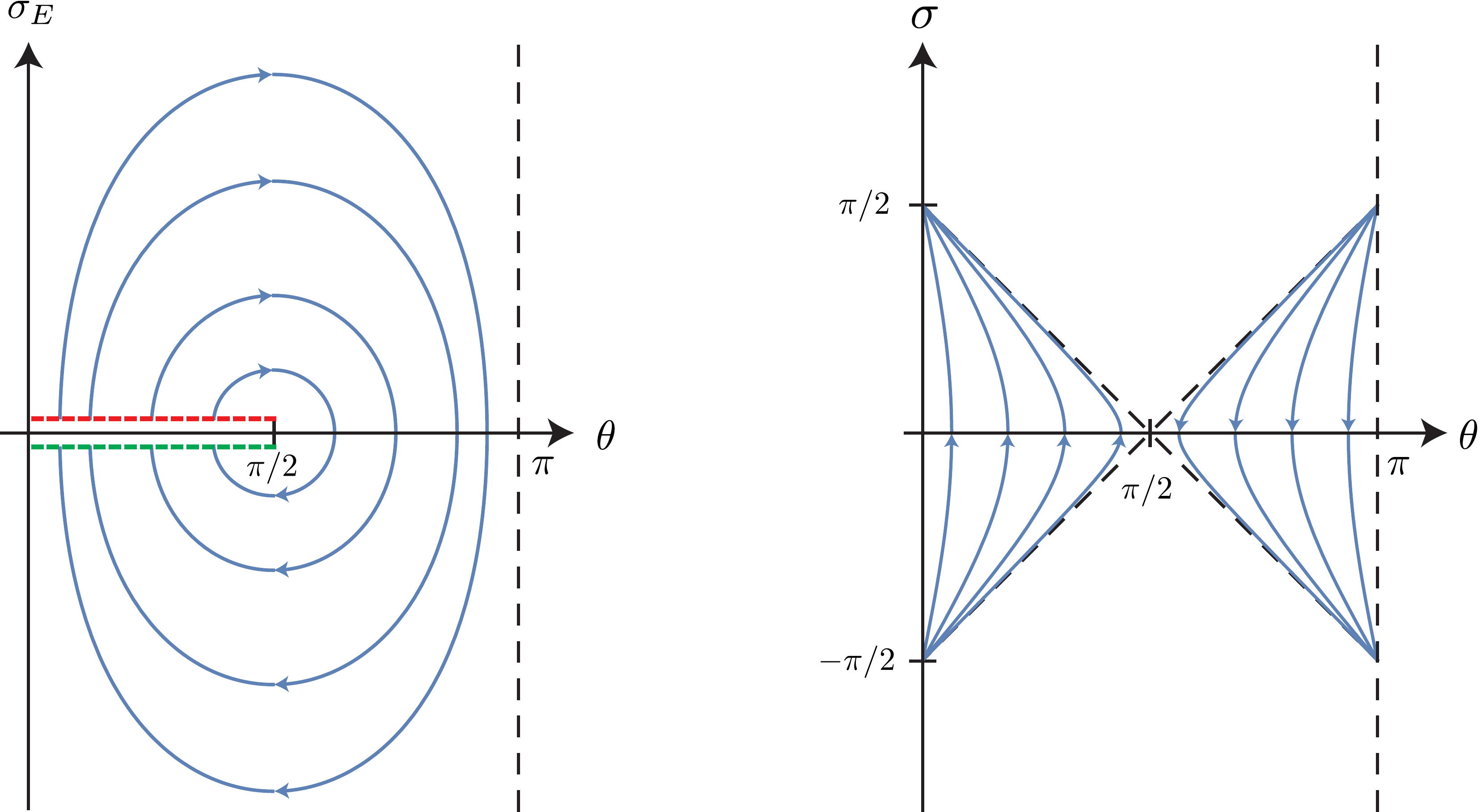}
\caption{On the left diagram we have the Euclidean flow generated by (\ref{eq:70}) applied on the surface at $\sigma_E=0^+$ for $\tau\in(0,2\pi)$. When we analytically continue to Lorentzian time $\sigma_E=i\sigma$ we obtain the flow on the right diagram.}\label{fig:4}
\end{figure}

Since the geometrical setup is very simple and there are no operator insertions in the path integral, we might hope that the ``propagation" between the boundary conditions can be realized geometrically by an isometry of the space-time. This is certainly the case for the Rindler region in Minkowski, where such isometry simply corresponds to a rotation. Luckily, this is also the case in our setup, although the transformation is much less trivial and can be written as
\begin{equation}\label{eq:70}
\begin{aligned}
\coth(\sigma_E(\tau))&=
\frac{\cosh(\sigma_E)}
{\cos(\tau)\sinh(\sigma_E)+
\sin(\tau)\cos(\theta)}\ ,\\
\tan(\theta(\tau))&=
\frac{\sin(\theta)}
{\cos(\tau)\cos(\theta)-\sin(\tau)\sinh(\sigma_E)}\ ,
\end{aligned}
\end{equation}
where $\tau \in [0,2\pi]$ is the parameter in the transformation. The simplest way of obtaining this isometry is using the embedding description of Euclidean ${\rm AdS}_2$. It is also straightforward to check that the Euclidean metric (\ref{eq:87}) stays invariant. When applying this map to the surface~$\sigma_E=0^+$ in (\ref{eq:69}) we find that for $\tau=2\pi$ the surface is rotated in a non-trivial way to $\sigma_E=0^-$ (see left diagram in figure \ref{fig:4}), doing precisely as we require by the path integral in~(\ref{eq:69}). This means we can write the reduced density operator $\rho_{A_0}$ as
\begin{equation}\label{eq:89}
\rho_{A_0}=U(\tau=2\pi)\ ,
\end{equation}
where $U(\tau)$ is the unitary operator implementing the Euclidean transformation (\ref{eq:70}).
Analytically continuing back to Lorentzian time $\sigma=-i\sigma_{E}$ the isometry becomes
\begin{equation}\label{eq:71}
\begin{aligned}
\cot(\sigma(s))&=
\frac{\cos(\sigma)}
{\cosh(s)\sin(\sigma)+
\sinh(s)\cos(\theta)}\ ,\\
\tan(\theta(s))&=
\frac{\sin(\theta)}
{\cosh(s)\cos(\theta)
+\sinh(s)\sin(\sigma)}\ ,
\end{aligned}
\end{equation}
where we have defined a real Lorentzian parameter $s\equiv -i\tau$. The flow generated by this transformation for $s\in \mathbb{R}$ is plotted in the right diagram of figure \ref{fig:4}, where we see that its action is analogous to that of a boost in Minkowski: it is time-like in the wedge and maps~$\mathcal{D}(A_0)$ into itself. This is nothing more than the vacuum modular flow associated to the region. 

Using (\ref{eq:89}) we can write a concrete representation for $\rho_{A_0}$ as the operator implementing the transformation (\ref{eq:71}) with parameter $s=-i(2\pi)$. The generator of the transformation is easily written in terms of the stress tensor $T_{\mu \nu}$, so that we find
\begin{equation}\label{eq:73}
\rho_{A_0}=U(s=-2\pi i)=
  \frac{1}{Z}
  \exp\left[
  -i\left(-2\pi i\right)
  \int_{\Sigma} dS^\mu T_{\mu \nu}\xi^\nu
  \right]\ ,
\end{equation}
where $Z$ is a normalization constant and the integral gives the conserved charge associated to~(\ref{eq:71}). The integral is over a Cauchy surface $\Sigma$ in $\mathcal{D}(A_0)$ with future directed surface element~$dS^\mu$. The Killing vector $\xi^\mu$ is obtained by expanding (\ref{eq:71}) for small $s$ and has a simple expression when written in null coordinates $\theta_\pm=\theta\pm\sigma$
\begin{equation}\label{eq:74}
\xi^\mu=
  \cos(\theta_+)\partial_+-
  \cos(\theta_-)\partial_-\ .
\end{equation}
Its magnitude vanishes at the boundaries $\theta_\pm=\pi/2$. 

Let us write the integral in (\ref{eq:73}) explicitly. Although we can choose any Cauchy surface~$\Sigma$ in the region $\mathcal{D}(A_0)$ (left diagram in figure \ref{fig:8}) it is instructive to pick the future null horizon of $\mathcal{D}(A_0)$, which can be parametrized in terms of the affine parameter $\lambda$ in (\ref{eq:103}) with $\lambda\ge 0$. In this way we find
\begin{equation}\label{eq:107}
K_{A_0}=2\pi r_h^{d-2}
  \int_{S^{d-2}}d\Omega(\vec{x}_\perp)
  \int_0^{+\infty}d\lambda\,
  \lambda\,
  T_{\lambda \lambda}(\lambda,\vec{x}_\perp)+{\rm const}\ ,
\end{equation}
where the constant comes from $Z$ in (\ref{eq:73}). In writing this expression we have chosen the canonical normalization for the future directed normal null vector, given by $n^\mu=-2\sin^2(\theta)\delta^\mu_-$. The modular hamiltonian has a similar structure to that of the Rindler region in Minkowski.

Before moving on to the null deformed region let us mention that it is straightforward to generalize this construction to a wedge $\mathcal{D}(A_0)$ of any size, not necessarily $\pi/2$. We do this in appendix \ref{zapp:3} where we explicitly write the modular flow and modular hamiltonian for a wedge of size $\theta_0\in(0,\pi)$.

\subsubsection{First order null deformation}

We now compute the first order perturbation associated to the null deformation given in~(\ref{eq:91}). We do so by adapting the methods used for Minkowski in \cite{Faulkner:2016mzt,Balakrishnan:2020lbp,Banerjee:2011mg,Faulkner:2015csl} to the curved background ${\rm AdS}_2\times S^{d-2}$.

The basic idea is simple: since dealing with the vacuum reduced to the deformed region $\mathcal{D}(A)$ is complicated, we apply a diffeomorphism mapping $\mathcal{D}(A)\rightarrow \mathcal{D}(A_0)$. From (\ref{eq:88}) and~(\ref{eq:91}) we see this is given by $\tilde{\theta}_-=\theta_--\zeta^-(\vec{v}\,)$, written covariantly as $\tilde{x}^\mu=x^\mu-\zeta^\mu(\vec{v}\,)$. Since $\zeta^\mu$ is not a Killing vector, the metric in the coordinates $\tilde{x}^\mu$ is no longer ${\rm AdS}_2\times S^{d-2}$ but is given by
\begin{equation}\label{eq:93}
\tilde{g}_{\alpha \beta}=
  \frac{\partial x^\mu}
  {\partial \tilde{x}^{\alpha}}
  \frac{\partial x^\nu}
  {\partial \tilde{x}^{\beta}}
  g_{\mu \nu}=
  g_{\alpha \beta}
  +\mathcal{L}_\zeta(g_{\alpha \beta})+
  \mathcal{O}(\zeta^\mu)^2\ ,
\end{equation}
where~$g_{\mu \nu}$ is the~${\rm AdS}_2\times S^{d-2}$ metric and~$\mathcal{L}_\zeta(g_{\alpha \beta})=2\nabla_{(\alpha}\zeta_{\beta)}$ the Lie derivative along $\zeta^\mu$. This approach allows us to trade the deformed region~$\mathcal{D}(A)$ in the space-time~$g_{\mu \nu}$ by the simpler region~$\mathcal{D}(A_0)$ in the more complicated metric~$\tilde{g}_{\alpha \beta}$ (\ref{eq:93}).

The diffeomorphism is implemented in the Hilbert space of the Euclidean theory by a unitary operator $U(\zeta)$, that can be explicitly written as\footnote{There is no $i$ factor in the exponent since it is the Euclidean generator and no minus sign since the parameter in the diffeomorphism is $-\zeta^\mu(\vec{v}\,)$. See \cite{Faulkner:2016mzt,Banerjee:2011mg} for a path integral representation of $U(\zeta)$. }
\begin{equation}\label{eq:102}
U(\zeta)=
  \exp\left[
  \int_{\sigma_E=0}dS^\mu 
  T_{\mu \nu}(x)\zeta^\nu(\vec{v}\,)
  \right]\ ,
\end{equation}
where $dS^\mu$ is the induced surface element on $\sigma_E=0$, with future directed unit normal. Splitting the integral over $A_0$ and $A_0^c$ we can write it as $U(\zeta)=U_{A_0}\otimes U_{A_0^c}$. The reduced density operator $\rho_{A,g}$\footnote{The subscript $g$ indicates the background metric.} in the deformed region is mapped by $U_{A_0}$ according to
\begin{equation}\label{eq:110}
\rho_{A,g}=
  U^{-1}_{A_0}\,
  \rho_{A_0,\tilde{g}}\,
  U_{A_0}=
  \rho_{A_0,g}+
  \big[
  \rho_{A_0,g},
  \delta U_{A_0}
  \big]+
  \delta \rho_{A_0,g}+
  \mathcal{O}(\zeta^\mu)^2\ ,
\end{equation}
where in the second equality we have expanded both $U(\zeta)$ and $\rho_{A_0,\tilde{g}}$ to linear order in $\zeta^\mu$. While the first two terms on the right-hand side are simple and can be written from (\ref{eq:107}) and (\ref{eq:102}), the third is more complicated. It can still be written explicitly using the path integral representation of $\rho_{A_0,g}$ as (see eq. (21) in \cite{Faulkner:2015csl})
\begin{equation}\label{eq:111}
\delta \rho_{A_0,g}=
  \rho_{A_0,g}\int_{\mathcal{M}_E}
  d^dx\sqrt{g}
  \left(
  T^{\mu \nu}-\langle T^{\mu \nu} \rangle_0 \right)
  \nabla_{(\mu}\zeta_{\nu)}
  \ ,
\end{equation}
where the integral is over the Euclidean space-time~$\mathcal{M}_E$ with a branch cut along~$\theta\in(0,\pi/2)$. The vacuum expectation value of the stress tensor~$\langle T^{\mu \nu} \rangle_0$ comes from the variation of the normalization constant of~$\rho_{A_0,\tilde{g}}$, and we shall leave implicit in what follows.

\begin{figure}
\centering
\includegraphics[scale=0.45]{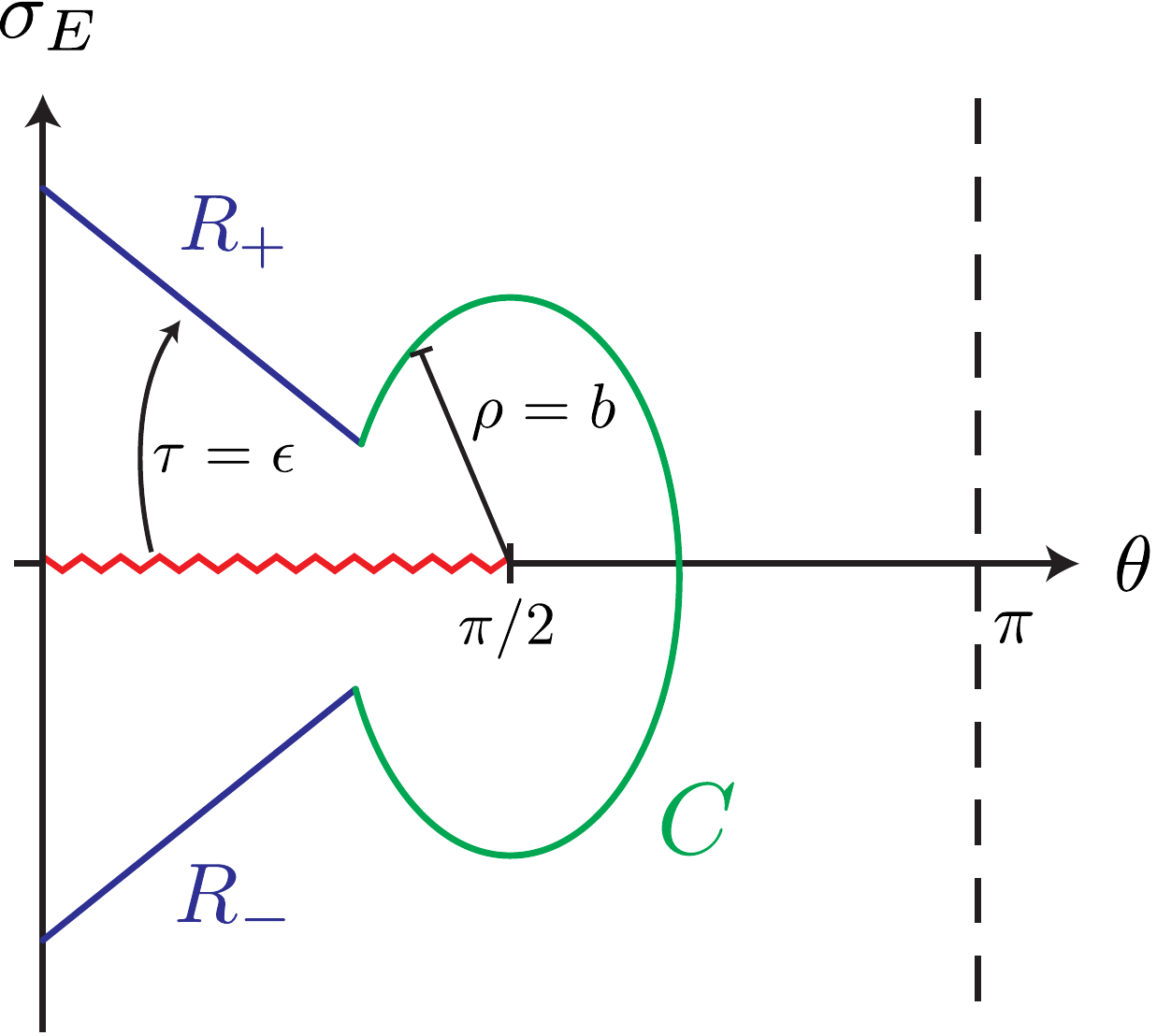}
\caption{Sketch of the boundary $\partial \mathcal{M}_E$ obtained from the regions $R_\pm$ (in blue) and $C$ (in green) described in (\ref{eq:96}) as $\epsilon,b\rightarrow 0$. The red wiggly line corresponds to the branch cut in the Euclidean space-time located at $\sigma_E=0$ and $\theta\in(0,\pi/2)$.}\label{fig:6}
\end{figure}

An analogous perturbative expansion to (\ref{eq:110}) holds for the modular hamiltonian $K_{A,g}$. This can be written explicitly using (\ref{eq:111}) and the smart identity proven in eq. (2.14) of~\cite{Balakrishnan:2020lbp},\footnote{The Baker-Campbell-Hausdorff formula used to derive eq. (2.14) in \cite{Balakrishnan:2020lbp} can be found in (4.173) of~\cite{CBHD}. One must be careful with the sign convention of the Bernoulli numbers~$B_n$, since they disagree between these references by a minus sign in~$B_1$. In the convention of~\cite{Balakrishnan:2020lbp} the~$B_n$ are written in terms of the Riemann zeta function as~$B_n=-n\zeta(1-n)$, which can be written as an integral using~(35) of \cite{Milgram_2013}.} which results in
\begin{equation}\label{eq:108}
K_{A,g}=
  U^{-1}_{A_0}\,
  K_{A_0,\tilde{g}}\,
  U_{A_0}=
  K_{A_0,g}+
  \big[
  K_{A_0,g},
  \delta U_{A_0}
  \big]+
  \delta K_{A_0,g}+
  \mathcal{O}(\zeta^\mu)^2\ ,
\end{equation}
where
\begin{equation}\label{eq:92}
\delta K_{A_0,g}=
  \int_{-\infty+i\alpha}^{+\infty+i\alpha}
  \frac{dz}{4\sinh^2(z/2)}
  \rho_{A_0}^{-\frac{iz}{2\pi}}
  \left[
  \int_{\mathcal{M}_E}d^dx
  \sqrt{g}\,
  T^{\mu \nu}(x)
  \nabla_{(\mu} \zeta_{\nu)}
  \right]
  \rho_{A_0}^{\,\frac{iz}{2\pi}}
  \ ,
\end{equation}
with $\alpha$ a free parameter in the range $\alpha\in(0,2\pi)$ that will be conveniently fixed further ahead. We crucially get complex powers of the reduced density operator $\rho_{A_0}$, which generates the modular flow of the undeformed region $\mathcal{D}(A_0)$ (sketched in figure~\ref{fig:4}) on the Hilbert space. Using the conservation of the stress tensor $\nabla_\mu (\sqrt{g} \,T^{\mu \nu})=0$, the integral over $\mathcal{M}_E$ can be reduced to its boundary $\partial \mathcal{M}_E$\footnote{In appendix \ref{zapp:2} we show that $\nabla_\mu \langle T_{\mu \nu} \rangle_0=0$, as required to obtain this relation.}
\begin{equation}\label{eq:92}
\delta K_{A_0}=
  \int_{-\infty+i\alpha}^{+\infty+i\alpha}
  \frac{dz}{4\sinh^2(z/2)}
  \int_{\partial \mathcal{M}_E}
  dS^\mu \zeta^\nu
  \rho_{A_0}^{-\frac{iz}{2\pi}}
  T_{\mu \nu}(x)
  \rho_{A_0}^{\,\frac{iz}{2\pi}}\ ,
\end{equation}
where we drop the sub-index $g$ for notation convenience. This is the non-trivial integral we must solve to compute the first order contribution to the modular hamiltonian. The boundary~$\partial \mathcal{M}_E$ does not get contributions from the \textit{conformal} boundary of Euclidean ${\rm AdS}_2$, but from the branch cut along $\sigma_E=0$ and $\theta\in(0,\pi/2)$ (red wiggly line in figure \ref{fig:6}).\footnote{What we usually called the AdS boundary is really a conformal boundary, meaning that the distance between any fixed point and $\theta=0,\pi$ is always infinity. When considering the region $\partial \mathcal{M}_E$ we can drop contributions from $\theta=0,\pi$ in the same way we do for spatial infinity in Minkowski.} 

To describe the region $\partial \mathcal{M}_E$ let us introduce two useful set of coordinates. First, we consider $\theta_\pm^E=\theta\mp i\sigma_E$ which upon analytic continuation yield the ordinary null coordinates $\theta_\pm=\theta\pm\sigma$. In addition, it is also convenient to take the parameter ${\tau\in[0,2\pi]}$ describing the Euclidean modular flow in (\ref{eq:70}) as a coordinate. An associated spatial coordinate which covers the entire Euclidean section can be chosen by parametrizing the surface $(\sigma_E=0,{\theta<\pi/2})$. This corresponds to the red or green lines in figure \ref{fig:4}, which propagate over the whole space-time as we vary $\tau\in[0,2\pi]$. A convenient parametrization of the initial surface at $\sigma_E=0$ is given by ${\rho(\theta)=\ln(\cot(\theta/2))\in \mathbb{R}_{\ge 0}}$. From (\ref{eq:70}) we see that the coordinates $(\tau,\rho)$ and $(\sigma_E,\theta)$ are related according to
\begin{equation}\label{eq:79}
\begin{cases}
\begin{aligned}
\,\,\tanh(\sigma_E)&=
  \sin(\tau)\tanh(\rho)\\
\cot(\theta)&=\cos(\tau)\sinh(\rho)
\end{aligned}
\end{cases}
\qquad \Longrightarrow \qquad
\tan(\theta_\pm^E)=
  \frac{\cos(\tau)\mp i\sin(\tau)\cosh(\rho)}
  {\sinh(\rho)}\ ,
\end{equation}
so that the Euclidean metric is
\begin{equation}\label{eq:80}
\frac{ds_E^2}{r_h^2}=
  \frac{d\theta_+^Ed\theta_-^E}
  {\sin^2(\theta)}
  +d\Omega_{d-2}^2=d\rho^2+
  \sinh^2(\rho)d\tau^2+d\Omega^2_{d-2}\ .
\end{equation}
Using these coordinates we can describe the boundary $\partial \mathcal{M}_E$, as the union of the surfaces $C$ and $R_\pm$ in the limit of $\epsilon,b\rightarrow 0$ 
\begin{equation}\label{eq:96}
\begin{aligned}
C&=
  \Big\lbrace  
  (\tau,\rho)\in [0,2\pi]\times \mathbb{R}_{\ge 0}:
  \quad
  2\pi-\epsilon\ge \tau\ge \epsilon\ ,
  \quad
  \rho=b\,
  \Big\rbrace\ ,\\
R_+&=
  \Big\lbrace 
  (\tau,\rho)\in [0,2\pi]\times \mathbb{R}_{\ge 0}:
  \quad
  \tau=\epsilon\ ,
  \quad
  \rho\ge b\,
  \Big\rbrace\ ,\\
R_-&=
  \Big\lbrace 
  (\tau,\rho)\in [0,2\pi]\times \mathbb{R}_{\ge 0}:
  \quad
  \tau=2\pi-\epsilon\ ,
  \quad
  \rho\ge b\,
  \Big\rbrace\ ,
\end{aligned}
\end{equation}
sketched in figure \ref{fig:6}. We now proceed to compute the contributions of these surfaces to the integral in $\delta K_{A_0}$ (\ref{eq:92}). We refer to $C$ and $R_\pm$ as ``branch point" and ``branch cut" contributions respectively.

\paragraph{Branch point contribution:}

Let us start by computing the branch point contribution, given by the integral over the (green) surface $C$ in figure \ref{fig:6}. The induced metric and inward pointing unit normal vector $n^\mu$ are 
$$\left.
\frac{ds^2}{r_h^2}\right|_{\rho=b}=
  \sinh^2(b)d\tau^2+d\Omega^2_{d-2}\ ,
  \qquad \qquad
  n^\mu=-
  \frac{1}{r_h}\delta^\mu_\rho
  =-
  \frac{1}{r_h}
  \left[
  \frac{\partial \theta^E_+}
  {\partial \rho}
  \delta^\mu_++
  \frac{\partial \theta^E_-}
  {\partial \rho}
  \delta^\mu_-
  \right]\ ,
$$
where the derivatives in the normal vector can be easily computed from (\ref{eq:79})
\begin{equation}\label{eq:125}
\frac{\partial \theta_\pm^E}
 {\partial \rho}=
 \frac{-1}
 {\cos(\tau)\cosh(\rho)\mp i \sin(\tau)}\ .
\end{equation}

Before using this to write the integral in (\ref{eq:92}), consider the action of the modular flow on the stress tensor, given by
\begin{equation}\label{eq:94}
\rho_{A_0}^{-\frac{iz}{2\pi}}
  T_{\mu \nu}(\theta_\pm^E)
  \rho_{A_0}^{\frac{iz}{2\pi}}=
  \left(
  \frac{\partial x^{\bar{\mu}}}
  {\partial x^\mu}
  \frac{\partial x^{\bar{\nu}}}
  {\partial x^\nu}
  \right)
  T_{\bar{\mu }\bar{\nu}}(\theta_\pm^E(z))\ ,
\end{equation}
where the indices $(\mu,\nu)$ correspond to $\theta_\pm^E$, while $(\bar{\mu},\bar{\nu})$ to the modular translated coordinates $\theta_\pm^E(z)$. The complex parameter $z$ along the integral in (\ref{eq:92}) can be written as $z=s+i\alpha$ with $s\in \mathbb{R}$. In the $(\tau,\rho)$ coordinates the modular flow is particularly simple and given by a rigid translation $\tau(z)=\tau+iz$, while $\rho$ is unchanged. Moreover, since $\alpha\in(0,2\pi)$ is a free parameter we can fix it to $\alpha=\tau$, so that the translated coordinate is $\tau(z)=is$, purely imaginary and independent of $\tau$. This crucially means the operator on the right-hand side of (\ref{eq:94}) is inserted only in Lorentzian time. For this value of $\alpha$ the modular flow in the $\theta_\pm^E$ coordinates can be found using (\ref{eq:79})
\begin{equation}\label{eq:95}
\tan(\theta_\pm^E(z))=
  \frac{\cosh(s)
  \pm 
  \sinh(s)\cosh(\rho)}
  {\sinh(\rho)}\ ,
\end{equation}
which is equivalent to the Lorentzian modular flow (\ref{eq:71}). The derivatives appearing in the Jacobian expression in (\ref{eq:94}) can be computed from
\begin{equation}\label{eq:100}
\frac{\partial \theta_\pm^E(z)}
  {\partial \theta^E_\pm}=
  \frac{\partial \rho}{\partial \theta_\pm^E}
  \frac{\partial \theta_\pm^E(z)}
  {\partial \rho}+
  \frac{\partial \tau}
  {\partial \theta_\pm^E}
  \frac{\partial \theta_\pm^E(z)}
  {\partial \tau}=
  \frac{\cos(\tau)\cosh(\rho)\mp
  i\sin(\tau)}
  {\cosh(s)\cosh(\rho)\pm \sinh(s)}\ ,
\end{equation}
where we used that $\theta_\pm^E(z)$ in (\ref{eq:95}) is independent of $\tau$.

\begin{figure}
\centering
\includegraphics[scale=0.33]{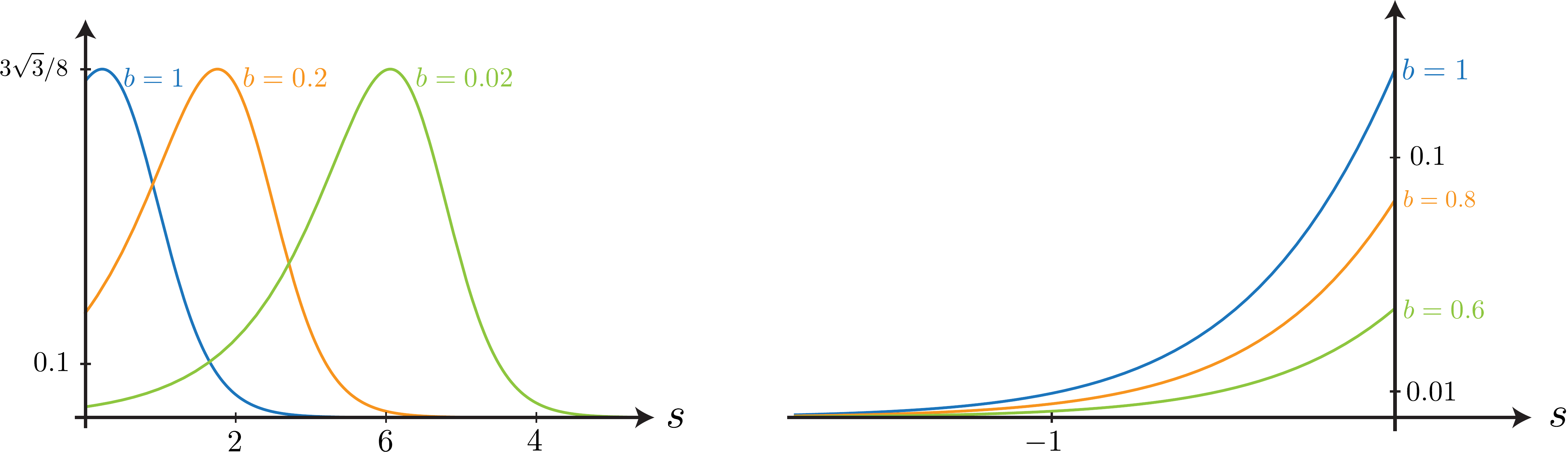}
\caption{Plot of the two kernels in the integral in (\ref{eq:97}) for several values of $b$. The left (right) diagram corresponds to the integral with positive (negative) $s$.}\label{fig:7}
\end{figure}

Putting everything together in (\ref{eq:92}) we write the contribution coming from the surface~$C$ as
\begin{equation}\label{eq:83}
\begin{aligned}
\delta K_{A_0}\big|_{C}=
  2\pi r_h^{d-2}
  \int_{S^{d-2}}d\Omega(\vec{v}\,)\,
  \zeta^-(\vec{v}\,)
  \int_{-\infty}^{+\infty}ds\,
  I(b,s)
  &\left[
  \frac{\sinh(b)T_{--}
  (\theta_\pm^E(z))}
  {(\cosh(b)\cosh(s)-\sinh(s))^2}
  \right.\\
  +&\left.
  \frac{\sinh(b)T_{+-}(\theta_\pm^E(z))}
  {\cosh^2(b)\cosh^2(s)-\sinh^2(s)}
  \right]
  \ ,
\end{aligned}
\end{equation}
where we have defined
$$I(b,s)=
  \int_{\epsilon}^{2\pi-\epsilon}
  \frac{d\tau}{2\pi}
  \frac{\cosh(b)\cos(\tau)+i\sin(\tau)}
  {4\sinh^2(\frac{s+i\tau}{2})}=
  \cosh^2(b/2)e^{- s}\Theta( s)
  +\sinh^2(b/2)e^{ s}\Theta(- s)-
  \cosh(b)
  \delta(s)\ .$$
This integral is solved by taking the limit $\epsilon\rightarrow 0$, changing the integration coordinate to $w=e^{i\tau}$ and computing a residue. Using this in (\ref{eq:83}), the integral in $s$ involving the component $T_{--}$ of the stress tensor is
\begin{equation}\label{eq:97}
\int_{0}^{+\infty}
  \frac{ds\,\sinh(b)
  \cosh^2(b/2)
  e^{-s}}
  {(\cosh(b)\cosh(s)-\sinh(s))^2}
  T_{--}
  (\theta_\pm^E(z))
  +
\int_{-\infty}^{0}
  \frac{ds\,\sinh(b)\sinh^2(b/2)e^s}
  {(\cosh(b)\cosh(s)-\sinh(s))^2}
  T_{--}
  (\theta_\pm^E(z))\ ,
\end{equation}	
where we have omitted the contribution coming from the Dirac delta since it vanishes in the limit $b\rightarrow 0$. Although the remaining terms also seem to vanish as $b\rightarrow 0$, we must be careful since the integration region goes to infinity. Plotting the two kernels in each integral for several values of $b$ we obtain the plots in figure \ref{fig:7}. The left (right) diagram corresponds to the integral with positive (negative) $s$.

While the diagram in the right goes to zero as $b\rightarrow 0$ for every value of $s$, the plot in the left contains a contribution that is not suppressed in the limit but merely translated to larger values of~$s$. The position of the maximum is given by ${s_{\rm max}(b)=\ln(\coth(b/2)/\sqrt{3})}$. Both integrals in $s$ coming from the contributions to $T_{+-}$ in (\ref{eq:83}) show the same behavior as the right diagram in figure \ref{fig:7}, and therefore vanish in the limit $b\rightarrow 0$. Thus, the only non-vanishing contribution to (\ref{eq:83}) is given by the first integral in (\ref{eq:97}). To extract the surviving piece as $b\rightarrow 0$, we redefine the integration variable to $u(s)=\theta^E_-(z)$ in~(\ref{eq:95})
$$u(s)=
  {\rm arctan}\left[
  \frac{\cosh(s)- \sinh(s)\cosh(b)}
  {\sinh(b)}
  \right]\ ,
  \quad {\rm with\,\,inverse} \quad
  s(u)=\ln\left[
  \frac{1-\sin(u)}{\tanh(b/2)\cos(u)}
  \right]\ .$$
Under this change of variables the first integral in (\ref{eq:97}) becomes
$$
\int_{0}^{+\infty}
  \frac{ds\,\sinh(b)
  \cosh^2(b/2)
  e^{-s}}
  {(\cosh(b)\cosh(s)-\sinh(s))^2}
  T_{--}
  (\theta_\pm^E(z))=
\frac{1}{2}
\int^{u_0(b)}_{-\pi/2}
  du
  \left(
  1+\sin(u)
  \right)
  T_{--}(\theta^E_-=u,\theta^E_+(u,b))\ .
$$
where~$u_0(b)={\rm arcccot}(\sinh(b))$. The kernel is now independent of $b$, meaning we can safely take~$b\rightarrow 0$, which gives~$u_0\rightarrow \pi/2$. Moreover, the complicated function~$\theta_+^E(u,b)$ simplifies to~$\theta^E_+\rightarrow \pi/2$. Putting everything together in (\ref{eq:83}) we find that the only surviving contribution as $\epsilon,b\rightarrow 0$ is
\begin{equation}\label{eq:135}
\delta K_{A_0}\big|_{C}=
  \pi r_h^{d-2}
  \int_{S^{d-2}}d\Omega(\vec{v}\,)\,
  \zeta^-(\vec{v}\,)
  \int^{\pi/2}_{-\pi/2}
  du
  \left(
  1+\sin(u)
  \right)
  T_{--}(\theta^E_-=u,\theta^E_+=\pi/2)\ .
\end{equation}
The $u$ integral is over the future horizon of the undeformed region $\mathcal{D}(A_0)$. Changing the integration variable to the affine parameter $\lambda$ in (\ref{eq:103}) we find
\begin{equation}\label{eq:104}
\delta K_{A_0}\big|_{C}=
  \pi r_h^{d-2}
  \int_{S^{d-2}}d\Omega(\vec{x}_\perp)\,
  \zeta^-(\vec{x}_\perp)
  \int_{0}^{+\infty}
  d\lambda\,
  T_{\lambda \lambda}(\lambda,\vec{x}_\perp)
  \ ,
\end{equation}
where we recognize half the ANEC operator.

\paragraph{Branch cut contribution:}

Let us now compute the two additional contributions coming from the integrals over the surfaces $R_\pm$ in figure \ref{fig:6}. The induced metric and unit normal vectors are easily found from (\ref{eq:80}) and (\ref{eq:96}) as
\begin{equation}\label{eq:98}
\left.
\frac{ds^2}{r_h^2}\right|_{\tau=\tau_0}=
  d\rho^2+d\Omega^2_{d-2}\ ,
  \qquad \qquad
  n_\pm^\mu=
  \mp \frac{\delta^\mu_\tau}
  {r_h\sinh(\rho)} \ ,
\end{equation}
where $n_\pm^\mu$ corresponds to the (inward) unit normal vector to $R_\pm$.

Similarly to (\ref{eq:94}) we can write the stress tensor insertions evolved under the modular flow, appropriately choosing the parameter $\alpha$ so that the resulting operator is inserted at~$\tau=0$. In each case we obtain the following expressions
\begin{equation}\label{eq:101}
\begin{aligned}
R_+&:\qquad
  \rho_{A_0}
  ^{-\frac{i(s+i\alpha_+)}{2\pi}}
  T_{\mu \nu}(\tau=\epsilon,\rho)
  \rho_{A_0}^{\frac{i(s+i\alpha_+)}
  {2\pi}} 
  =\left(
  \frac{\partial x^{\bar{\mu}}}
  {\partial x^\mu}
  \frac{\partial x^{\bar{\nu}}}
  {\partial x^\nu}
  \right)
  \rho_{A_0}
  ^{-\frac{is}{2\pi}}
  T_{\bar{\mu }\bar{\nu}}
  (\tau=0,\rho)
  \rho_{A_0}^{\frac{is}
  {2\pi}}
  \ ,\\
R_-&:\qquad
  \rho_{A_0}
  ^{-\frac{i(s+i\alpha_-)}{2\pi}}
  T_{\mu \nu}(\tau=2\pi-\epsilon,\rho)
  \rho_{A_0}^{\frac{i(s+i\alpha_-)}
  {2\pi}} 
  =\left(
  \frac{\partial x^{\bar{\mu}}}
  {\partial x^\mu}
  \frac{\partial x^{\bar{\nu}}}
  {\partial x^\nu}\right)
  \rho_{A_0}
  ^{-\frac{is}{2\pi}}
  T_{\bar{\mu }\bar{\nu}}
  (\tau=0,\rho)
  \rho_{A_0}^{\frac{is}
  {2\pi}}
  \ ,
\end{aligned}
\end{equation}
where $\alpha_+=\epsilon$ and $\alpha_-=2\pi-\epsilon$. We have only applied the modular flow explicitly in the Euclidean direction, while left the flow in $s$ still written in terms of a complex power of $\rho_{A_0}$. 

The Jacobian matrices on the right-hand side of (\ref{eq:101}) are computed from (\ref{eq:100}) with $s=0$ and $\tau=\alpha_\pm$. Their behavior for small $\epsilon$ is always given by $\partial \theta_\pm^E(z)/\partial \theta_\pm^E=1+\mathcal{O}(\epsilon)$, where higher orders in $\epsilon$ drop out since they do not contribute in the limit $\epsilon\rightarrow 0$. This means the Jacobian matrices appearing in (\ref{eq:101}) are effectively equal to the identity and can be ignored.

Putting everything together, the contributions from $R_\pm$ to $\delta K_{A_0}$ in (\ref{eq:92}) are given by
\begin{equation}
\begin{aligned}
\delta K_{A_0}\big|_{R_+}&=
  r^{d-1}_h
  \int_{S^{d-2}}d\Omega\,
  \zeta^\nu(\vec{v}\,)
  \int_b^{+\infty}d\rho\,
  n_-^\mu 
  \int_{-\infty}
  ^{+\infty}
  \frac{-ds}{4\sinh^2(\frac{s+i\epsilon}
  {2})}
  \rho_{A_0}^{-\frac{is}{2\pi}}
  T_{\mu \nu}(\tau=0,\rho)
  \rho_{A_0}^{\,\frac{is}{2\pi}}\ ,
  \\
\delta K_{A_0}\big|_{R_-}&=
  r^{d-1}_h
  \int_{S^{d-2}}d\Omega\,
  \zeta^\nu(\vec{v}\,)
  \int_b^{+\infty}d\rho\,
  n_-^\mu 
  \int_{-\infty}
  ^{+\infty}
  \frac{ds}{4\sinh^2(\frac{s-i\epsilon}{2})}
  \rho_{A_0}^{-\frac{is}{2\pi}}
  T_{\mu \nu}(\tau=0,\rho)
  \rho_{A_0}^{\,\frac{is}{2\pi}}\ ,
\end{aligned}
\end{equation}
where the overall sign difference comes from $n_+^\mu=-n_-^\mu$. While there is no easy way of solving each integral separately, the contribution of both regions $R_+\cup R_-$ results in a path in the complex $s$ plane which simply picks the residue of the double pole at $s=0$. This can be easily computed and written as
$$
\delta K_{A_0}\big|_{R_+\cup R_-}=
  r_h^{d-1}
  \int_{S^{d-2}}d\Omega\,
  \zeta^\nu(\vec{v}\,)
  \int_0^{+\infty}d\rho\,
  n_-^\mu 
  \big[
  T_{\mu \nu}(\tau=0,\rho),
  K_{A_0}
  \big]\ .
$$
where we have safely taken the $\epsilon,b\rightarrow 0$ limits. The unit normal vector $n^\mu$ now points vertically upward in the Euclidean time direction $\partial_{\sigma_E}$. The resulting integral only involves the stress tensor over the undeformed region $A_0$ and we recognize it as the exponent of the unitary operator $U_{A_0}(\zeta)$ in (\ref{eq:102}), \textit{i.e.}
\begin{equation}\label{eq:105}
\delta K_{A_0}\big|_{R_+\cup R_-}=
  \left[
  \int_{A_0}dS^\mu
  T_{\mu \nu}(x)\zeta^\nu(\vec{v}\,),
  K_{A_0}
  \right]=
  \big[
  \delta U_{A_0},K_{A_0}
  \big]\ .
\end{equation}

\paragraph{All first order contributions:}

We can now put everything together in (\ref{eq:108}) to obtain an explicit expression for the modular hamiltonian $K_A$. Using (\ref{eq:104}) and (\ref{eq:105}) we find
$$
K_{A}=
  K_{A_0}+
  \pi r_h^{d-2}
  \int_{S^{d-2}}d\Omega(\vec{x}_\perp)\,
  \zeta^-(\vec{x}_\perp)
  \int_{0}^{+\infty}
  d\lambda\,
  T_{\lambda \lambda}(\lambda,\vec{x}_\perp)+
  \mathcal{O}(\zeta^\mu)^2\ ,
$$
which is precisely the result given in (\ref{eq:106}) used to prove the ANEC. Notice that the terms coming from the branch cut $R_+\cup R_-$ in (\ref{eq:105}) crucially canceled the contributions from the variation of the unitary $U_{A_0}$ in (\ref{eq:108}). This concludes the proof of the achronal ANEC for arbitrary QFTs in ${\rm AdS}_2\times S^{d-2}$.

\subsection{Proof for conformal theories}
\label{sub:2.3}

In this subsection we give an alternative proof of the achronal ANEC in ${\rm AdS}_2\times S^{d-2}$ that is valid for CFTs. Using a conformal transformation from Minkowski to the near horizon geometry we show that the full modular hamiltonian $\hat{K}_A$ is given by (\ref{eq:106}) to all orders in $\zeta^\mu$, which from (\ref{eq:109}) implies the ANEC. This subsection heavily relies on the calculations of \cite{Rosso:2019txh}.

Let us start by describing the setup in Minkowski, by taking Cartesian coordinates ${X^\mu=(T,X,\vec{Y})}$ and parametrizing the null plane $X-T=0$ as
\begin{equation}\label{eq:150}
X^\mu(\lambda,\vec{x}_\perp)=
  (\lambda,\lambda,\vec{x}_\perp)\ ,
  \qquad \qquad
  (\lambda,\vec{x}_\perp)
  \in \mathbb{R}\times \mathbb{R}^{d-2}\ ,
\end{equation}
where $\lambda$ is an affine parameter. For the Rindler region $\mathcal{R}_0$ described as $X_\pm\equiv X\pm T\ge 0$, we consider a null deformation given by the vector $\zeta^\mu=R(\vec{x}_\perp)\delta^\mu_+$. The vacuum full modular hamiltonian associated to the null deformed region $\mathcal{R}$ can be computed to every order in $\zeta^\mu$ and is given by \cite{Casini:2017roe,Balakrishnan:2020lbp}
\begin{equation}\label{eq:151}
\hat{K}_{\mathcal{R}}=2\pi
\int_{\mathbb{R}^{d-2}}d\vec{x}_\perp
\int_{-\infty}^{+\infty}d\lambda\,
(\lambda-R(\vec{x}_\perp))
\bar{T}_{\lambda \lambda}(\lambda,\vec{x}_\perp)\ .
\end{equation}
The idea is to apply a conformal transformation mapping the null plane (\ref{eq:150}) to the future horizon of $\mathcal{D}(A)$ in (\ref{eq:91}) and obtain the operator $\hat{K}_A$. See section 3 of \cite{Rosso:2019txh} for other examples and a more detailed analysis involving conformal transformations of this modular hamiltonian.

A conformal transformation is applied by first considering a change of coordinates $X^\mu\rightarrow x^\mu$ which puts the Minkowski metric in the form $ds^2_{\rm Mink}=w^2(x^\mu)ds^2$, followed by a  Weyl rescaling removing the factor $w^2(x^\mu)$. The end result is a transformation from Minkowski to some other space-time $ds^2$. When mapping the geodesics in the null plane (\ref{eq:150}) there is no need to change the parametrization coordinates $(\lambda,\vec{x}_\perp)$. It is however important to note that $\lambda$ might not be affine in the new space-time (see subsection 2.1 of \cite{Rosso:2019txh}). The conformal transformation is implemented on the Hilbert space $\bar{\mathcal{H}}$ of the Minkowski CFT by a unitary operator $U:\bar{\mathcal{H}}\rightarrow \mathcal{H}$. The stress tensor in $\bar{T}_{\lambda \lambda}$ in (\ref{eq:151}) transforms according to (see subsection 2.2 of \cite{Rosso:2019txh})
\begin{equation}\label{eq:152}
U\bar{T}_{\lambda \lambda}U^\dagger
  =
  \frac{T_{\lambda \lambda}-
  \langle 
  T_{\lambda \lambda}
  \rangle_0}{w(\lambda)^{d-2}}\ ,
\end{equation}
where $w(\lambda)$ is the conformal factor evaluated along on the null plane (\ref{eq:150}). The term $\langle T_{\lambda \lambda} \rangle_0=\bra{0}T_{\lambda \lambda}\ket{0}$, where $\ket{0}$ the vacuum state of the mapped CFT, arises from the anomalous transformation of the stress tensor for even $d$.

Let us now specialize to the conformal transformation relating Minkowski to ${\rm AdS}_2\times S^{d-2}$. We split the mapping in two steps $(A)$ and $(B)$, given by 
\begin{equation}\label{eq:120}
\mathbb{R}\times \mathbb{R}^{d-1}
  \qquad 
  \xrightarrow[]{(A)} 
  \qquad
  \mathbb{R}\times S^{d-1}
  \qquad 
  \xrightarrow[]{(B)} 
  \qquad
  {\rm AdS}_2\times S^{d-2}\ ,
\end{equation}
where $\mathbb{R}\times S^{d-1}$ is the Lorentzian cylinder. This way of applying the transformation has the advantage that the first step $(A)$ was already considered in subsection 3.1 of \cite{Rosso:2019txh}.\footnote{The convenience of the mapping described in subsection 3.1 of \cite{Rosso:2019txh} is that it first applies a special conformal transformation mapping the Minkowski null plane to the Minkowski null cone. The necessity of considering the Lorentzian cylinder comes from the fact that special conformal transformations are not globally well defined in Minkowski, see \cite{Rosso:2019txh} for details.} Writing the metric in the cylinder as
\begin{equation}\label{eq:35}
\frac{ds_{\rm cyl}^2}{r_h^2}=-d\sigma^2+d\theta^2+
  \sin^2(\theta)d\Omega^2_{d-2}(\vec{v}\,)
  \ ,
\end{equation}
with $d\Omega_{d-2}$ in (\ref{eq:87}), it was shown that the null plane in (\ref{eq:150}) is mapped to the following surface in the cylinder\footnote{These expressions are given in (3.17) and (3.15) of \cite{Rosso:2019txh} after making some redefinitions to match with the conventions used in this paper. The time coordinate $\sigma$ in each case are related according to ${\sigma_{\rm here}=\sigma_{\rm there}/r_h+\pi/2}$. The function $p(\vec{x}_\perp)$, defining the stereographic coordinates on $S^{d-2}$ in (3.8) of \cite{Rosso:2019txh}, is considered with $R=1/2$, so that $2p(\vec{x}_\perp)=|\vec{x}_\perp|^2+1$, matching with our definition in (\ref{eq:87}). Finally, here we are using a different affine parameter $\lambda$, that is related by an affine transformation according to the one used in \cite{Rosso:2019txh} $\lambda_{\rm here}=(\lambda_{\rm there}+p(\vec{x}_\perp))/2p(\vec{x}_\perp)$. }
\begin{equation}\label{eq:36}
x^\mu(\lambda,\vec{x}_\perp)=
  (\theta_+,\theta_-,\vec{v}\,)=
  \left(
  \pi/2,2\,
  {\rm arccot}(\lambda)-\pi/2
  ,\vec{x}_\perp
  \right)\ ,
\end{equation}
where the conformal factor $w_A(x)$ defined from~${ds^2_{\rm Mink}=w^2_A(x)ds^2_{\rm cyl}}$ and evaluated on the null surface is
$$
w^2_{A}(\lambda)=\left(
  \frac{1+|\vec{x}_\perp|^2}{2r_h}
  \right)^2
  (1+\lambda^2)\ .
$$
Once we have the surface in the cylinder it is straightforward to apply the mapping $(B)$ in~(\ref{eq:120}). As noted in \cite{Galloway:2020xfz}, writing the metric in the cylinder (\ref{eq:35}) as
$$\frac{ds_{\rm cyl}^2}{r_h^2}=
  \sin^2(\theta)
  \left[
  \frac{-d\sigma^2+d\theta^2}
  {\sin^2(\theta)}+
  d\Omega_{d-2}^2(\vec{v}\,)
  \right]\ ,$$
we can perform a simple Weyl rescaling $w^2_B(x^\mu)=\sin^2(\theta)$ to obtain ${\rm AdS}_2\times S^{d-2}$. The overall conformal factor resulting from both transformations in (\ref{eq:120}) evaluated on the null plane is
\begin{equation}\label{eq:121}
w^2(\lambda)=
  w^2_A(\lambda)w_B^2(\lambda)=
  \left(
  \frac{1+|\vec{x}_\perp|^2}{2r_h}
  \right)^2
  (1+\lambda^2)
  \sin^2(\theta(\lambda))=
  \left(
  \frac{1+|\vec{x}_\perp|^2}{2r_h}
  \right)^2\ ,
\end{equation}
that is independent of $\lambda$! This is crucial, since it implies that the affine parameter in the null plane $\lambda$ is also affine for the surface (\ref{eq:36}) in ${\rm AdS}_2\times S^{d-2}$. This becomes evident after comparing (\ref{eq:36}) with (\ref{eq:103}). 

Using this on the transformation of the stress tensor (\ref{eq:152}), we can map the Rindler null deformed modular hamiltonian (\ref{eq:151}) to $\hat{K}_A$ in ${\rm AdS}_2\times S^{d-2}$ as
\begin{equation}\label{eq:153}
\hat{K}_A=
  U\hat{K}_{\mathcal{R}}
  U^\dagger=
  2\pi r_h^{d-2}\int_{\mathbb{R}^{d-2}} d\vec{x}_\perp
  \left(
  \frac{2}{1+|\vec{x}_\perp|^2}
  \right)^{d-2}
  \int_{-\infty}^{+\infty}d\lambda\,
  (\lambda-R(\vec{x}_\perp))
  T_{\lambda \lambda}(\lambda,\vec{x}_\perp)\ .
\end{equation}
We have used that since the geodesics (\ref{eq:36}) only have a non-trivial motion along the directions of ${\rm AdS}_2$, the vacuum contribution $\langle T_{\lambda \lambda} \rangle_0$ vanishes (see appendix \ref{zapp:2} for details). Moreover, notice that the transverse integral over $\vec{x}_\perp$ rearranged itself in just the appropriate way to give the integration measure over the unit sphere $S^{d-2}$ in stereographic coordinates (\ref{eq:87}). 

To compare with the previous result obtained for $\hat{K}_A$ in (\ref{eq:112}), we must relate the function $R(\vec{x}_\perp)$ and the null vector $\zeta^\mu=\zeta^-(\vec{v}\,)\delta^\mu_-$, characterizing $\mathcal{D}(A)$ in (\ref{eq:91}). Expanding (\ref{eq:36}) for small $\lambda=R(\vec{x}_\perp)$ we find
$$\theta_-(\lambda,\vec{x}_\perp)\big|_{\lambda=R(\vec{x}_\perp)}=
  \pi/2-2R(\vec{x}_\perp)+\mathcal{O}(R(\vec{x}_\perp))^2\ .$$
Comparing with (\ref{eq:91}) we identify $\zeta^-(\vec{x}_\perp)=-2R(\vec{x}_\perp)$. Putting everything together we can write (\ref{eq:153}), the null deformed full modular hamiltonian in ${\rm AdS}_2\times S^{d-2}$, as
$$\hat{K}_A=
  \hat{K}_{A_0}
  +
  \pi r_h^{d-2}\int_{S^{d-2}}d\Omega(\vec{x}_\perp)
  \zeta^-(\vec{x}_\perp)
  \mathcal{E}(\vec{x}_\perp)\ .$$
with $\hat{K}_{A_0}$ given by the first term in (\ref{eq:153}). This shows that for conformal theories, the result in (\ref{eq:112}) to first order in perturbation theory is exact to every order. Moreover, monotonicity of relative entropy (\ref{eq:109}) gives an independent proof of the achronal ANEC for CFTs in ${\rm AdS}_2\times S^{d-2}$.

\subsection{Proof for free scalar}
\label{sub:2.4}

In this subsection we give a third (and final) proof of the achronal ANEC in ${\rm AdS}_2\times S^{d-2}$ for the simple case of a free scalar field. This provides a sanity check for the previous more general and abstract proofs. We follow the simple approach used in \cite{Klinkhammer:1991ki} for Minkowski, which involves showing that $\mathcal{E}(\vec{x}_\perp)=WW^\dagger\ge 0$ for some operator $W$.\footnote{I am grateful to Joan La Madrid for discussions and collaboration regarding this subsection.} 

An arbitrary free scalar in a curved manifold is characterized by the action
\begin{equation}\label{eq:124}
I[\phi]=
  -\frac{1}{2}\int
  d^dx\sqrt{-g}\left[
  g^{\mu \nu}
  \left(\partial_\mu \phi\right)
  \left(\partial_\mu \phi\right)+
  \left(
  m_0^2+\xi\mathcal{R}
  \right)\phi^2
  \right]\ ,
\end{equation}
where $m_0$ is the bare mass and $\xi$ the non-minimal coupling to the space-time geometry. Varying the action with respect to the metric it is straightforward to compute the stress tensor and find
\begin{equation}\label{eq:51}
T_{\mu \nu}=
  \left(\partial_\mu \phi\right)
  \left(\partial_\nu \phi\right)+
  \xi 
  \left(
  \mathcal{R}_{\mu \nu}
  -
  \nabla_{\mu }\nabla_\nu
  \right)\phi^2+
  g_{\mu \nu}\left(
  2\xi-
  1/2
  \right)
  \Big[
  \left(\partial^\alpha \phi\right)
  \left(\partial_\alpha \phi\right)+
  (m_0^2+\xi\mathcal{R})\phi^2
  \Big]
  \ ,
\end{equation}
where we have used the equations of motion to write it in this way. 

The tangent vector $k^\mu$ to the complete achronal null geodesics is obtained by differentiating~(\ref{eq:103}) with respect to $\lambda$. Projecting the stress tensor in the $k^\mu$ direction, we can drop the third term. Since $k^\mu$ only has non-trivial components in the ${\rm AdS}_2$ sector, which is maximally symmetric, we have $R_{\lambda \lambda}\equiv 0$. The covariant derivatives in the second term are replaced by ordinary $\lambda$ derivatives after using that $k^\mu$ satisfies the geodesic equation $k^\nu\nabla_\nu k^\mu=0$. The null component of the stress tensor is written as
$$T_{\lambda \lambda}(\lambda,\vec{x}_\perp)=
  \phi'(\lambda)^2-\xi
  \partial_\lambda^2\phi(\lambda)^2\ .$$
Integrating over $\lambda \in \mathbb{R}$ to get the ANEC operator, the second term gives a boundary contribution which drops and we find
\begin{equation}\label{eq:156}
\mathcal{E}(\vec{x}_\perp)\equiv
  \int_{-\infty}^{+\infty}d\lambda\,
  T_{\lambda \lambda}(\lambda,\vec{x}_\perp)=
  \int_{-\infty}^{+\infty}d\lambda
  \,\phi'(\lambda)^2\ .
\end{equation}
All classical scalar fields in ${\rm AdS}_2\times S^{d-2}$ trivially satisfy the achronal ANEC.

This property is not obviously true after we quantize the theory, since the the operator $\phi'(\lambda)^2$ only makes sense after regularization, which spoils its positivity. Given that the scalar is free and hermitian it can be written as $\phi(x^\mu)=\phi^+(x^\mu)+\phi^-(x^\mu)$ where $\phi^+(x^\mu)=\phi^-(x^\mu)^\dagger$ is expanded in terms of creation and annihilation operators as
\begin{equation}\label{eq:44}
\phi^+(x^\mu)=
  \sum_{i}H_i(x^\mu)a_i\ ,
  \qquad \qquad
  \left[
  a_i,a_{i'}^\dagger
  \right]=\delta_{ii'}\ .
\end{equation}
The label $i$ goes over the linearly independent and orthogonal functions $H_i(x^\mu)$ solving the equations of motion. Using this expansion we can write the ANEC operator in (\ref{eq:156}) as
\begin{equation}\label{eq:43}
\mathcal{E}(\vec{x}_\perp)=
  \int_{-\infty}^{+\infty}d\lambda\,
  \langle T_{\lambda \lambda} \rangle_0+
  \int_{-\infty}^{+\infty}d\lambda 
  \left[
  (\partial_\lambda\phi^+)^\dagger
  (\partial_\lambda\phi^+)+
  (\partial_\lambda\phi^+)
  (\partial_\lambda\phi^+)
  +
  {\rm h.c.} 
  \right]
  \ ,
\end{equation}
where ${\rm h.c.}$ is the hermitian conjugate and we have identified the vacuum contribution as
$$\langle T_{\lambda \lambda} \rangle_0=
  \left[
  (\partial_\lambda \phi^+),
  (\partial_\lambda \phi^+)^\dagger
  \right]\ .$$
This is divergent and requires a regularization procedure. After regularization, the vacuum contribution vanishes $\langle T_{\lambda \lambda} \rangle_0=0$ due to the general arguments given in appendix \ref{zapp:2}. While the first term in the second integral (\ref{eq:43}) is an explicitly positive operator, the other is not. This means the ANEC holds if and only if the integral over $\lambda$ of this additional term vanishes, which gives the following condition
\begin{equation}\label{eq:46}
\mathcal{E}(\vec{x}_\perp)\ge 0
\qquad \Longleftrightarrow \qquad
  C_{ii'}=
  \int_{-\infty}^{+\infty}d\lambda\,
  H_i'(\lambda)
  H_{i'}'(\lambda)
  =0\ ,
  \qquad \forall\,\,(i,i')\ .
\end{equation}

The proof of the ANEC is reduced to computing some integrals. Let us do this by first writing $H(x^\mu)$ obtained from solving the equation of motion, given by
\begin{equation}\label{eq:47}
\left(
\nabla^2_{{\rm AdS}_2}
+\nabla^2_{S^{d-2}}
\right)H(x^\mu)=
  (m_0^2+\xi\mathcal{R})
  H(x^\mu)\ .
\end{equation}
Writing an ansatz with $H(x^\mu)=f(\sigma,\theta)Y_\ell(\vec{v}\,)$ with $Y_\ell(\vec{v}\,)$, the eigenfunctions of the $S^{d-2}$ Laplacian, we find
\begin{equation}\label{eq:45}
  \nabla^2_{{\rm AdS}_2}f(\sigma,\theta)=
  \mu^2
  f(\sigma,\theta)\ ,
  \qquad {\rm where} \qquad
  \mu^2=
  \frac{\ell(\ell+d-3)}{r_h^2}
  +(m_0^2+\xi\mathcal{R})\ ,
\end{equation}
with $\ell \in \mathbb{N}_0$. The function $f(\sigma,\theta)$ satisfies the differential equation of a scalar field in ${\rm AdS}_2$ with an effective mass $\mu^2$. The solution can be written as
$$f_{n,\Delta}(\sigma,\theta)=
  \sin^\Delta(\theta)
  F_{21}\left(
  -n,\Delta+n,1/2,\cos^2(\theta)
  \right)
  e^{-i(\Delta +2n)\sigma}\ ,
$$
where $n\in \mathbb{N}_0$ and
\begin{equation}\label{eq:48}
\Delta=
  \frac{1+\sqrt{1+(2r_h\mu )^2}}{2}\ge 1\ .
\end{equation}

To prove the achronal ANEC using (\ref{eq:46}) we can forget about the dependence of $H(x^\mu)$ on the $S^{d-2}$ coordinates, since the complete achronal null geodesic in (\ref{eq:36}) has fixed values of~$\vec{x}_\perp$. Moreover, it is convenient to translate the geodesics in (\ref{eq:36}) by redefining $\sigma\rightarrow \sigma-\pi/2$. Changing the integration variable in (\ref{eq:46}) to $\beta(\lambda)={\rm arccot}(\lambda)$ we get
\begin{equation}\label{eq:50}
C_{(n,\Delta)(n',\Delta')}=
\int_{0}^{\pi}d\beta\,
  \sin^2(\beta)
  f'_{n,\Delta}(-\beta,\beta)
  f'_{n',\Delta'}(-\beta,\beta)\ ,
\end{equation}
where the derivatives are now with respect to $\beta$. The ANEC is proven by showing $C_{(n,\Delta)(n',\Delta')}=0$.

Let us start by considering the simple case in which $n=n'=0$, so that the integral can be easily written and solved analytically as
\begin{equation}\label{eq:49}
C_{(0,\Delta)(0,\Delta')}=\Delta\Delta'
\int_0^\pi d\beta\,
\sin(\beta)^{\Delta+\Delta'}
e^{-i\beta(2+\Delta+\Delta')}=0\ .
\end{equation}
For the first few values of $(n,n')$ the integral can still be solved analytically and shown to vanish, as we have explicitly checked for all combinations involving $n,n'<3$. For higher values the analytic computation becomes very complicated and its convenient to integrate numerically. We have checked that the integral still vanishes with a numerical precision of up to eleven digits, for $\Delta=\Delta'$ and all combinations $(n,n')$ up to $n,n'\le 10$ and for $\Delta=1/2$ to $\Delta=20$ in half-integer steps. An analytic result which holds for arbitrary values of $(n,n')$ can be obtained for the particular case of $\Delta=\Delta'=1$, where the integral simplifies to
$$C_{(n,1)(n',1)}=
  (-1)^{n+n'}
  (1+2n)
  (1+2n')
  \int_0^\pi
  d\beta\,
  \sin^2(\beta)
  e^{i4\beta (1+n+n')}
  =0\ .$$
We have also solved the integral numerically for random values of all the parameters and found always a vanishing answer. Overall, we have found enough analytic and numerical evidence to safely conclude the integral in (\ref{eq:50}) is identically zero, meaning the achronal ANEC holds for any free scalar in ${\rm AdS}_2\times S^{d-2}$. 

While there might be a general analytic result showing the integral vanishes in full generality, we have not been able to find it. Using the equations of motion to simplify and solve the integral has not been useful. An experimental observation is that after solving the indefinite integral the result seems to always be given by $\sin(\beta)^{1+\Delta+\Delta'}g_{nn'}(\beta)$ where $g_{nn'}(\beta)$ is a regular function at $\beta=0,\pi$. The integral vanishes after evaluating at $\beta=0,\pi$.

\section{Achronal ANEC for maximally symmetric space-times}
\label{sec:3}

In this section we show how the previous calculations can be generalized to prove the ANEC for arbitrary QFTs in de Sitter and anti-de Sitter. More precisely we find that the computation of the null deformed modular hamiltonian in subsection \ref{sub:2.2} can be adapted to these space-times. This generalizes the recent proof of the ANEC for CFTs in (A)dS given in \cite{Rosso:2019txh}.

\subsection{De Sitter}

Let us start by considering an arbitrary QFT defined on de Sitter, which is described in global coordinates as
\begin{equation}\label{eq:132}
\frac{ds^2}{L^2}=\frac{-d\sigma^2+
  d\theta^2+\sin^2(\theta)d\Omega_{d-2}^2
  }{\cos^2(\sigma)}\ ,
\end{equation}
where $L$ is the de Sitter length scale and $\theta\in[0,\pi]$. The time coordinate is constrained to $|\sigma|<\pi/2$, with the spatial dS boundaries located at $\sigma=\pm \pi/2$. 

Following the calculations in subsection \ref{sub:2.2}, we consider the vacuum state $\ket{0}$ and reduce it to the half-space $A_0$ at $\sigma=0$. Since the topology of dS is that of $\mathbb{R}\times S^{d-1}$, the half-space in this case corresponds to the spherical cap given by $\theta\in[0,\pi/2]$. Its causal domain $\mathcal{D}(A_0)$ is easily described in terms of the null coordinates $\theta_\pm=\theta\pm\sigma$ as
$$\mathcal{D}(A_0)=\left\lbrace
  (\sigma,\theta,\vec{v}\,)
  \in 
  (-\pi/2,\pi/2)
  \times [0,\pi]\times
  \mathbb{R}^{d-2}:
  \quad
  \theta_+<\pi/2\ ,
  \quad
  \theta_-<\pi/2\,
  \right\rbrace\ .$$
This region is equivalent to the static patch of de Sitter. Its null deformation in the direction~$\theta_-$ can be parametrized by the vector $\zeta^\mu=\zeta^-(\vec{v}\,)\delta^\mu_-$ as
\begin{equation}\label{eq:133}
\mathcal{D}(A)=\left\lbrace
  (\sigma,\theta,\vec{v}\,)
  \in 
  (-\pi/2,\pi/2)
  \times [0,\pi]\times
  \mathbb{R}^{d-2}:
  \quad
  \theta_+<\pi/2\ ,
  \quad
  \theta_-<\pi/2+\zeta^-(\vec{v}\,)\,
  \right\rbrace\ .
\end{equation}
The diagrams of these regions are the same as the ones given in figure \ref{fig:8} for ${\rm AdS}_2\times S^{d-2}$. The difference is that the time coordinate is restricted to $|\sigma|<\pi/2$ and there are no boundaries at $\theta=0,\pi$, since in (\ref{eq:132}) these correspond to the North and South pole of the spatial $S^{d-1}$.\footnote{Since the time coordinate in dS is constrained to $|\sigma|<\pi/2$ the region $\mathcal{D}(A)$ for $\zeta^-(\vec{v}\,)$ intersects with the spatial boundary at $\sigma=-\pi/2$, see figure \ref{fig:8}. The full modular hamiltonian we use to prove the ANEC is not affected by this.} Let us now compute the first order contribution to the modular hamiltonian associated to~$\mathcal{D}(A)$.

\paragraph{Undeformed region:}

We start by considering the vacuum modular hamiltonian associated to the undeformed region $\mathcal{D}(A_0)$. Analytically continuing to Euclidean time $\sigma_E=i\sigma$ the de Sitter metric (\ref{eq:132}) becomes  
\begin{equation}\label{eq:138}
\frac{ds^2_{E}}{L^2}=
  \frac{d\sigma_E^2+d\theta^2+
  \sin^2(\theta)d\Omega_{d-2}^2}
  {\cosh^2(\sigma_E)}\ .
\end{equation}
Since the function in the denominator does not vanish, the Euclidean time is free to take any real value $\sigma_E\in \mathbb{R}$. 

As in (\ref{eq:72}), the path integral over the region $\sigma_E<0$  gives a representation of the vacuum state~$\ket{0}$, so that the reduced density operator $\rho_{A_0}$ is obtained from the path integral with the boundary conditions given in~(\ref{eq:69}). We should now look for an isometry of the Euclidean manifold~(\ref{eq:138}) that smoothly maps between the surfaces~$\theta\in[0,\pi/2)$ at $\sigma_E=0^\pm$. Quite surprisingly, the appropriate isometry is exactly the same as in ${\rm AdS}_2\times S^{d-2}$ (given by (\ref{eq:70})) whose action in the~$(\sigma_E,\theta)$ plane is shown in figure \ref{fig:4}. It is straightforward to check that~(\ref{eq:138}) is invariant under this transformation. The fact that the modular flow of both these space-times is the same is not trivial to us.

The modular hamiltonian is obtained from the generator of the Lorentzian isometry (\ref{eq:71}) according to (\ref{eq:73})
\begin{equation}\label{eq:157}
K_{A_0}=
  2\pi\int_{\Sigma}
  dS^\mu T_{\mu \nu}\xi^\nu+
  {\rm const}\ ,
  \qquad \qquad
  \xi^\mu=\cos(\theta_+)\partial_+-
  \cos(\theta_-)\partial_-\ ,
\end{equation}
where $\Sigma$ is a Cauchy surface on $\mathcal{D}(A_0)$. We can write this explicitly by taking $\Sigma$ as the future null horizon of $\mathcal{D}(A_0)$, which can be parametrized as
\begin{equation}\label{eq:126}
x^\mu(\lambda,\vec{x}_\perp)=
  (\theta_+,\theta_-,\vec{v}\,)=(
  \pi/2,2\,{\rm arccot}(\lambda)-\pi/2,
  \vec{x}_\perp)\ ,
  \qquad \qquad
  (\lambda,\vec{x}_\perp)\in 
  \mathbb{R}\times \mathbb{R}^{d-2}\ ,
\end{equation}
where for fixed $\vec{x}_\perp$ the parameter $\lambda$ is affine in the dS metric (\ref{eq:132}). This description coincides with the null surface in (\ref{eq:103}) for ${\rm AdS}_2\times S^{d-2}$, where $\lambda$ is also affine. While the coordinate description of the null curves coincides, the geodesics travel along very different space-times. From (\ref{eq:157}), the modular hamiltonian for $\mathcal{D}(A_0)$ is written along this null surface as
$$K_{A_0}=
  2\pi L^{d-2}
  \int_{S^{d-2}}
  d\Omega(\vec{x}_\perp)
  \int_0^{+\infty}
  d\lambda\,\lambda\,
  T_{\lambda \lambda}(\lambda,\vec{x}_\perp)+{\rm const}\ ,$$
which has the same structure as for ${\rm AdS}_2\times S^{d-2}$ in (\ref{eq:107}). 

\paragraph{First order deformation:}

Given that the modular flow and affine parameter $\lambda$ on de Sitter and ${\rm AdS}_2\times S^{d-2}$ coincide for the undeformed region, the computation of the first order null deformation is almost exactly equivalent the one in subsection \ref{sub:2.2}. The series expansion in~$\zeta^-(\vec{v}\,)$ is given by (\ref{eq:108}) with $\delta K_{A_0,g}$ in (\ref{eq:92}), which we rewrite here for convenience
\begin{equation}\label{eq:139}
\delta K_{A_0}=
  \int_{-\infty+i\alpha}^{+\infty+i\alpha}
  \frac{dz}{4\sinh^2(z/2)}
  \int_{\partial \mathcal{M}_E}
  dS^\mu \zeta^\nu
  \rho_{A_0}^{-\frac{iz}{2\pi}}
  T_{\mu \nu}(x)
  \rho_{A_0}^{\,\frac{iz}{2\pi}}\ .
\end{equation}
Since the undeformed modular flow is the same, so is its action on the stress tensor $T_{\mu \nu}(x)$. The difference comes from the integral $\partial \mathcal{M}_E$, which can be described in terms of the surfaces $C\cup R_\pm$ (\ref{eq:96}) plotted in figure \ref{fig:6}. Writing the Euclidean dS metric in the coordinates $(\tau,\rho)$ in (\ref{eq:79}) we find
$$\frac{ds^2_E}{L^2}=
  \frac{\sinh^2(\rho)d\tau^2+d\rho^2+d\Omega_{d-2}^2}
  {\cosh^2(\rho)}\ .$$
This differs only by the overall factor $\cosh^2(\rho)$ with respect to the metric in ${\rm AdS}_2\times S^{d-2}$ (\ref{eq:80}), which contributes to $dS^\mu$ in (\ref{eq:139}).

The integral over $C$ in (\ref{eq:96}) is given by $\rho=b$ with $b\rightarrow 0$. Since $\cosh(b)\rightarrow 1$, the calculation leading to (\ref{eq:104}) is identical and gives the same result
$$\delta K_{A_0}\big|_{C}=\pi L^{d-2}
  \int_{S^{d-2}}d\Omega(\vec{x}_\perp)
  \zeta^-(\vec{x}_\perp)
  \int_0^{+\infty}d\lambda\,
  T_{\lambda \lambda}(\lambda,\vec{x}_\perp)\ .$$
For the integral over the surface $R_\pm$ (given essentially by $\tau=0^\pm$) the factor $\cosh^2(\rho)$ contributes in a non-trivial way to give the induced surface element of dS at $\sigma_E=0$, which is different from that on ${\rm AdS}_2\times S^{d-2}$. However, this is exactly what we require so that $\delta K_{A_0}\big|_{R_+\cup R_-}$ in (\ref{eq:105}) gives $\delta U_{A_0}$, but with the unitary $U_{A_0}$ defined as in (\ref{eq:102}) on the surface $\sigma_E=0$ in de Sitter. This contribution is essential to cancel the commutator in (\ref{eq:108}), so that the end result for the null deformed modular hamiltonian in dS gives
\begin{equation}\label{eq:158}
K_{A}=
  K_{A_0}+
  \pi L^{d-2}
  \int_{S^{d-2}}d\Omega(\vec{x}_\perp)\,
  \zeta^-(\vec{x}_\perp)
  \int_{0}^{+\infty}
  d\lambda\,
  T_{\lambda \lambda}(\lambda,\vec{x}_\perp)+
  \mathcal{O}(\zeta^\mu)^2\ .
\end{equation}
Using relative entropy, the same calculation leading to (\ref{eq:109}) gives the ANEC for a QFT in de Sitter. This modular hamiltonian agrees with the result of \cite{Rosso:2019txh}, which computed the operator for arbitrary CFTs to every order in the deformation parameter $\zeta^\mu$, using a conformal transformation as in subsection \ref{sub:2.3}.

\subsection{Anti-de Sitter}

\begin{figure}
\centering
\includegraphics[scale=0.30]{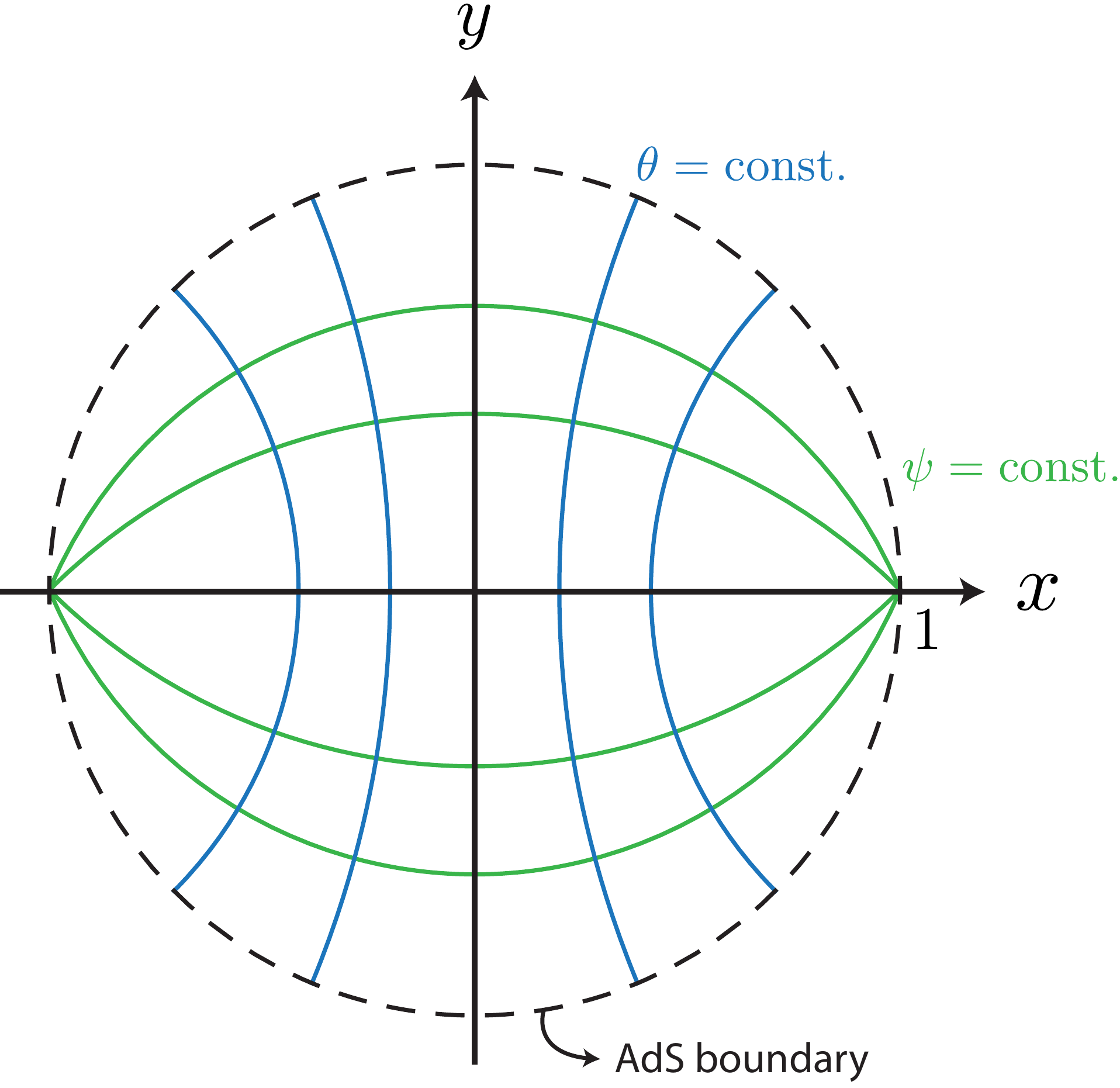}
\caption{Constant time section of ${\rm AdS}_3$ given by the region $x^2+y^2< 1$. We have plotted the constant $\theta$ and $\psi$ trajectories obtained from (\ref{eq:131}). The half space $A_0$ defined from $x>0$ corresponds to $\theta<\pi/2$.}\label{fig:9}
\end{figure}

An analogous construction can be considered for a QFT in anti-de Sitter. There is a more subtle aspect when it comes to finding an appropriate set of coordinates to describe the half-space region of AdS, given by half a cross section of the solid cylinder. The easiest way of parametrizing this region, is by considering some sort of spatial Cartesian coordinate $\vec{x}\in \mathbb{R}^{d-1}$, such that the boundary is located at $|\vec{x}|=1$. These coordinates can be defined from the embedding description of AdS, given by the surface
\begin{equation}\label{eq:129}
-(X^0)^2-(X^1)^2+
  \sum_{i=2}^{d}(X^i)^2=-L^2\ ,
\end{equation}
in the space $\mathbb{R}^2\times \mathbb{R}^{d-1}$
$$ds^2=-(dX^0)^2
  -(dX^1)^2
  +\sum_{i=2}^{d}(dX^i)^2
  \ .$$
The constraint (\ref{eq:129}) is automatically satisfied if we define the coordinates $x^\mu=(\sigma,\vec{x}\,)$ as
\begin{equation}\label{eq:130}
X^0=L\left(
  \frac{1+|\vec{x}\,|^2}
  {1-|\vec{x}\,|^2}
  \right)\sin(\sigma)\ ,
  \quad \quad
  X^1=L\left(
  \frac{1+|\vec{x}\,|^2}
  {1-|\vec{x}\,|^2}
  \right)
  \cos(\sigma)\ ,
  \quad \quad
  X^i=L
  \frac{2x^i}{1-|\vec{x}\,|^2}\ .
\end{equation}
where $\sigma \in \mathbb{R}$ and $\vec{x}\in \mathbb{R}^{d-1}$ with $|\vec{x}|<1$. The induced metric on $\mathbb{R}^2\times \mathbb{R}^{d-1}$ gives the metric in global AdS
\begin{equation}\label{eq:127}
\frac{ds^2}{L^2}=
  -\left(
  \frac{1+|\vec{x}|^2}
  {1-|\vec{x}|^2}
  \right)^2d\sigma^2+
  \frac{4|d\vec{x}|^2}
  {(1-|\vec{x}|^2)^2}\ .
\end{equation}
The boundary is located at $|\vec{x}|=1$, with the interior of the solid cylinder described by the Cartesian coordinates $\vec{x}$. This unusual way of writing the AdS metric allows for a simple description of the half space. Picking an arbitrary direction in $\vec{x}$ we write $\vec{x}=(x,\vec{y}\,)$ with $\vec{y}\in \mathbb{R}^{d-2}$, so that the half-space $A_0$ at $\sigma=0$ is given by
$$A_0=\left\lbrace 
  (\sigma,x,\vec{y}\,)
  \in 
  \mathbb{R}\times 
  \mathbb{R}\times 
  \mathbb{R}^{d-2}:
  \quad
  x^2+|\vec{y}\,|^2< 1\ ,
  \quad
  \sigma=0 \ ,
  \quad
  x> 0\,
  \right\rbrace\ .$$
A constant time surface for ${\rm AdS}_3$ is plotted in figure \ref{fig:9}, where $A_0$ corresponds to the $x>0$ region.  

Although the half-space $A_0$ has a very simple description in these coordinates, they are not well suited to describe its causal domain $\mathcal{D}(A_0)$, given that null geodesics in (\ref{eq:127}) are not given by simple straight lines. We can fix this by applying another change of coordinates $(x,|\vec{y}\,|)\rightarrow (\theta,\psi)$, defined as\footnote{These relations are inspired by holographic calculations of entanglement entropy in appendix B of \cite{Rosso:2019txh}.}
\begin{equation}\label{eq:131}
\begin{aligned}
x&=\frac{\cos(\theta)}
{1+\sin(\theta)\sin(\psi)}\ ,\\
|\vec{y}\,|&=
\frac{\sin(\theta)\cos(\psi)}
{1+\sin(\theta)\sin(\psi)}\ ,
\end{aligned}
\quad
\qquad {\rm with\,\,inverse} 
\qquad
\quad
\begin{aligned}
\cos(\theta)&=
\frac{2x}{1+(x^2+|\vec{y}\,|^2)}\ ,\\
\cot(\psi)&=
\frac{2|\vec{y}\,|}
{1-(x^2+|\vec{y}\,|^2)}\ ,
\end{aligned}
\end{equation}
where $\theta\in(0,\pi)$ and $\psi\in(0,\pi/2]$. Taking stereographic coordinates $\vec{v}\in \mathbb{R}^{d-3}$ to describe the angular direction of $\vec{y}$, the AdS metric (\ref{eq:127}) becomes
\begin{equation}\label{eq:137}
\frac{ds^2}{L^2}=
  \frac{1}{\sin^2(\psi)}\left[
  \frac{-d\sigma^2+d\theta^2}
  {\sin^2(\theta)}+
  d\psi^2+\cos^2(\psi)d\Omega_{d-3}^2(\vec{v}\,)
  \right]\ .
\end{equation}
Although hardly recognizable, this metric describes global AdS. There are two ways we can approach the boundary, obtained by taking $\psi\rightarrow 0$ or $\theta\rightarrow 0,\pi$, which from (\ref{eq:131}) corresponds to $x^2+|\vec{y}\,|^2\rightarrow 1$ and $x \rightarrow\pm 1$ respectively. The $x$ direction is mainly controlled by $\theta$, with the half-space $x>0$ given by $\theta<\pi/2$. In figure \ref{fig:9} we plot the constant $(\theta,\psi)$ trajectories on a fixed time slice of ${\rm AdS}_3$.

The crucial aspect of (\ref{eq:137}) is that null curves in the $(\sigma,\theta)$ direction are straight lines. This is precisely what is needed to describe the causal domain of the half-space $A_0$, that is given by
$$\mathcal{D}(A_0)=\left\lbrace
  (\sigma,\theta,\psi,\vec{v}\,)
  \in 
  \mathbb{R}
  \times (0,\pi)
  \times (0,\pi/2]
  \times \mathbb{R}^{d-3}:
  \quad
  \theta_+< \pi/2\ ,
  \quad
  \theta_-< \pi/2\,
  \right\rbrace\ ,$$
where $\theta_\pm=\theta\pm \sigma$. The transverse space is parametrized by $(\psi,\vec{v}\,)$, where $\vec{v}\in \mathbb{R}^{d-3}$ has one dimension less than in the previous cases in ${\rm AdS}_2\times S^{d-2}$ and de Sitter. It is now easy to consider the null deformed half-space, parametrized by $\zeta^\mu=\zeta^-(\psi,\vec{v}\,)\delta^\mu_-$ as
\begin{equation}\label{eq:141}
\mathcal{D}(A)=\left\lbrace
  (\sigma,\theta,\psi,\vec{v}\,)
  \in 
  \mathbb{R}
  \times (0,\pi)
  \times (0,\pi/2]
  \times \mathbb{R}^{d-3}:
  \quad
  \theta_+< \pi/2\ ,
  \quad
  \theta_-< \pi/2+\zeta^-(\psi,\vec{v}\,)\,
  \right\rbrace\ .
\end{equation}
Plotting these regions in the $(\sigma,\theta)$ plane we obtain exactly the same diagrams as in figure \ref{fig:8}, with the AdS boundary located at $\theta=0,\pi$. 

Since the sector $(\sigma,\theta)$ of the AdS metric in (\ref{eq:137}) is exactly the same as for ${\rm AdS}_2$, the calculation of the vacuum modular hamiltonian associated to $\mathcal{D}(A)$ is extremely similar to that in ${\rm AdS}_2\times S^{d-2}$. Let us sketch the calculation, highlighting the most salient differences with respect to the construction in subsection \ref{sub:2.2}.

\paragraph{Undeformed region:}

The vacuum modular hamiltonian is obtained in exactly the same way as for ${\rm AdS}_2\times S^{d-2}$ in subsection \ref{sub:2.2}: we analytically continue to Euclidean time $\sigma_E=i\sigma$, describe $\rho_{A_0}$ from the path integral in (\ref{eq:69}) and look for an isometry of the Euclidean manifold which generates the appropriate flow. Since the $(\sigma,\theta)$ dependence of the AdS metric (\ref{eq:137}) is the same as that of ${\rm AdS}_2\times S^{d-2}$ in (\ref{eq:87}), the appropriate isometry is the same as the one given by (\ref{eq:70}), which generates the flow shown in figure \ref{fig:4}. The modular hamiltonian is related to the generator of this isometry (\ref{eq:73}) according to
\begin{equation}\label{eq:140}
K_{A_0}=
  2\pi\int_{\Sigma}
  dS^\mu T_{\mu \nu}\xi^\nu+{\rm const.}\ ,
  \qquad \qquad
  \xi^\mu=\cos(\theta_+)\partial_+-
  \cos(\theta_-)\partial_-\ .
\end{equation}
We take $\Sigma$ as the future null horizon of $\mathcal{D}(A_0)$, which we can parametrize in terms of an affine parameter $\lambda\in \mathbb{R}$ as
\begin{equation}\label{eq:142}
x^\mu(\lambda,\psi,\vec{x}_\perp)=
  (\theta_+,\theta_-,\psi,\vec{v}\,)=
  (\pi/2,2\,{\rm arccot}(\lambda)-\pi/2,
  \psi,\vec{x}_\perp)\ ,
\end{equation}
where the transverse space is parametrized by $(\psi,\vec{x}_\perp)\in (0,\pi/2]\times \mathbb{R}^{d-3}$. Notice that $\vec{x}_\perp$ has one less component in comparison with the previous cases. The modular hamiltonian (\ref{eq:140}) along this surface can be written and we find  
$$K_{A_0}=2\pi L^{d-2}
  \int_{S^{d-3}}
  d\Omega(\vec{x}_\perp)
  \int_0^{\pi/2}
  d\psi
  \frac{\cos(\psi)^{d-3}}
  {\sin(\psi)^{d-4}}
  \int_0^{+\infty}d\lambda\,
  \lambda  \,
  T_{\lambda \lambda}
  (\lambda,\psi,\vec{x}_\perp)+
  {\rm const}\ .$$
While the integral over the transverse space $(\psi,\vec{x}_\perp)$ is different, the $\lambda$ sector is equivalent to the previous cases.

\paragraph{First order deformation:}

The first order contribution in the deformation parameter~$\zeta^-(\psi,\vec{v}\,)$ in (\ref{eq:141}) follows exactly the same as in subsection \ref{sub:2.2} for ${\rm AdS}_2\times S^{d-2}$. The integral we must solve is again given by (\ref{eq:139}). The boundary of the Euclidean AdS manifold~$\partial \mathcal{M}_E$ written in the coordinates $(\tau,\rho)$ in (\ref{eq:79}) is given by
$$
\frac{ds^2_E}{L^2}=
  \frac{d\rho^2+\sinh^2(\rho)d\tau^2} 
  {\sin^2(\psi)}+
\frac{d\psi^2+\cos^2(\psi)d\Omega^2_{d-3}(\vec{v}\,)}
  {\sin^2(\psi)}\ .
$$
Since the $(\tau,\rho)$ dependence of the metric is the same as for ${\rm AdS}_2\times S^{d-2}$ we can easily solve the integral in (\ref{eq:139}) and find the following result for the null deformed modular hamiltonian on AdS
\begin{equation}\label{eq:159}
K_A=
  K_{A_0}+
  \pi L^{d-1}
  \int_{S^{d-3}}
  d\Omega(\vec{x}_\perp)
  \int_0^{\pi/2}
  d\psi
  \frac{\cos(\psi)^{d-3}}{\sin(\psi)^{d-4}}
  \zeta^-(\psi,\vec{x}_\perp)
  \int_0^{+\infty}d\lambda\,
  T_{\lambda \lambda}
  (\lambda,\psi,\vec{x}_\perp)+
  \mathcal{O}(\zeta^\mu)^2\ .
\end{equation}
The crucial aspect of this relation is that we recover half the ANEC operator in $\lambda$. Writing the full modular hamiltonian $\hat{K}_A=K_A-K_{A^c}$ and using relative entropy, the same calculation leading to (\ref{eq:109}) gives the ANEC for a QFT in AdS for the null geodesics in (\ref{eq:142}).

\section{Constraint for incomplete achronal geodesics}
\label{sec:4}

As mentioned in the introduction and discussed in section 8 of \cite{Witten:2019qhl}, complete achronal null geodesics are very rare. As a result, the ANEC does not constraint typical null geodesics that arise in generic space-times. This raises the question of whether we can relax some of the conditions of the achronal ANEC and still obtain a useful constraint.

For CFTs in the Lorentzian cylinder $\mathbb{R}\times S^{d-1}$ this question was addressed in \cite{Rosso:2019txh}. Writing the metric as
$$ds^2=-d\sigma^2+d\theta^2+
  \sin^2(\theta)d\Omega_{d-2}^2(\vec{v}\,)\ ,$$
consider the following family of null geodesics
\begin{equation}\label{eq:155}
x^\mu(\lambda,\vec{x}_\perp)=
  (\theta_+,\theta_-,\vec{v}\,)=
  (\pi/2,\pi/2-2\lambda,\vec{x}_\perp)\ ,
  \qquad \qquad
  (\lambda,\vec{x}_\perp)\in 
  [-\pi/2,\pi/2]\times
  \mathbb{R}^{d-2}\ .
\end{equation}
These curves go between antipodal points of the spatial sphere $S^{d-1}$, from the South to the North pole at $\theta=\pi,0$ respectively. They are achronal but not complete,  since the affine parameter $\lambda$ has a finite range $|\lambda|\le \pi/2$. We say these geodesics are \textit{maximally extended}, meaning that further extending the curves results in chronal trajectories. In \cite{Rosso:2019txh} these geodesics where shown to satisfy the following condition
\begin{equation}\label{eq:154}
\int_{-\pi/2}^{\pi/2}d\lambda\,
  \cos^d(\lambda)\big(
  T_{\lambda \lambda}-
  \langle T_{\lambda \lambda} \rangle_0
  \big)\ge 0\ ,
\end{equation}
where $\langle T_{\lambda \lambda} \rangle_0$ gives the vacuum expectation value. This subtraction ensures the constraint is not violated by a trivial Casimir contribution to the energy. The positivity condition involves a non-local operator with the kernel $\cos^d(\lambda)$, that is positive, smooth and vanishes at the boundary of the integral. It is unclear whether this result can be extended to arbitrary QFTs in the Lorentzian cylinder $\mathbb{R}\times S^{d-1}$.\footnote{One of the proofs of (\ref{eq:154}) given in \cite{Rosso:2019txh} involves the same method using modular Hamiltonians and relative entropy used in subsection \ref{sub:2.3} to prove the achronal ANEC for CFTs. However, a crucial difference in this case is that the modular hamiltonian of the undeformed region in the cylinder (a spherical cap of size $\pi/2$) is related to a \textit{conformal} Killing vector $\xi^\mu$, rather than an ordinary Killing vector $\xi^\mu$ as in (\ref{eq:74}). For this reason, it does not seem likely that (\ref{eq:154}) can be extended beyond CFTs using this approach.\label{foot:1}} 

In this section we show the bound generalizes to CFTs in more general backgrounds, by deriving (\ref{eq:154}) for ${\rm AdS}_2\times S^{d-2}$. The motion of the null geodesics in this space-time is more interesting than in the cylinder, since more complicated trajectories than (\ref{eq:155}) are allowed (see (\ref{eq:18}) and figure \ref{fig:3} below).

\subsection{Extremal horizons}
\label{sub:4.1}

To derive the bound (\ref{eq:154}) for CFTs in ${\rm AdS}_2\times S^{d-2}$ we follow a simple procedure. Starting from the ANEC in Minkowski we apply a conformal transformation to ${\rm AdS}_2\times S^{d-2}$ which gives (\ref{eq:154}). The appropriate map is a variation of the one considered in subsection \ref{sub:2.3} to compute the null deformed modular hamiltonian.

We start by taking Cartesian coordinates ${X^\mu=(T,X,\vec{Y})}$ in Minkowski and parametrizing the null geodesics as
\begin{equation}\label{eq:123}
X^\mu(\lambda,\vec{x}_\perp,t_0)= 
  (\lambda+t_0,\lambda,\vec{x}_\perp)\ ,
  \qquad \qquad
  \lambda \in \mathbb{R}\ ,
\end{equation}
where $\lambda$ is an affine parameter for fixed $(\vec{x}_\perp,t_0)\in \mathbb{R}^{d-2}\times \mathbb{R}$. We have introduced the parameter~$t_0$ that will have a non-trivial effect on the geodesics after the mapping. The ANEC holds along any of these null trajectories in Minkowski \cite{Faulkner:2016mzt,Hartman:2016lgu,Longo:2018obd}
\begin{equation}\label{eq:114}
\int_{-\infty}^{+\infty}
  d\lambda\,
  \bar{T}_{\lambda \lambda}
  \ge 0\ ,
\end{equation}
where we add a bar for operators defined in Minkowski. The conformal transformation is implemented in the Hilbert space by the unitary operator $U:\bar{\mathcal{H}}\rightarrow \mathcal{H}$.

The first step is to change the spatial coordinates in Minkowski, from Cartesian $(X,\vec{Y})$ to spherical $(r,\vec{v}\,)$, defined as
$$r=(
  X^2+|\vec{Y}|^2
  )^{1/2}\ ,
  \qquad \qquad
  \vec{v}=
  \frac{\vec{Y}}
  {X+(X^2+|\vec{Y}|^2)^{1/2}}\ ,$$
so that the Minkowski metric becomes $ds_{\rm Mink}^2=-dT^2+dr^2+r^2d\Omega^2_{d-2}(\vec{v}\,)$. By further redefining $r\pm T=\tan(\theta_\pm/2)$ with $\theta_\pm=\theta\pm \sigma$ we get
\begin{equation}\label{eq:136}
ds^2_{\rm Mink}=
  w^2(\sigma,\theta)
  \left[
  \frac{-d\sigma^2+d\theta^2}
  {\sin^2(\theta)}+
  d\Omega^2_{d-2}(\vec{v}\,)
  \right]\ ,
\end{equation}	
where we identified the conformal factor as
$$
w^2(\sigma,\theta)=
  \left(
  \frac{\sin(\theta)}
  {2\cos(\theta_+/2)
  \cos(\theta_-/2)}
  \right)^2\ .
$$
The conformal transformation to ${\rm AdS}_2\times S^{d-2}$ is completed by performing the Weyl rescaling. We can easily work out how the Minkowski null geodesics (\ref{eq:123}) are mapped to the ${x^\mu=(\theta_+,\theta_-,\vec{v}\,)}$ coordinates. Instead of using the parameter $\lambda$ it is convenient to redefine it to
\begin{equation}\label{eq:21}
\lambda(\alpha)=
  |\vec{x}_\perp|
  \tan(\alpha)\ ,
  \qquad \qquad
  |\alpha|\le \pi/2\ ,
\end{equation}
so that the null geodesics are given by
\begin{equation}\label{eq:18}
x^\mu(\alpha,\vec{x}_\perp,t_0)
=
  \left(
  \theta_+(\alpha),\theta_-(\alpha), 
  \frac{\vec{x}_\perp}
  {|\vec{x}_\perp|}
  \frac{\cos(\alpha)}
  {1+\sin(\alpha)}
  \right)\ ,
\end{equation}
with the functions $\theta_\pm(\alpha)$ defined from
$$
  \tan(\theta_\pm/2)=
  |\vec{x}_\perp|
  \left(
  \frac{1\pm \sin(\alpha)}
  {\cos(\alpha)}
  \right)
  \pm t_0
  \ .
$$
It is straightforward to check that these curves satisfy the geodesic equation with affine parameter $\alpha$.

Let us analyze the geodesics (\ref{eq:18}). For arbitrary values of $(t_0,\vec{x}_\perp)$ the curves have a non-trivial motion along all the directions in AdS$_2\times S^{d-2}$, with their initial and final spatial points determined by $t_0$ according to
$$(\theta,|\vec{v}\,|)
  \big|_{\rm initial}=
  (\pi/2+{\rm arctan}(t_0),\infty)\ ,
  \qquad \qquad
  (\theta,|\vec{v}\,|)
  \big|_{\rm final}=
  (\pi/2-{\rm arctan}(t_0),0)\ .
  $$
The geodesics travel between the antipodal points of the $S^{d-2}$, since $|\vec{v}\,|=0,\infty$ correspond to the South and North pole of the sphere. The geodesics are not complete, since they can be extended beyond their initial and final points, as is manifest in the restricted range of the affine parameter $\alpha$ in (\ref{eq:21}). However, extending the trajectories any further results in a violation of their achronality. 

\begin{figure}
\centering
\includegraphics[scale=0.28]{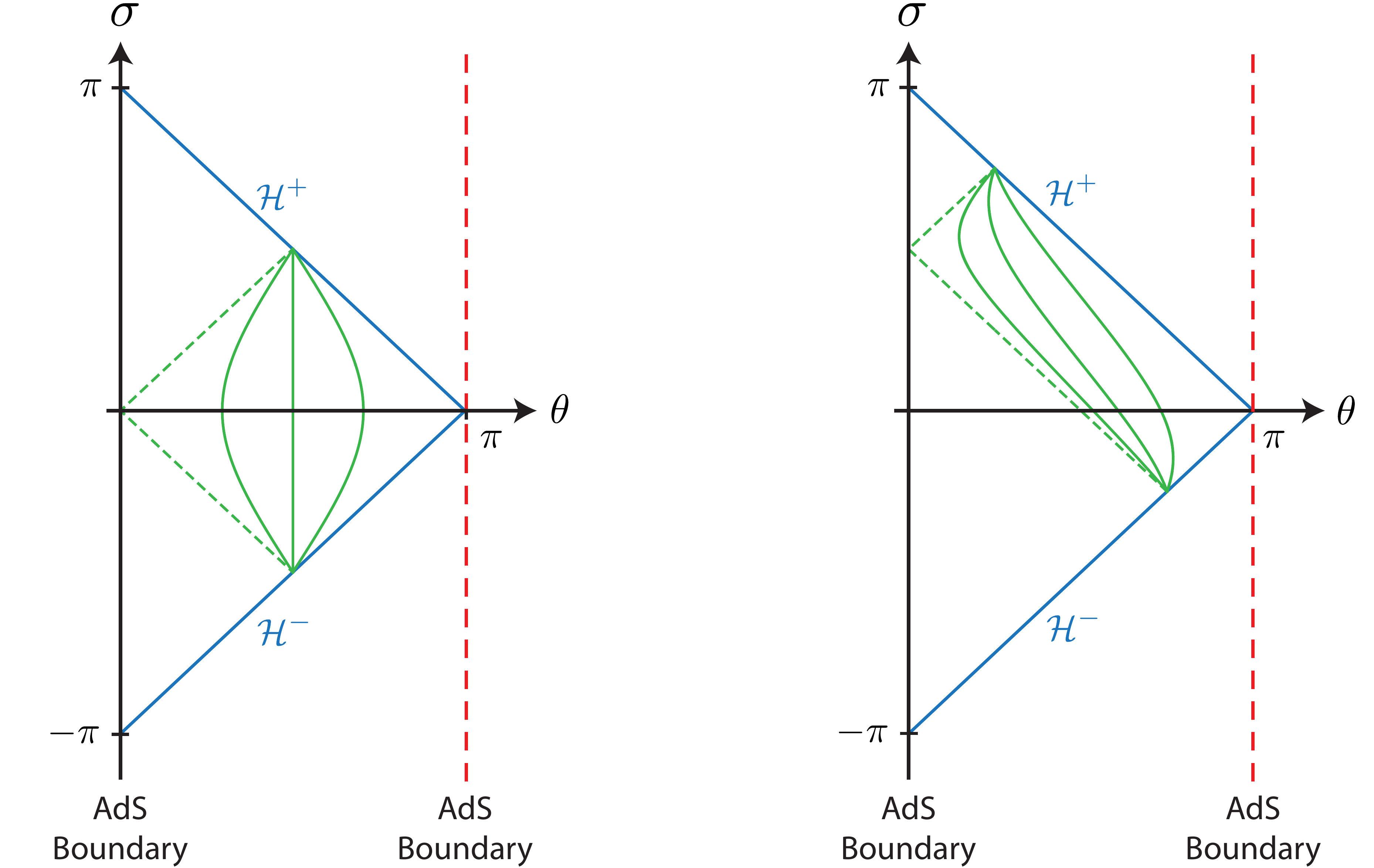}
\caption{Null geodesics in (\ref{eq:18}) in the $(\sigma,\theta)$ plane describing the AdS$_2$ sector. The left diagram is for $t_0=0$ while the right $t_0=1$. For each case we have plotted several geodesics obtained by taking different values of $\vec{x}_\perp$. As $|\vec{x}_\perp|$ goes to infinity, the geodesics get closer to the Poincare horizon $\mathcal{H}^\pm$ of ${\rm AdS}_2$.}\label{fig:3}
\end{figure}

The complete trajectory is in general very complicated and can be visualized more easily by numerically plotting in the $(\sigma,\theta)$ plane, as seen in figure \ref{fig:3}. The left and right diagrams correspond to geodesics with $t_0=0$ and $t_0=1$ respectively, for several values of $|\vec{x}_\perp|$. These diagrams are not capturing the motion in the $S^{d-2}$ sector, that is non-trivial. All curves travel between the past and future Poincare horizons $\mathcal{H}^\pm$, with their initial and final points determined by $t_0$. The magnitude of $\vec{x}_\perp$ controls how close to the ${\rm AdS}_2$ boundary the curves travels, with $|\vec{x}_\perp|=0,\infty$ reaching the boundary. 

We can now map the Minkowski ANEC in (\ref{eq:114}) to ${\rm AdS}_2\times S^{d-2}$, using the transformation of the stress tensor in (\ref{eq:152}). The conformal factor (\ref{eq:136}) evaluated along the null curves gives ${w(\alpha)=|\vec{x}_\perp|/\cos(\alpha)}$, so that (\ref{eq:114}) becomes
\begin{equation}\label{eq:17}
0\le
\int_{-\infty}^{+\infty}d\lambda\,U
\bar{T}_{\lambda \lambda}
U^\dagger=
  \frac{1}{|\vec{x}_\perp|^{d-1}}
  \int_{-\pi/2}^{\pi/2}
  d\alpha
  \cos^{d}(\alpha)
  \big(
  T_{\alpha\alpha}-
  \langle
  T_{\alpha\alpha}
  \rangle_0
  \big)\ ,
\end{equation}
where we have changed the integration variable to the affine parameter $\alpha$ in (\ref{eq:21}). This is the constraint (\ref{eq:154}) previously obtained for CFTs in the cylinder. Notice that the mapping of the operator becomes ill-defined for geodesics with $|\vec{x}_\perp|=0,\infty$, where the geodesics reach the AdS boundary.

\appendix
\addtocontents{toc}{\protect\setcounter{tocdepth}{1}}

\section{A novel conformal transformation}
\label{zapp:1}

The results derived in subsections \ref{sub:2.3} and \ref{sub:4.1} relied on a simple conformal transformation relating ${\rm AdS}_2\times S^{d-2}$ to the Lorentzian cylinder $\mathbb{R}\times S^{d-1}$, recently noted in \cite{Galloway:2020xfz}. These manifolds are related by a simple Weyl rescaling
\begin{equation}\label{eq:143}
ds^2=
  -d\sigma^2+d\theta^2+\sin^2(\theta)d\Omega_{d-2}^2=
  \sin^2(\theta)\left[
  \frac{-d\sigma^2+d\theta^2}
  {\sin^2(\theta)}+
  d\Omega_{d-2}^2
  \right]\ ,
\end{equation}
where between brackets we recognize ${\rm AdS}_2$ in global coordinates. A similar transformation was also used in \cite{Hofman:2008ar} to relate Minkowski to ${\rm AdS}_2\times S^{d-2}$ in Poincare coordinates. In this appendix we show how these maps can be generalized according to
\begin{equation}\label{eq:12}
  \mathbb{R}\times \Sigma^{(k)}
  \qquad \longrightarrow \qquad
  {\rm AdS}_n^{(k)}\times S^{d-n}\ ,
\end{equation}
where $n\in \mathbb{N}_{\ge 2}$ and $\Sigma^{(k)}$ corresponds to a maximally symmetric space with zero ($k=0$), positive ($k=1$) or negative ($k=-1$) constant curvature. Accordingly, the AdS factor on the right is given in flat, spherical or hyperbolic slicing.

\subsection{AdS Poincare patch}

Let us start by considering the case $k=0$ where the transformation is given by
$$
  \mathbb{R}\times \mathbb{R}^{d-1}
  \qquad \longrightarrow \qquad
  {\rm AdS}_n^{(k=0)}\times S^{d-n}\ ,
$$
with ${\rm AdS}_n^{(k=0)}$ corresponding to the Poincare patch. Since the transformation for generic $n$ can be a little confusing, let us warm up by considering the first few values of $n$. For $n=2$ this mapping was previously noted in \cite{Hofman:2008ar}.

\paragraph{Two-dimensional AdS:}

Taking $n=2$ we can proceed in an analogous way to (\ref{eq:143}) and rewrite the Minkowski metric $\mathbb{R}\times \mathbb{R}^{d-1}$ as
$$ds^2=
  -dt^2+dr^2+r^2d\Omega_{d-2}^2=
  (r/L)^2\left[
  \frac{-dt^2+dr^2}{(r/L^2)}+
  L^2d\Omega_{d-2}^2
  \right]\ .
  $$
Applying a Weyl rescaling with $w(r)=r/L$ we find
$$
\frac{ds^2}{w^2(r)}=
\frac{-dt^2+dr^2}{(r/L^2)}+
  L^2d\Omega_{d-2}^2\ ,
$$  
which is explicitly given by ${\rm AdS}_2\times S^{d-2}$ in Poincare coordinates $(t,r)$. From now on we set $L=1$ to simplify the notation.

\paragraph{Three-dimensional AdS:}

To obtain the $n=3$ case we first write the Minkowski metric as
$$ds^2=
  -dt^2+dr^2+r^2\left(
  d\theta_1^2+\sin^2(\theta_1)
  d\Omega_{d-3}^2
  \right)=
  w^2(r,\theta_1)\left[
  \frac{-dt^2+dr^2+
  r^2d\theta_1^2}
  {r^2\sin^2(\theta_1)}+
  d\Omega_{d-3}^2
  \right]\ ,
  $$ 
where we defined $w^2(r,\theta_1)=r^2\sin^2(\theta_1)$. Although after the Weyl rescaling we have the appropriate contribution for the $S^{d-3}$, it is not clear that the first factor corresponds to the Poincare patch of ${\rm AdS}_3$. We can put it in a more familiar form by defining the new coordinates $(z,x)$ as
$$z(r,\theta_1)=
  r\sin(\theta_1)\ ,
  \qquad \qquad
  x(r,\theta_1)=
  r\cos(\theta_1)\ .$$
Since $\theta_1\in[0,\pi]$, the range of the new coordinates is $(z,x)\in \mathbb{R}_{\ge 0}\times \mathbb{R}$. Implementing the transformation we find
$$\frac{ds^2}{w^2(r,\theta_1)}=
  \frac{-dt^2+dz^2+dx^2}
  {z^2}+
  d\Omega_{d-3}^2\ ,$$
which corresponds to ${\rm AdS}_3\times S^{d-3}$ in Poincare coordinates.

\paragraph{n-dimensional AdS:}

Now that we have some intuition, let us consider the transformation for arbitrary values of $n\in \mathbb{N}_{\ge 2}$. To do so, we first write the metric of a unit sphere $S^\ell$ in the usual spherical angles $(\theta_1,\dots,\theta_{\ell-1},\phi)$, where $\theta_i\in[0,\pi]$ and $\phi\in[0,2\pi)$ is the only periodic angle. The metric on the unit sphere is given by
\begin{equation}\label{eq:7}
d\Omega^2_{\ell}=
  \sum_{j=1}^{\ell-1}
  \Big(
  \prod_{i=1}^{j-1}
  \sin^2(\theta_i)
  \Big)
  d\theta_j^2+
  \prod\limits_{i=1}^{\ell-1}
  \sin^2(\theta_i)
  d\phi^2\ ,
\end{equation}
where we are using a convention in which $\prod_{i=1}^0a_i\equiv 1$. Let us now split the sum in the first term in two pieces, up to some term $m$
$$d\Omega^2_{\ell}=
  \sum_{j=1}^{m}
  \Big(
  \prod_{i=1}^{j-1}
  \sin^2(\theta_i)
  \Big)
  d\theta_j^2+
  \prod_{i=1}^{m}
  \sin^2(\theta_i)
  \left[
  \sum_{j=m+1}^{\ell-1}
  \Big(
  \prod_{i=m+1}^{j-1}
  \sin^2(\theta_i)
  \Big)
  d\theta_j^2+
  \prod\limits_{i=m+1}^{\ell-1}
  \sin^2(\theta_i)
  d\phi^2
  \right]\ .$$
Comparing with (\ref{eq:7}), we recognize the term between square brackets as the line element of a unit sphere $S^{\ell-m}$, so that we have the following relation
\begin{equation}\label{eq:8}
d\Omega^2_{\ell}=
  \sum_{j=1}^{m}
  \Big(
  \prod_{i=1}^{j-1}
  \sin^2(\theta_i)
  \Big)
  d\theta_j^2+
  \prod_{i=1}^{m}
  \sin^2(\theta_i)
  d\Omega^2_{\ell-m}\ .
\end{equation}
This way of writing the metric of $S^\ell$ is very useful for our purposes. Writing the Minkowski metric in spherical coordinates and using (\ref{eq:8}) with $\ell=d-2$ and $m=n-2$ we find
$$ds^2=
  -dt^2+dr^2+r^2
  \sum_{j=1}^{n-2}
  \Big(
  \prod_{i=1}^{j-1}
  \sin^2(\theta_i)
  \Big)
  d\theta_j^2+r^2
  \prod_{i=1}^{n-2}
  \sin^2(\theta_i)
  d\Omega^2_{d-n}
  \  .$$
To obtain a factor $S^{d-n}$, we perform a Weyl rescaling with a conformal factor $w(r,\theta_i)=r\prod_{i=1}^{n-2}\sin(\theta_i)$, so that we find
\begin{equation}\label{eq:9}
\frac{ds^2}
  {w^2(r,\theta_i)}=
  \frac{-dt^2+dr^2+r^2
  \sum_{j=1}^{n-2}
  \Big(
  \prod_{i=1}^{j-1}
  \sin^2(\theta_i)
  \Big)
  d\theta_j^2}
  {r^2
  \prod_{i=1}^{n-2}
  \sin^2(\theta_i)}
  +
  d\Omega^2_{d-n}
  \  .
\end{equation}
This gives the desired line element on the unit sphere $S^{d-n}$. We can put the first factor in a nicer form by performing the following change of coordinates
\begin{equation}\label{eq:10}
z=r\prod_{j=1}^{n-2}\sin(\theta_j)\ ,
\qquad \qquad
x_i=r\cos(\theta_i)
 \prod_{j=1}^{i-1}\sin(\theta_j)\ ,
 \qquad i=1,\dots,n-2\ .
\end{equation}
This is essentially the same relation going from spherical coordinates $(r,\theta_1,\dots,\theta_{n-2})$ to Cartesian $(z,x_1,\dots,x_{n-2})$ in flat space. The only difference is that all the angles $\theta_i$ are in the range $\theta_i\in[0,\pi]$, there is no azimuthal angle $\phi\in[0,2\pi)$. This results in $z\ge 0$, which is exactly what we require for AdS, as the metric (\ref{eq:9}) in these coordinates becomes
$$\frac{ds^2}
  {w^2(r,\theta_i)}=
  \frac{-dt^2+dz^2+
  |d\vec{x}\,|^2}
  {z^2}+
  d\Omega_{d-n}^2\ ,$$
that we immediately recognize as ${\rm AdS}_n\times S^{d-n}$ in Poincare coordinates.  

\subsection{Global AdS}

A similar construction can be considered for the $k=1$ case, where the transformation is given by
$$\mathbb{R}\times S^{d-1}
  \qquad \longrightarrow \qquad
  {\rm AdS}_n^{(k=1)}\times S^{d-n}\ ,$$
where ${\rm AdS}_n^{(k=1)}$ corresponds to global coordinates on AdS. For the $n=2$ case the transformation was considered in \cite{Galloway:2020xfz} and is given by the Weyl rescaling in (\ref{eq:143}). Let us see how it works for general $n$.

\paragraph{Three-dimensional AdS:}

For the case with $n=3$ let us write the metric in the Lorentzian cylinder $\mathbb{R}\times S^{d-1}$ as
$$ds^2=
  -d\sigma^2+
  d\theta_1^2+
  \sin^2(\theta_1)
  \left(d\theta_2^2+
  \sin^2(\theta_2)
  d\Omega_{d-3}^2
  \right)=
  w^2(\theta_i)
  \left[
  \frac{-d\sigma^2+d\theta_1^2+\sin^2(\theta_1)d\theta_2^2}
  {\sin^2(\theta_1)\sin^2(\theta_2)}+
  d\Omega_{d-3}^2
  \right]\ ,$$
where we have defined $w^2(\theta_i)=\sin^2(\theta_1)\sin^2(\theta_2)$. The Weyl rescaling gives the appropriate factor for $S^{d-3}$, while the following change of coordinates
\begin{equation}
\begin{aligned}
\cos^2(\rho)&=\sin^2(\theta_1)\sin^2(\theta_2)\ ,\\
\tan^2(\alpha_1)&=
\tan^2(\theta_1)\cos^2(\theta_2)\ ,
\end{aligned}
\qquad {\rm with\,\,inverse} \qquad
\begin{aligned}
\cos^2(\theta_1)&=\sin^2(\rho)\cos^2(\alpha_1)\ ,\\
\cot^2(\theta_2)&=\tan^2(\rho)
\sin^2(\alpha_1)\ ,
\end{aligned}
\end{equation}
gives the metric
$$
\frac{ds^2}{w^2(\theta_i)}=
\frac{-d\sigma^2+d\rho^2+\sin^2(\rho)d\alpha_1^2}
  {\cos^2(\rho)}+
  d\Omega_{d-3}^2\ .
$$
We recognize this space-time as global ${\rm AdS}_3\times S^{d-3}$, where the range of the new coordinates is given by $\alpha_1\in [0,2\pi)$ and $\rho\in[0,\pi/2)$, with the AdS boundary at $\rho=\pi/2$. 

\paragraph{n-dimensional AdS:}

Let us now consider the transformation for any $n\in \mathbb{N}_{\ge 2 }$. Using (\ref{eq:8}) with $\ell=d-1$ and $m=n-1$, the metric in the Lorentzian cylinder can be written as
$$ds^2=
  -d\sigma^2+d\Omega^2_{d-1}=
  -d\sigma^2+
  \sum_{j=1}^{n-1}
  \Big(
  \prod_{i=1}^{j-1}
  \sin^2(\theta_i)
  \Big)
  d\theta_j^2
  +
  \prod_{i=1}^{n-1}
  \sin^2(\theta_i)
  d\Omega^2_{d-n}\ .$$
The conformal factor which gives a factor of $S^{d-n}$ in the metric is given by $w(\theta_i)=\prod_{i=1}^{n-1}\sin(\theta_i)$, so that the metric becomes
\begin{equation}\label{eq:11}
\frac{ds^2}
  {w^2(\theta_i)}=
  \frac{-d\sigma^2+
  \sum_{j=1}^{n-1}
  \Big(
  \prod_{i=1}^{j-1}
  \sin^2(\theta_i)
  \Big)
  d\theta_j^2}
  {\prod_{i=1}^{n-1}
  \sin^2(\theta_i)}
  +
  d\Omega^2_{d-n}\ .
\end{equation}
We now apply a change of coordinates from $(\theta_1,\dots,\theta_{n-1})$ to $(\rho,\alpha_1,\dots,\alpha_{n-2})$ which generalizes~(\ref{eq:47}) according to
\begin{equation}\label{eq:145}
\begin{aligned}
\cos^2(\rho)&=\prod_{i=1}^{n-1}\sin^2(\theta_i)\ ,\\
\tan^2(\alpha_i)&=
\frac{\tan^2(\theta_i)\cos^2(\theta_{i+1})}
  {\cos^2(\alpha_{i+1})}\ , \qquad
i=1,\dots,n-2\ ,
\end{aligned}
\end{equation}
where $\alpha_{n-1}\equiv 0$. Using $\cos^2(\alpha_{i+1})=1/(1+\tan^2(\alpha_{i+1}))$ gives a recursion relation for the coordinates $\alpha_i$. Although we do not recognize this as a standard change of coordinates, it is straightforward to check for arbitrary values of $n$ that the resulting metric in (\ref{eq:11}) is given by
$$
\frac{ds^2}
  {w^2(\theta_i)}=
  \frac{-d\sigma^2+
  d\rho^2+\sin^2(\rho)
  d\Omega_{n-2}^2(\alpha_i)
  }
  {\cos^2(\rho)}
  +
  d\Omega^2_{d-n}\ ,
$$
where
$$
d\Omega^2_{n-2}(\alpha_i)=
\sum_{j=1}^{n-3}
\left(
\prod_{i=1}^{j-1}
\sin^2(\alpha_i)
\right)
d\alpha_j^2+
\prod_{i=1}^{n-3}
\sin^2(\alpha_i)
d\alpha_{n-2}^2\ ,
$$
is the line element of a unit sphere $S^{n-2}$. The coordinate $\alpha_{n-2}$ is the only periodic one with a range given by $\alpha_{n-2}\in[0,2\pi)$. The resulting metric is ${\rm AdS}_n\times S^{d-n}$ in global coordinates, with the boundary at $\rho=\pi/2$.

\subsection{Rindler AdS}

Finally, let us consider the $k=-1$ case, given by the transformation
$$\mathbb{R}\times \mathbb{H}^{d-1}
  \qquad \longrightarrow \qquad
  {\rm AdS}_n^{(k=-1)}
  \times S^{d-n}\ ,$$
where $\mathbb{H}$ is the hyperbolic plane and the AdS factor is in hyperbolic slicing, sometimes called Rindler AdS.

\paragraph{Two-dimensional AdS:}

Let us start by considering the case with $n=2$, where we write the metric in $\mathbb{R}\times \mathbb{H}^{d-1}$ as
$$ds^2=
  -d\tau^2+
  du^2+\sinh^2(u)d\Omega_{d-2}^2=
  \sinh^2(u)\left[
  \frac{-d\tau^2+du^2}{\sinh^2(u)}
  +d\Omega_{d-2}^2
  \right]\ ,$$
where $u> 0$. Applying a Weyl rescaling with $w^2(u)=\sinh^2(u)$ we find
$$
\frac{ds^2}{w^2(u)}=
\frac{-d\tau^2+du^2}{\sinh^2(u)}
  +d\Omega_{d-2}^2\ .
$$
The first factor corresponds to Rindler ${\rm AdS}_2$, as might be clearer by redefining the spatial coordinate to $\sinh^2(u)=1/(r^2-1)$.

\paragraph{Three-dimensional AdS:}

For $n=3$ we write the metric in $\mathbb{R}\times \mathbb{H}^{d-1}$ as
$$ds^2=
  -d\tau^2+du^2+\sinh^2(u)(
  d\theta_1^2+\sin^2(\theta_1)d\Omega_{d-3}^2
  )\ ,$$
and apply a Weyl transformation given by $w(u,\theta_i)=\sinh(u)\sin(\theta_1)$
$$\frac{ds^2}{w^2(u,\theta_i)}=
  \frac{-d\tau^2+du^2+
  \sinh^2(u)
  d\theta_1^2}{\sinh^2(u)
  \sin^2(\theta_1)}+
  d\Omega_{d-3}^2
  \ .$$
Changing coordinates to $(\varrho,\xi)$ defined from
\begin{equation}\label{eq:144}
\begin{aligned}
\sinh^2(\varrho)&=
\sinh^2(u)\sin^2(\theta_1)\ ,\\
\tanh^2(\xi)&=
\tanh^2(u)\cos^2(\theta_1)\ ,
\end{aligned}
\qquad {\rm with\,\, inverse}\qquad
\begin{aligned}
\cosh^2(u)&=
\cosh^2(\varrho)
\cosh^2(\xi)\ ,\\
\tan^2(\theta_1)&=
\frac{\tanh^2(\varrho)}
{\sinh^2(\xi)}\ ,
\end{aligned}
\end{equation}
the metric becomes
$$\frac{ds^2}{w^2(u,\theta_i)}=
  \frac{-d\tau^2+d\varrho^2+
  \cosh^2(\varrho)d\xi^2
  }
  {\sinh^2(\varrho)}+
  d\Omega_{d-3}^2
  \ .$$
We recognize this as ${\rm AdS}_3\times S^{d-3}$ in Rindler coordinates, where $\varrho> 0$ and $\xi\in \mathbb{R}$.

\paragraph{n-dimensional AdS:}

For arbitrary values of $n\in \mathbb{N}_{\ge 2}$ we write the metric in $\mathbb{R}\times \mathbb{H}^{d-1}$ as
$$
\begin{aligned}
ds^2&=
  -d\tau^2+
  du^2+\sinh^2(u)d\Omega^2_{d-2}\ ,
  \\
  &=
  -d\tau^2+du^2
  +\sinh^2(u)
  \sum_{j=1}^{n-2}
  \Big(
  \prod_{i=1}^{j-1}
  \sin^2(\theta_i)
  \Big)
  d\theta_j^2+
  \sinh^2(u)
  \prod_{i=1}^{n-2}
  \sin^2(\theta_i)
  d\Omega^2_{d-n}\ ,
\end{aligned}
$$
where in the second equality we have used (\ref{eq:8}) with $\ell=d-2$ and $m=n-2$. Dividing the metric by the conformal factor $w^2(u,\theta_i)=\sinh^2(u)\prod_{i=1}^{n-2}\sin^2(\theta_i)$ we get
$$
\frac{ds^2}{w^2(u,\theta_i)}=
  \frac{-d\tau^2+du^2
  +\sinh^2(u)
  \sum_{j=1}^{n-2}
  \Big(
  \prod_{i=1}^{j-1}
  \sin^2(\theta_i)
  \Big)
  d\theta_j^2}
  {\sinh^2(u)
  \prod_{i=1}^{n-2}
  \sin^2(\theta_i)}
  +
  d\Omega^2_{d-n}\ .
$$  
The appropriate change of coordinates which generalizes (\ref{eq:144}) from $(u,\theta_1,\dots,\theta_{n-2})$ to $(\varrho
,\xi,\alpha_1,\dots,\alpha_{n-3})$ is given by
$$
\begin{aligned}
\sinh^2(\varrho)&=
\sinh^2(u)\prod_{i=1}^{n-2}
\sin^2(\theta_i)\ ,\\
\tanh^2(\xi)&=
  \frac{\tanh^2(u)\cos^2(\theta_1)}
  {\cos^2(\alpha_1)}\ ,\\
\tan^2(\alpha_i)&=
\frac{\tan^2(\theta_i)
  \cos^2(\theta_{i+1})}
  {\cos^2(\alpha_{i+1})}\ ,
  \qquad \qquad
  i=1,\dots,n-3\ .
\end{aligned}
$$
Similarly to (\ref{eq:145}) the angles $\alpha_i$ are defined by a recursion relation, where we define $\alpha_{n-2}\equiv 0$. For arbitrary values of $n$ it is straightforward to check that the metric becomes
$$
\frac{ds^2}{w^2(u,\theta_i)}=
  \frac{-d\tau^2+
  d\varrho^2+\cosh^2(\varrho)
  dH^2_{n-2}
  }
  {\sinh^2(\varrho)}+
  d\Omega^2_{d-n}\ ,$$
where the metric of a unit hyperboloid $dH^2_{n-2}$ is given by 
$$dH^2_{n-2}=
  d\xi^2+\sinh^2(\xi)
  \sum_{j=1}^{n-3}
  \left(
  \prod_{i=1}^{j-1}
  \sin^2(\alpha_i)
  \right)
  d\alpha_j^2=
  d\xi^2+\sinh^2(\xi)
  d\Omega_{n-3}^2\ .
  $$
This gives the space-time ${\rm AdS}_n\times S^{d-n}$ in Rindler coordinates.

\section{Symmetry and the vacuum stress tensor}
\label{zapp:2}

There are several instances in the main text where we make some assumptions regarding the vacuum stress tensor of a QFT in ${\rm AdS}_2\times S^{d-2}$. In this appendix we justify such claims by analyzing how $\langle T_{\mu \nu} \rangle_0$ is constrained by symmetry considerations.

We start by considering the simpler case of a QFT in Minkowski. Without any assumptions, the vacuum expectation value of the stress tensor is given in terms of  an arbitrary tensor function $A_{\mu\nu}(x)$
\begin{equation}\label{eq:117}
\bra{0}T_{\mu \nu}(x)\ket{0}=
A_{\mu\nu}(x)\ .
\end{equation}
Let us show how symmetry constraints $A_{\mu \nu}(x)$. Any space-time symmetry of the metric (in this case $\eta_{\mu \nu}$) manifests itself in the Hilbert space as an invariance of the vacuum state $\ket{0}$. For instance, since the Minkowski metric is invariant under space-time translations $x^\mu \rightarrow x^\mu+a^\mu$, the state $\ket{0}$ is invariant under the unitary operator $U(a)$ implementing this symmetry, \textit{i.e.} $U(a)\ket{0}=\ket{0}$. Using this in (\ref{eq:117}) it is straightforward to show that $A_{\mu \nu}(x)$ is independent of the space-time coordinates. In a similar way, invariance of the metric under Lorentz transformation $\Lambda\in{\rm SO}(d-1,1)$ implies
$$A_{\mu \nu}=
  \bra{0}U^\dagger(\Lambda)
  T_{\mu \nu}(x)
  U(\Lambda)\ket{0}=
  \Lambda^\rho_{\,\,\,\mu}
  \Lambda^\sigma_{\,\,\,\nu}
  \bra{0}
  T_{\rho \sigma}(\Lambda^{-1}x)
  \ket{0}=
  \Lambda^\rho_{\,\,\,\mu}
  \Lambda^\sigma_{\,\,\,\nu}
  A_{\rho \sigma}\ .
  $$
This condition completely fixes the tensor $A_{\mu \nu}$ (up to an overall constant) to be equal to the Minkowski metric, \textit{i.e.} $\bra{0}T_{\mu \nu}(x)\ket{0}=a_0\,\eta_{\mu \nu}$.

The vacuum stress tensor is highly constrained in Minkowski due to the fact that it is a maximally symmetry space-time, \textit{i.e.} it admits a maximal number of $d(d+1)/2$ independent Killing vectors. The same is true for QFTs in the other Lorentzian maximally symmetric space-times (anti-)de Sitter, where the vacuum stress tensor is also proportional to the metric.

Let us now consider a QFT in ${\rm AdS}_2\times S^{d-2}$. In this case, the space-time is the product of two maximally symmetric manifolds. Using the Killing vectors in each factor in an analogous way to the Minkowski case we find
$$\bra{0}
  T_{\mu \nu}(x)
  \ket{0}=
\left(
\begin{array}{c|c}
a_0 g_{ij} & N_{iA}(x) \\
\hline
N_{Ai}(x) & b_0 g_{AB}
\end{array}
\right)\ ,$$
where the indices $(i,j)$ and $(A,B)$ run over ${\rm AdS}_2$ and $S^{d-2}$ respectively. Although the off-diagonal contributions $N_{iA}(x)$ are naively not fixed by symmetry, they actually vanish. This is because $\langle T_{iA}(x) \rangle_0$ has a single index in the ${\rm AdS}_2$ direction, meaning it transforms as a vector under its isometries. A non-vanishing vacuum expectation value would have a preferred direction in ${\rm AdS}_2$ and be inconsistent with the symmetries of the vacuum. Putting everything together, symmetry considerations alone constraint the vacuum stress tensor of any QFT in ${\rm AdS}_2\times S^{d-2}$ to
\begin{equation}\label{eq:116}
\bra{0}T_{\mu \nu}(x)\ket{0}=
  a_0\,g_{ij}+b_0\,g_{AB}\ ,
\end{equation}
where $a_0$ and $b_0$ are arbitrary constants.

For conformal field theories we can show this explicitly. The vacuum stress tensor for a conformally flat background is obtained from eq. (21) of \cite{Herzog:2013ed}, that is written in terms of contractions of the Riemann tensor. Instead of using the expression for arbitrary dimensions, let us consider the four dimensional case, which captures the essential features and has the following simple expression
\begin{equation}\label{eq:115}
\langle T_{\mu \nu} \rangle_0
\propto
g_{\mu \nu}\left(
\frac{1}{2}\mathcal{R}^2-\mathcal{R}^2_{\lambda\rho}
\right)+
2\mathcal{R}_{\mu}^{\,\,\,\lambda}\mathcal{R}_{\lambda \nu}-
\frac{4}{3}\mathcal{R}\mathcal{R}_{\mu \nu}\ .
\end{equation}
To compute this for ${\rm AdS}_2\times S^{d-2}$ we use that the Riemann tensor of any product manifold decomposes accordingly \cite{inproceedings}
\begin{equation}\label{eq:118}
\mathcal{R}_{\mu \nu\rho \sigma}=
  \mathcal{R}_{ijkl}^{\rm AdS_{2}}+
  \mathcal{R}_{ABCD}^{S^{d-2}}\  .
\end{equation}
Moreover, since both manifolds are maximally symmetric both terms are proportional to the metric in each factor. Using this in (\ref{eq:115}) we obtain a decomposition of the vacuum stress tensor that is in agreement with (\ref{eq:116}).

The result in (\ref{eq:116}) has several interesting consequences. Since the connection $\Gamma^\mu_{\nu \rho}$ also admits a decomposition as the Riemann tensor in (\ref{eq:118}), the covariant derivative of the vacuum stress tensor becomes 
$$\nabla_\mu 
  \langle T_{\mu \nu} \rangle_0=
  a_0 \nabla_{\mu}g_{ij}+
  b_0 \nabla_{\mu}g_{AB}=
  a_0 \nabla_{k}g_{ij}+
  b_0 \nabla_{C}g_{AB}=0 \ .$$
We use this when deriving the achronal ANEC for ${\rm AdS}_2\times S^{d-2}$ in subsection \ref{sub:2.2}.

Another important application is obtained by considering a null geodesic $x^\mu(\lambda)$ moving entirely in ${\rm AdS}_2$, \textit{i.e.} $dx^\mu/d\lambda=dx^i/d\lambda$. Projecting the vacuum stress tensor along this direction gives
$$\langle  T_{\lambda \lambda} \rangle_0=
  \frac{dx^\mu }{d\lambda}
  \frac{dx^\nu }{d\lambda}
  \langle T_{\mu \nu} \rangle_0=a_0
  \frac{dx^i}{d\lambda}
  \frac{dx^j}{d\lambda}g_{ij}=0\ .$$
Notice that as soon as we consider a null geodesic that also moves in the $S^{d-2}$ direction, this property is no longer true, since $a_0\neq b_0$ in general.

\section{Modular flow for arbitrary wedge}
\label{zapp:3}

\begin{figure}
\centering
\includegraphics[scale=0.35]{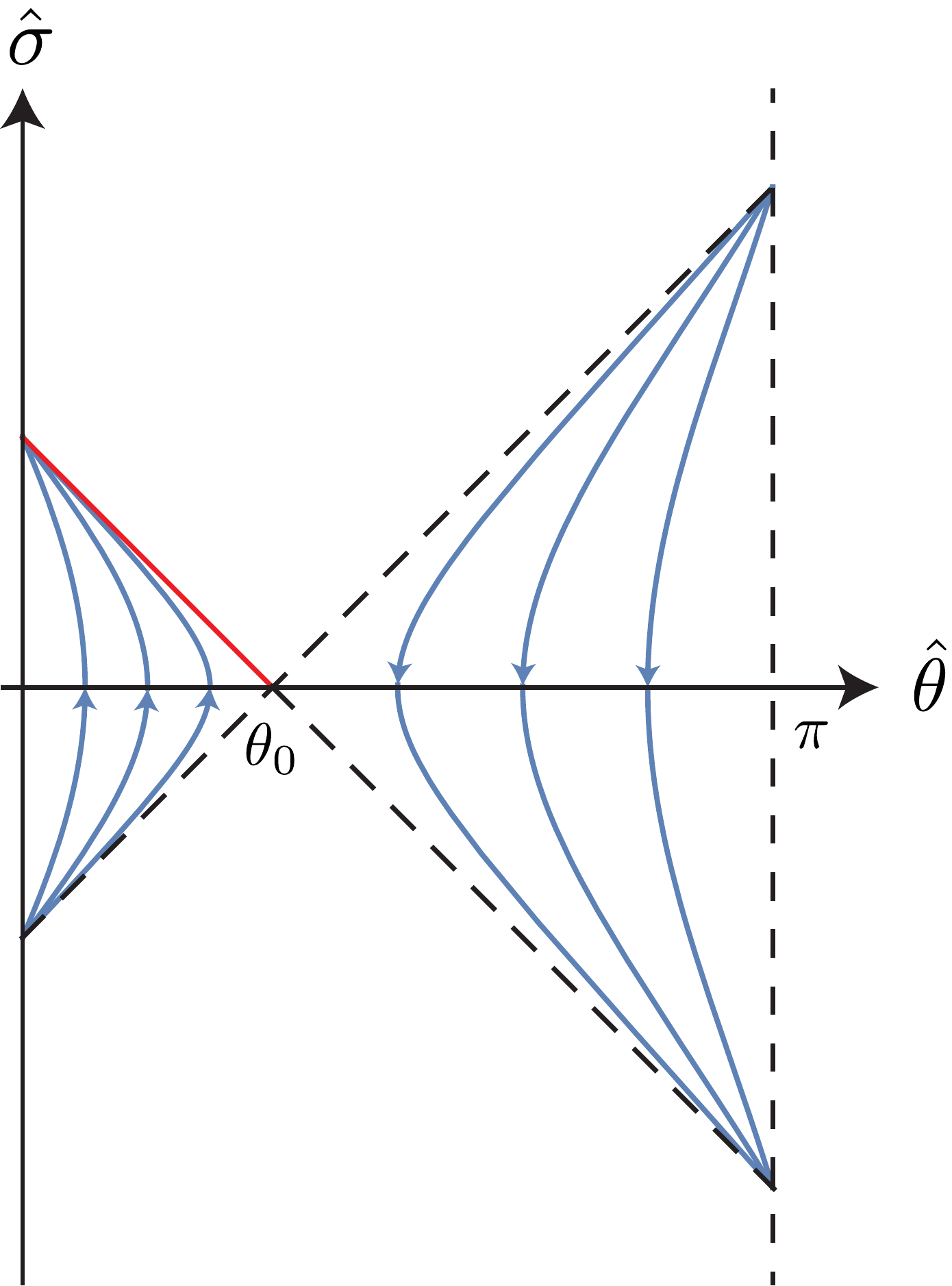}
\caption{Modular flow for a wedge in AdS$_2\times S^{d-2}$ with $\theta_0\neq \pi/2$ obtained from (\ref{eq:75}). In the left wedge $\theta<\theta_0$ we indicate its future horizon in red, described by (\ref{eq:113}).}\label{fig:5}
\end{figure}

In this appendix we generalize the computation in subsection \ref{sub:2.2} of the modular flow of a wedge of size $\pi/2$ in ${\rm AdS}_2\times S^{d-2}$ (\ref{eq:88}), to a wedge of arbitrary size $\theta_0\in(0,\pi)$. We do this by using an isometry of AdS$_2$. Since this it is a maximally symmetric space-time, it has three independent Killing vectors: rigid time translations in $\sigma$, the transformation given in~(\ref{eq:71}) and an additional one, which can be easily found from the embedding description and compactly written in terms of the null coordinates as
$$
\tan(\hat{\theta}_\pm)=
\frac{\sin(\theta_\pm)\sin(\theta_0)}
{\cos(\theta_\pm)+\cos(\theta_0)}\ ,
  \qquad {\rm with\,\,inverse} \qquad
  \tan(\theta_\pm)=
  \frac{\sin(\hat{\theta}_\pm)
  \sin(\theta_0)}
  {\cos(\hat{\theta}_\pm)-\cos(\theta_0)}\ .
$$
This transformation maps the wedge $\mathcal{D}(A_0)$ of size $\pi/2$ to an arbitrary wedge $\mathcal{D}(\hat{A}_0)$ of size~$\theta_0$ in the $\hat{\theta}_\pm$ coordinates. We can use this to map the modular flow of the wedge of size $\pi/2$ in~(\ref{eq:71}) to arbitrary $\theta_0\in(0,\pi)$. In practice it is much simpler to first map the Killing vector in (\ref{eq:74})
\begin{equation}\label{eq:76}
\xi^\mu=
  \frac{\cos(\hat{\theta}_+)-
  \cos(\theta_0)}{\sin(\theta_0)}
  \hat{\partial}_+-
  \frac{\cos(\hat{\theta}_-)-
  \cos(\theta_0)}{\sin(\theta_0)}
  \hat{\partial}_-\ ,
\end{equation}
and then solve for its integral curves, that are given by
\begin{equation}\label{eq:75}
\frac{\tan(\theta_\pm(s)/2)}
  {\tan(\theta_0/2)}=
  \frac{\sin
  \Big(
  \frac{
  \hat{\theta}_\pm+\theta_0}{2}
  \Big)
  +e^{\mp s}\sin\left(
  \frac{\hat{\theta}_\pm -\theta_0}{2}
  \right)}
  {\sin\left(
  \frac{\hat{\theta}_\pm +\theta_0}{2}\right)
  -e^{\mp s}\sin\left(
  \frac{\hat{\theta}_\pm-\theta_0}{2}
  \right)}
  \ ,
\end{equation}
with $s\in \mathbb{R}$. In figure \ref{fig:5} we plot the trajectories $\hat{\theta}_\pm(s)$ and see that they correspond to the modular flow associated to a wedge of arbitrary size $\theta_0$.

Using the Killing vector in (\ref{eq:76}), we can easily write the modular hamiltonian using (\ref{eq:73}). We choose the Cauchy surface $\Sigma$ as the future null horizon of $\mathcal{D}(\hat{A}_0)$ (marked in red in figure~\ref{fig:5}), that is described by an affine parameter $\lambda$ according to
\begin{equation}\label{eq:113}
x^\mu(\lambda,\vec{x}_\perp)=
  (\hat{\theta}_+,
  \hat{\theta}_-,
  \vec{v}\,)=
  \left(
  \theta_0,
  2\,{\rm arccot}(\lambda)-\theta_0
  ,\vec{x}_\perp
  \right)
  \ ,\quad
  \qquad
  \lambda\in [\lambda_0,+\infty)
  \ ,
\end{equation}
with $\lambda_0=\cot(\theta_0)$. Using this in (\ref{eq:73}), the modular hamiltonian can be written as
$$
K_{\hat{A}_0}=
  2\pi r_h^{d-2}
  \int_{S^{d-2}}
  d\Omega(\vec{x}_\perp)
  \int_{\lambda_0}^{+\infty}
  d\lambda\,
  (\lambda-\lambda_0)
  T_{\lambda \lambda}(\lambda,\vec{x}_\perp)+{\rm const}\ .
$$
This generalizes (\ref{eq:107}) to a wedge of arbitrary size $\theta_0\in(0,\pi)$.

\bibliography{sample}

\providecommand{\href}[2]{#2}\begingroup\raggedright\begin{thebibliography}{10}

\bibitem{Fewster:2012yh}
C.~J. Fewster, {\it {Lectures on quantum energy inequalities}},
  \href{http://arxiv.org/abs/1208.5399}{{\tt arXiv:1208.5399}}.

\bibitem{Hawking:1971vc}
S.~Hawking, {\it {Black holes in general relativity}},  {\em Commun. Math.
  Phys.} {\bf 25} (1972) 152--166.

\bibitem{Penrose:1964wq}
R.~Penrose, {\it {Gravitational collapse and space-time singularities}},  {\em
  Phys. Rev. Lett.} {\bf 14} (1965) 57--59.

\bibitem{Witten:2019qhl}
E.~Witten, {\it {Light Rays, Singularities, and All That}},
  \href{http://arxiv.org/abs/1901.03928}{{\tt arXiv:1901.03928}}.

\bibitem{Epstein:1965zza}
H.~Epstein, V.~Glaser, and A.~Jaffe, {\it {Nonpositivity of energy density in
  Quantized field theories}},  {\em Nuovo Cim.} {\bf 36} (1965) 1016.

\bibitem{Cordova:2017zej}
C.~Cordova, J.~Maldacena, and G.~J. Turiaci, {\it {Bounds on OPE Coefficients
  from Interference Effects in the Conformal Collider}},  {\em JHEP} {\bf 11}
  (2017) 032, [\href{http://arxiv.org/abs/1710.03199}{{\tt arXiv:1710.03199}}].

\bibitem{Hawking:1974sw}
S.~Hawking, {\it {Particle Creation by Black Holes}},  {\em Commun. Math.
  Phys.} {\bf 43} (1975) 199--220. [Erratum: Commun.Math.Phys. 46, 206 (1976)].

\bibitem{Hawking:1976ra}
S.~Hawking, {\it {Breakdown of Predictability in Gravitational Collapse}},
  {\em Phys. Rev. D} {\bf 14} (1976) 2460--2473.

\bibitem{Mathur:2009hf}
S.~D. Mathur, {\it {The Information paradox: A Pedagogical introduction}},
  {\em Class. Quant. Grav.} {\bf 26} (2009) 224001,
  [\href{http://arxiv.org/abs/0909.1038}{{\tt arXiv:0909.1038}}].

\bibitem{Bousso:2015mna}
R.~Bousso, Z.~Fisher, S.~Leichenauer, and A.~C. Wall, {\it {Quantum focusing
  conjecture}},  {\em Phys. Rev.} {\bf D93} (2016), no.~6 064044,
  [\href{http://arxiv.org/abs/1506.02669}{{\tt arXiv:1506.02669}}].

\bibitem{Koeller:2015qmn}
J.~Koeller and S.~Leichenauer, {\it {Holographic Proof of the Quantum Null
  Energy Condition}},  {\em Phys. Rev.} {\bf D94} (2016), no.~2 024026,
  [\href{http://arxiv.org/abs/1512.06109}{{\tt arXiv:1512.06109}}].

\bibitem{Balakrishnan:2017bjg}
S.~Balakrishnan, T.~Faulkner, Z.~U. Khandker, and H.~Wang, {\it {A General
  Proof of the Quantum Null Energy Condition}},
  \href{http://arxiv.org/abs/1706.09432}{{\tt arXiv:1706.09432}}.

\bibitem{Ceyhan:2018zfg}
F.~Ceyhan and T.~Faulkner, {\it {Recovering the QNEC from the ANEC}},
  \href{http://arxiv.org/abs/1812.04683}{{\tt arXiv:1812.04683}}.

\bibitem{Leichenauer:2018obf}
S.~Leichenauer, A.~Levine, and A.~Shahbazi-Moghaddam, {\it {Energy density from
  second shape variations of the von Neumann entropy}},  {\em Phys. Rev.} {\bf
  D98} (2018), no.~8 086013, [\href{http://arxiv.org/abs/1802.02584}{{\tt
  arXiv:1802.02584}}].

\bibitem{Balakrishnan:2019gxl}
S.~Balakrishnan, V.~Chandrasekaran, T.~Faulkner, A.~Levine, and
  A.~Shahbazi-Moghaddam, {\it {Entropy Variations and Light Ray Operators from
  Replica Defects}},  \href{http://arxiv.org/abs/1906.08274}{{\tt
  arXiv:1906.08274}}.

\bibitem{Graham:2007va}
N.~Graham and K.~D. Olum, {\it {Achronal averaged null energy condition}},
  {\em Phys. Rev.} {\bf D76} (2007) 064001,
  [\href{http://arxiv.org/abs/0705.3193}{{\tt arXiv:0705.3193}}].

\bibitem{Faulkner:2016mzt}
T.~Faulkner, R.~G. Leigh, O.~Parrikar, and H.~Wang, {\it {Modular Hamiltonians
  for Deformed Half-Spaces and the Averaged Null Energy Condition}},  {\em
  JHEP} {\bf 09} (2016) 038, [\href{http://arxiv.org/abs/1605.08072}{{\tt
  arXiv:1605.08072}}].

\bibitem{Hartman:2016lgu}
T.~Hartman, S.~Kundu, and A.~Tajdini, {\it {Averaged Null Energy Condition from
  Causality}},  {\em JHEP} {\bf 07} (2017) 066,
  [\href{http://arxiv.org/abs/1610.05308}{{\tt arXiv:1610.05308}}].

\bibitem{Longo:2018obd}
R.~Longo, {\it {Entropy distribution of localised states}},  {\em Commun. Math.
  Phys.} {\bf 373} (2019), no.~2 473--505,
  [\href{http://arxiv.org/abs/1809.03358}{{\tt arXiv:1809.03358}}].

\bibitem{Wald:1991xn}
R.~M. Wald and U.~Yurtsever, {\it {General proof of the averaged null energy
  condition for a massless scalar field in two-dimensional curved space-time}},
   {\em Phys. Rev. D} {\bf 44} (1991) 403--416.

\bibitem{Fewster:2006uf}
C.~J. Fewster, K.~D. Olum, and M.~J. Pfenning, {\it {Averaged null energy
  condition in spacetimes with boundaries}},  {\em Phys. Rev.} {\bf D75} (2007)
  025007, [\href{http://arxiv.org/abs/gr-qc/0609007}{{\tt gr-qc/0609007}}].

\bibitem{Wall:2009wi}
A.~C. Wall, {\it {Proving the Achronal Averaged Null Energy Condition from the
  Generalized Second Law}},  {\em Phys. Rev.} {\bf D81} (2010) 024038,
  [\href{http://arxiv.org/abs/0910.5751}{{\tt arXiv:0910.5751}}].

\bibitem{Wall:2011hj}
A.~C. Wall, {\it {A proof of the generalized second law for rapidly changing
  fields and arbitrary horizon slices}},  {\em Phys. Rev.} {\bf D85} (2012)
  104049, [\href{http://arxiv.org/abs/1105.3445}{{\tt arXiv:1105.3445}}].
  [erratum: Phys. Rev.D87,no.6,069904(2013)].

\bibitem{Kontou:2012ve}
E.-A. Kontou and K.~D. Olum, {\it {Averaged null energy condition in a
  classical curved background}},  {\em Phys. Rev.} {\bf D87} (2013), no.~6
  064009, [\href{http://arxiv.org/abs/1212.2290}{{\tt arXiv:1212.2290}}].

\bibitem{Kontou:2015yha}
E.-A. Kontou and K.~D. Olum, {\it {Proof of the averaged null energy condition
  in a classical curved spacetime using a null-projected quantum inequality}},
  {\em Phys. Rev.} {\bf D92} (2015) 124009,
  [\href{http://arxiv.org/abs/1507.00297}{{\tt arXiv:1507.00297}}].

\bibitem{Rosso:2019txh}
F.~Rosso, {\it {Global aspects of conformal symmetry and the ANEC in dS and
  AdS}},  \href{http://arxiv.org/abs/1912.08897}{{\tt arXiv:1912.08897}}.

\bibitem{Freivogel:2018gxj}
B.~Freivogel and D.~Krommydas, {\it {The Smeared Null Energy Condition}},  {\em
  JHEP} {\bf 12} (2018) 067, [\href{http://arxiv.org/abs/1807.03808}{{\tt
  arXiv:1807.03808}}].

\bibitem{Leichenauer:2018tnq}
S.~Leichenauer and A.~Levine, {\it {Upper and Lower Bounds on the Integrated
  Null Energy in Gravity}},  {\em JHEP} {\bf 01} (2019) 133,
  [\href{http://arxiv.org/abs/1808.09970}{{\tt arXiv:1808.09970}}].

\bibitem{Friedman:1993ty}
J.~L. Friedman, K.~Schleich, and D.~M. Witt, {\it {Topological censorship}},
  {\em Phys. Rev. Lett.} {\bf 71} (1993) 1486--1489,
  [\href{http://arxiv.org/abs/gr-qc/9305017}{{\tt gr-qc/9305017}}]. [Erratum:
  Phys.Rev.Lett. 75, 1872 (1995)].

\bibitem{Hofman:2008ar}
D.~M. Hofman and J.~Maldacena, {\it {Conformal collider physics: Energy and
  charge correlations}},  {\em JHEP} {\bf 05} (2008) 012,
  [\href{http://arxiv.org/abs/0803.1467}{{\tt arXiv:0803.1467}}].

\bibitem{Tipler:1978zz}
F.~J. Tipler, {\it {Energy conditions and spacetime singularities}},  {\em
  Phys. Rev.} {\bf D17} (1978) 2521--2528.

\bibitem{Borde:1987qr}
A.~Borde, {\it {Geodesic focusing, energy conditions and singularities}},  {\em
  Class. Quant. Grav.} {\bf 4} (1987) 343--356.

\bibitem{Hofman:2009ug}
D.~M. Hofman, {\it {Higher Derivative Gravity, Causality and Positivity of
  Energy in a UV complete QFT}},  {\em Nucl. Phys. B} {\bf 823} (2009)
  174--194, [\href{http://arxiv.org/abs/0907.1625}{{\tt arXiv:0907.1625}}].

\bibitem{Chowdhury:2012km}
D.~Chowdhury, S.~Raju, S.~Sachdev, A.~Singh, and P.~Strack, {\it {Multipoint
  correlators of conformal field theories: implications for quantum critical
  transport}},  {\em Phys. Rev. B} {\bf 87} (2013), no.~8 085138,
  [\href{http://arxiv.org/abs/1210.5247}{{\tt arXiv:1210.5247}}].

\bibitem{Cordova:2017dhq}
C.~Cordova and K.~Diab, {\it {Universal Bounds on Operator Dimensions from the
  Average Null Energy Condition}},  {\em JHEP} {\bf 02} (2018) 131,
  [\href{http://arxiv.org/abs/1712.01089}{{\tt arXiv:1712.01089}}].

\bibitem{Delacretaz:2018cfk}
L.~V. Delacrétaz, T.~Hartman, S.~A. Hartnoll, and A.~Lewkowycz, {\it
  {Thermalization, Viscosity and the Averaged Null Energy Condition}},  {\em
  JHEP} {\bf 10} (2018) 028, [\href{http://arxiv.org/abs/1805.04194}{{\tt
  arXiv:1805.04194}}].

\bibitem{Belin:2019mnx}
A.~Belin, D.~M. Hofman, and G.~Mathys, {\it {Einstein gravity from ANEC
  correlators}},  {\em JHEP} {\bf 08} (2019) 032,
  [\href{http://arxiv.org/abs/1904.05892}{{\tt arXiv:1904.05892}}].

\bibitem{Maldacena:2018gjk}
J.~Maldacena, A.~Milekhin, and F.~Popov, {\it {Traversable wormholes in four
  dimensions}},  \href{http://arxiv.org/abs/1807.04726}{{\tt
  arXiv:1807.04726}}.

\bibitem{Blanco:2013lea}
D.~D. Blanco and H.~Casini, {\it {Localization of Negative Energy and the
  Bekenstein Bound}},  {\em Phys. Rev. Lett.} {\bf 111} (2013), no.~22 221601,
  [\href{http://arxiv.org/abs/1309.1121}{{\tt arXiv:1309.1121}}].

\bibitem{Casini:2017roe}
H.~Casini, E.~Teste, and G.~Torroba, {\it {Modular Hamiltonians on the null
  plane and the Markov property of the vacuum state}},  {\em J. Phys.} {\bf
  A50} (2017), no.~36 364001, [\href{http://arxiv.org/abs/1703.10656}{{\tt
  arXiv:1703.10656}}].

\bibitem{Balakrishnan:2020lbp}
S.~Balakrishnan and O.~Parrikar, {\it {Modular Hamiltonians for Euclidean Path
  Integral States}},  \href{http://arxiv.org/abs/2002.00018}{{\tt
  arXiv:2002.00018}}.

\bibitem{Galloway:2020xfz}
G.~J. Galloway, M.~Graf, and E.~Ling, {\it {A conformal infinity approach to
  asymptotically $\text{AdS}_2\times S^{n-1}$ spacetimes}},
  \href{http://arxiv.org/abs/2003.00093}{{\tt arXiv:2003.00093}}.

\bibitem{Iizuka:2019ezn}
N.~Iizuka, A.~Ishibashi, and K.~Maeda, {\it {Conformally invariant averaged
  null energy condition from AdS/CFT}},
  \href{http://arxiv.org/abs/1911.02654}{{\tt arXiv:1911.02654}}.

\bibitem{Myers:1986un}
R.~C. Myers and M.~Perry, {\it {Black Holes in Higher Dimensional
  Space-Times}},  {\em Annals Phys.} {\bf 172} (1986) 304.

\bibitem{Spradlin:1999bn}
M.~Spradlin and A.~Strominger, {\it {Vacuum states for AdS(2) black holes}},
  {\em JHEP} {\bf 11} (1999) 021,
  [\href{http://arxiv.org/abs/hep-th/9904143}{{\tt hep-th/9904143}}].

\bibitem{Witten:2018lha}
E.~Witten, {\it {APS Medal for Exceptional Achievement in Research: Invited
  article on entanglement properties of quantum field theory}},  {\em Rev. Mod.
  Phys.} {\bf 90} (2018), no.~4 045003,
  [\href{http://arxiv.org/abs/1803.04993}{{\tt arXiv:1803.04993}}].

\bibitem{Nishioka:2018khk}
T.~Nishioka, {\it {Entanglement entropy: holography and renormalization
  group}},  {\em Rev. Mod. Phys.} {\bf 90} (2018), no.~3 035007,
  [\href{http://arxiv.org/abs/1801.10352}{{\tt arXiv:1801.10352}}].

\bibitem{Rangamani:2016dms}
M.~Rangamani and T.~Takayanagi, {\it {Holographic Entanglement Entropy}},  {\em
  Lect. Notes Phys.} {\bf 931} (2017) pp.1--246,
  [\href{http://arxiv.org/abs/1609.01287}{{\tt arXiv:1609.01287}}].

\bibitem{Banerjee:2011mg}
S.~Banerjee, {\it {Wess-Zumino Consistency Condition for Entanglement
  Entropy}},  {\em Phys. Rev. Lett.} {\bf 109} (2012) 010402,
  [\href{http://arxiv.org/abs/1109.5672}{{\tt arXiv:1109.5672}}].

\bibitem{Faulkner:2015csl}
T.~Faulkner, R.~G. Leigh, and O.~Parrikar, {\it {Shape Dependence of
  Entanglement Entropy in Conformal Field Theories}},  {\em JHEP} {\bf 04}
  (2016) 088, [\href{http://arxiv.org/abs/1511.05179}{{\tt arXiv:1511.05179}}].

\bibitem{CBHD}
A.~Bonfiglioli and R.~Fulci, {\em Topics in Noncommutative Algebra: The Theorem
  of Campbell, Baker, Hausdorff and Dynkin}.
\newblock Springer-Verlag Berlin Heidelberg, 2012.

\bibitem{Milgram_2013}
M.~S. Milgram, {\it Integral and series representations of riemann’s zeta
  function and dirichlet’s eta function and a medley of related results},
  {\em Journal of Mathematics} {\bf 2013} (2013) 1–17.

\bibitem{Klinkhammer:1991ki}
G.~Klinkhammer, {\it {Averaged energy conditions for free scalar fields in flat
  space-times}},  {\em Phys. Rev.} {\bf D43} (1991) 2542--2548.

\bibitem{Herzog:2013ed}
C.~P. Herzog and K.-W. Huang, {\it {Stress Tensors from Trace Anomalies in
  Conformal Field Theories}},  {\em Phys. Rev.} {\bf D87} (2013) 081901,
  [\href{http://arxiv.org/abs/1301.5002}{{\tt arXiv:1301.5002}}].

\bibitem{inproceedings}
A.~Sadighi, R.~Chavosh~Khatamy, and M.~Toomanian, {\it On the product of
  symmetric riemannian manifolds},  07, 2017.

\end{thebibliography}\endgroup
\bibliographystyle{JHEP}

\end{document}